\documentclass[superscriptaddress,twocolumn,10pt,prx,aps,showpacs,longbibliography,floatfix]{revtex4-2}

\usepackage[dvipsnames]{xcolor}

\definecolor{darkred}{rgb}{0.6,0.05,0.05}
\definecolor{darkgreen}{rgb}{0.05,0.6,0.05}
\definecolor{darkblue}{rgb}{0.05,0.05,0.6}
\usepackage[colorlinks]{hyperref}
\hypersetup{colorlinks,
linkcolor=darkred,
filecolor=darkgreen,
urlcolor=darkblue,
citecolor=darkgreen}

\usepackage{mathtools}
\usepackage{amsmath}
\allowdisplaybreaks
\usepackage{ bbold }
\usepackage{amssymb}
\usepackage[normalem]{ulem}
\usepackage{amsthm}
\usepackage{xfrac}
\usepackage{dsfont}
\usepackage{siunitx}
\usepackage{physics}
\usepackage{graphicx}
\usepackage{hyperref}
\usepackage[capitalise]{cleveref}
\usepackage{dcolumn}
\usepackage{bm}
\usepackage{layouts}
\usepackage{comment}

\newcommand{\LL}{\mathcal{L}}

\begin{document}

\author{Filippo Ferrari}
\affiliation{Laboratory of Theoretical Physics of Nanosystems (LTPN), Institute of Physics, \'{E}cole Polytechnique F\'{e}d\'{e}rale de Lausanne (EPFL), 1015 Lausanne, Switzerland}
\affiliation{Center for Quantum Science and Engineering, \\ \'{E}cole Polytechnique F\'{e}d\'{e}rale de Lausanne (EPFL), CH-1015 Lausanne, Switzerland}
\author{Luca Gravina}
\affiliation{Laboratory of Theoretical Physics of Nanosystems (LTPN), Institute of Physics, \'{E}cole Polytechnique F\'{e}d\'{e}rale de Lausanne (EPFL), 1015 Lausanne, Switzerland}
\affiliation{Center for Quantum Science and Engineering, \\ \'{E}cole Polytechnique F\'{e}d\'{e}rale de Lausanne (EPFL), CH-1015 Lausanne, Switzerland}
\author{Debbie Eeltink}
\affiliation{Laboratory of Theoretical Physics of Nanosystems (LTPN), Institute of Physics, \'{E}cole Polytechnique F\'{e}d\'{e}rale de Lausanne (EPFL), 1015 Lausanne, Switzerland}
\affiliation{Center for Quantum Science and Engineering, \\ \'{E}cole Polytechnique F\'{e}d\'{e}rale de Lausanne (EPFL), CH-1015 Lausanne, Switzerland}
\author{Pasquale Scarlino}
\affiliation{Center for Quantum Science and Engineering, \\ \'{E}cole Polytechnique F\'{e}d\'{e}rale de Lausanne (EPFL), CH-1015 Lausanne, Switzerland}
\affiliation{Hybrid Quantum Circuits Laboratory (HQC), Institute of Physics, \'{E}cole Polytechnique F\'{e}d\'{e}rale de Lausanne (EPFL), 1015 Lausanne, Switzerland}
\author{Vincenzo Savona}
\email[E-mail: ]{vincenzo.savona@epfl.ch}
\affiliation{Laboratory of Theoretical Physics of Nanosystems (LTPN), Institute of Physics, \'{E}cole Polytechnique F\'{e}d\'{e}rale de Lausanne (EPFL), 1015 Lausanne, Switzerland}
\affiliation{Center for Quantum Science and Engineering, \\ \'{E}cole Polytechnique F\'{e}d\'{e}rale de Lausanne (EPFL), CH-1015 Lausanne, Switzerland}
\author{Fabrizio Minganti}
\email[E-mail: ]{fabrizio.minganti@epfl.ch}
\affiliation{Laboratory of Theoretical Physics of Nanosystems (LTPN), Institute of Physics, \'{E}cole Polytechnique F\'{e}d\'{e}rale de Lausanne (EPFL), 1015 Lausanne, Switzerland}
\affiliation{Center for Quantum Science and Engineering, \\ \'{E}cole Polytechnique F\'{e}d\'{e}rale de Lausanne (EPFL), CH-1015 Lausanne, Switzerland}

\title{Dissipative Quantum Chaos unveiled by Stochastic Quantum Trajectories
}

\date{\today}

\begin{abstract}
We define quantum chaos and integrability in open quantum many-body systems as a dynamical property of single stochastic realizations, referred to as quantum trajectories.
This definition relies on the predictions of random matrix theory applied to the subset of the Liouvillian eigenspectrum involved in each quantum trajectory.
Our approach, which we name \textit{spectral statistics of quantum trajectories} (SSQT), enables a natural distinction between transient and steady-state quantum chaos as general phenomena in  open setups.
We test the generality and reliability of the SSQT criterion on several dissipative systems, further showing that an open system with a chaotic structure can evolve towards either a chaotic or integrable steady state. 
We apply our theoretical framework to two driven-dissipative bosonic systems. 
First, we study the driven-dissipative Bose-Hubbard model, a paradigmatic example of a quantum simulator, clarifying the interplay of integrability, transient, and steady-state chaos across its phase diagram. 
Our analysis shows the existence of an emergent dissipative quantum chaotic phase, whereas the classical and semi-classical limits display an integrable behavior. 
In this regime, chaos arises from the quantum and classical fluctuations associated with the dissipation mechanisms.
Second, we investigate dissipative quantum chaos in the dispersive readout of a transmon qubit: a measurement technique ubiquitous in superconducting-based quantum hardware.
Through the SSQT, we distinguish several regimes where the performance of the measurement instrument can be connected to the integrable or chaotic nature of the underlying driven-dissipative bosonic system.
Our work offers a general understanding of the integrable and chaotic dynamics of open quantum systems and paves the way for the investigation of dissipative quantum chaos and its consequences on state-of-the-art noisy intermediate-scale quantum devices.
\end{abstract}

\maketitle

\tableofcontents 

\section{Introduction} \label{sec:introduction}

The study of chaos and integrability in open quantum many-body systems is central in many research areas, from high-energy physics to quantum optics, from quantum technologies to condensed matter. 
In an open quantum system, the degrees of freedom of the system are coupled to those of the surrounding environment, resulting in an effective dynamics that departs significantly from the paradigm of unitary quantum mechanics \cite{breuer_theory_2007, rivas_open_2012}.
The properties of open quantum systems have long been used for the preparation and stabilization of quantum states, with both fundamental \cite{witthaut_2008, diehl_quantum_2008, polkovnikov_colloquium_2011, qisert_quantum_2015} and technological \cite{Verstraete_quantum_2009, devoret_confining2015, touzard_2018, chamberland_building2022, lescanne_exponential_2020} purposes.
Progress in this field includes the theoretical and experimental study of dissipative phase transitions \cite{carmichael_breakdown_2015, bartolo_exact_2016, macieszczak_towards_2016, fink_observation_2017, fitzpatrick_observation_2017, biondi_nonequilibrium_2017, minganti_spectral_2018, fink_signatures_2018, di_candia_critical_2023, gravina_critical2023}, the development of protocols for error correction and suppression \cite{grimsmo_quantum_2021, hillmann_performance_2022, Lidar_NS1998, Knill_EC2000, Albert_Geometry2016, Lidar_NSBook2014, Kempe_NS2001}, the stabilization and control of entangled, topological and localized phases \cite{gong_topological_2018, kawabata_symmetry_2019, hamazaki_non-hermitian_2019, ippoliti_entanglement_2021, hamazaki_lindbladian_2022, kawabata_entanglement_2023}, the characterization of dissipative quantum chaos (DQC) \cite{akemann_universal_2019, denisov_universal_2019, can_spectral_2019, hamazaki_universality_2020, sa_complex_2020, sa_spectral_2020, Sa_2020, li_spectral_2021, dahan_classical_2022, rubio-garcia_integrability_2022, prasad_dissipative_2022, garcia-garcia_symmetry_2022, costa_spectral2023, sa_symmetry_2023, kawabata_symmetry2023, kawabata_singular2023, matsoukas-roubeas_quantum_2024}. 

Chaos in classical systems is understood as the sensitivity of the dynamics on the choice of initial conditions, and has been characterized through several quantifiers such as a positive Lyapunov exponent \cite{strogatz_nonlinear_2018}. 
Quantum chaos, both in isolated and open systems, is usually understood within the framework of the \textit{quantum chaos conjecture} \cite{bohigas_characterization_1984, haake_quantum_2001, dalessio_quantum_2016}.
For open systems, the original Grobe-Haake-Sommer (GHS) conjecture \cite{grobe_quantum_1988} states: ``the distribution of the smallest [eigenvalue] distances display linear and cubic repulsion under the conditions of classically regular and chaotic motion, respectively".
That is, given the hypothesis that (a) the system admits a meaningful classical limit, and this limit exhibits classical chaos, then one conjectures that (b) the spectral properties of the time evolution generator match the universal predictions of random matrix theory.
In the absence of (a), however, the predictions of random matrix theory are still used as a criterion to define quantum chaos \cite{kos_manybody2018} due to their phenomenological success in predicting the properties of quantum systems lacking a meaningful classical counterpart \cite{ponte_manybody_2015, serbyn_spectral_2016, bordia_periodically_2017, abanin_colloquium_2019}.
Following this, general criteria for probing DQC have been developed \cite{sa_complex_2020}, and symmetry-based systematic classifications of DQC have been elaborated \cite{hamazaki_universality_2020, garcia-garcia_symmetry_2022, kawabata_symmetry2023, sa_symmetry_2023}.

As we show below, however, a definition like (b) based uniquely on the spectral properties of the evolution generator is likely to fail in predicting the nature of the dynamics.
In closed quantum systems, for example, the same Hamiltonian may generate either regular or chaotic motion depending on the initial condition.
In open quantum systems, ever since the introduction of the GHS conjecture, chaos has been treated as an inherently  state-dependent \textit{and} transient feature.
Dissipative dynamics irreversibly leads to an often unique steady state ~\footnote{If the Hilbert space is finite and the Liouvillian superoperator is time-independent, the existence of at least one steady state is guaranteed \cite{rivas_open_2012}. If the system is not invariant under any strong Liouvillian symmetry, the steady state is also unique \cite{albert_symmetries_2014}. In this work, we consider open quantum systems admitting a unique steady state.} 
describing the long-time physical properties of the system. 
Thus, the definition of chaos provided by Ref.~\cite{grobe_quantum_1988} addresses the transient dynamics but \textit{not} the steady state.

Understanding the chaotic or integrable nature of a state that does not evolve in time seems indeed contradictory.
As we will show, however, such a question is not only well-posed, but strictly related to fundamental properties of quantum hardware.

\subsection{A criterion to describe chaos in open quantum systems}

Current approaches to DQC can provide ambiguous answers in defining the transient dynamics and the steady state of open quantum systems as regular or chaotic.
A rigorous characterization of DQC remains a major theoretical challenge.
The first result of this work is to provide an operational definition of quantum chaos in open systems by considering the density matrix as the average over a statistical ensemble of single dynamical realizations.
Specifically, we introduce the \textit{spectral statistics of quantum trajectories} (SSQT): 
a criterion applying the universal predictions of random matrix theory to the stochastic dynamics of single quantum trajectories \cite{wiseman_quantum_2009, jacobs_2014, daley_quantum_2014, weimer_simulation_2021}.

The SSQT leads to a rigorous and model-independent definition of DQC, and can unambiguously identify chaotic or integrable dynamics both in the transient and in the steady state of an open quantum system. 
When applied to the transient dynamics, the characterization resulting from the SSQT generalizes those in Refs. \cite{akemann_universal_2019, sa_complex_2020, garcia-garcia_symmetry_2022, kawabata_symmetry2023},
correctly reproducing the dependence of the chaotic behavior on the choice of the initial state.
Moreover, the SSQT defines chaos and integrability also in the steady state of an open system. Differently from Hamiltonian or transient dynamics, the information about the initial condition is lost in the steady
state, and the presence of chaos depends only on the structure of the time evolution generator. 
Our criterion thus resolves the ambiguities discussed above and offers a general understanding of the chaotic and integrable properties of Lindblad dynamics.

\subsection{Dissipative quantum chaos in bosonic-based architectures}

The second result of this work is to investigate quantum chaos in coupled driven-dissipative nonlinear bosonic oscillators.
These systems are the focus of intense investigations, as they provide a realistic model for several quantum technological platforms \cite{Nori_quantum2014, lehur_manybody_2016, altman_quantum_2021}, such as superconducting circuits \cite{krantz_quantum_2019, blais_circuit_2021}, semiconductor artificial structures \cite{burkard_superconductor_2020, garcia_de_arquer_semiconductor_2021}, optomechanical resonators \cite{aspelmeyer_cavity_2014}, and trapped ions \cite{muller_engineered_2012, bruzewicz_trapped-ion_2019}. These systems have been theoretically and experimentally designed to
perform quantum computing tasks \cite{mirrahimi_dynamically_2014, devoret_confining2015, puri_engineering_2017, touzard_2018, puri_stabilized_2019, lescanne_exponential_2020, gautier_2022, chamberland_building2022, gravina_critical2023} and function as sensing devices \cite{heugel_quantum2019, di_candia_critical_2023}.
It is therefore of paramount importance to understand and characterize quantum chaos while looking for effects that may be detrimental to the coherent manipulation and storage of quantum information.
As we discuss, the application of bare spectral methods to these systems leads to ambiguous results in the prediction of the occurrence of chaos. 
When applying the SSQT these ambiguities are relieved, allowing to straightforwardly predict the manifestation and effects of quantum chaos.

We apply the SSQT to two nonlinear bosonic models.
One is a Bose-Hubbard dimer, i.e., the building block of chains of coupled nonlinear bosonic resonators. 
We demonstrate how chaos emerges and modifies the properties of this quantum simulator.
The second system we consider is the circuit quantum electrodynamics setup enabling the dispersive readout of a superconducting transmon qubit, a problem of central importance towards the development of fast, reliable, and error-correctable quantum hardware.
We demonstrate that the application of the SSQT is essential for accurately understanding the different operational regimes of the readout. Furthermore, we illustrate how different features of DQC can either enhance or diminish the performance of the quantum device.

\subsection{Breakdown of the quantum-to-classical conjecture}

The third result of the paper concerns the breakdown of the quantum-to-classical correspondence.
As explained above, while a correspondence between classical and quantum chaos is expected \cite{grobe_quantum_1988}, our study provides an explicit example of an emergent DQC phase whose classical counterpart is integrable.

For the driven-dissipative Bose-Hubbard model, we find a large portion of parameter space whose steady state is chaotic, while both the classical and semiclassical limits predict integrable dynamics.
We show that the emergence of this quantum but not classical chaos coincides with sub-Poissonian fluctuations of the bosonic field, which do not admit a classical counterpart \cite{scully_quantum_1997}.
We attribute this lack of correspondence to the quantum nature of dissipation.

\subsection{Structure of the paper}

The paper is organized as follows. 
In Sec.~\ref{sec:motivation} we lay out the motivation of this work and present an overview of the main results. 
In Sec.~\ref{sec:SSQT}  we formalize the SSQT criterion.
In Sec.~\ref{sec:quantum_simulation} we present a numerical investigation of DQC in the driven-dissipative Bose-Hubbard model.
In Sec.~\ref{sec:quantum_information} we present a study on the effects of DQC in superconducting-based quantum technologies.
In Sec.~\ref{sec:conclusion} we draw the conclusions and discuss the outlook of the work. 
A technical overview of the methods is presented in Appendix~\ref{sec:appendix}. A comparison between the SSQT and other criteria is analyzed on several Liouvillian models in Appendix~\ref{Sec:Comparison}. A study of the driven Bose-Hubbard Hamiltonian is realized in Appendix~\ref{sec:Ham}. A discussion on superconducting circuits is presented in Appendices~\ref{sec:Transmon_appendix} and \ref{sec:Readout_appendix}.

\section{Motivation}
\label{sec:motivation}

Open Markovian systems are governed by a (Gorini-Kossakowski-Sudarshan)-Lindblad master equation \cite{lindblad_generators_1976,Gorini1976}
\begin{equation}\label{eqs:lindblad_general}
\begin{split}
    \frac{\partial\hat{\rho}}{\partial t} & = -i[\hat{H}, \hat{\rho}] + \sum_{\mu} \gamma_\mu 
    \mathcal{D}[\hat{L}_\mu] \hat{\rho},     
\end{split}
\end{equation}
where $\hat{\rho}$ is the reduced density matrix of the system, obtained by tracing out the degrees of freedom of the environment. 
The Hamiltonian $\hat{H}$ describes the unitary evolution of the system, while the Lindblad dissipator $\mathcal{D}[\hat{L}_\mu]$, defined as 
\begin{equation}\label{eqs:lindblad_dissipator}
    \mathcal{D}[\hat{L}_\mu] \hat{\rho} = \hat{L}_\mu \hat{\rho} \hat{L}_\mu^\dagger - \frac{\hat{L}_\mu^\dagger \hat{L}_\mu \hat{\rho} + \hat{\rho} \hat{L}_\mu^\dagger \hat{L}_\mu}{2},
\end{equation}
describes the non-unitary action of the jump operator $\hat{L}_\mu$ at a rate $\gamma_\mu$.
Equation~\eqref{eqs:lindblad_general} can be recast in the compact form $\partial \hat{\rho}/\partial t = \LL \hat{\rho}$
where $\LL$ is the non-Hermitian Liouvillian superoperator.
As $\hat\rho(t) = \exp(\LL t) \hat\rho(0)$, 
the eigendecomposition of $\LL$ fully characterizes the dynamics of the density matrix.
The right eigenoperators $\hat{\eta}_j$ and left eigenoperators $\hat{\sigma}_j$ of $\LL$ \cite{breuer_theory_2007} are defined by
\begin{equation}\label{Eq:Liouvillian_eigenvalues}
\mathcal{L}\hat{\eta}_j = \lambda_j\hat{\eta}_j,\,\,\,\,\,\,\,\,\,\mathcal{L}^{\dagger}\hat{\sigma}_j = \lambda_j^*\hat{\sigma}_j,
\end{equation}
where $\lambda_j$ are complex, ${\rm Re} (\lambda_j)\leq0$, and  $\operatorname{Tr}(\hat{\sigma}_j^{\dagger}\hat{\eta}_k) = \delta_{jk}$ \footnote{In all numerical simulations presented in the paper, we first diagonalize the Liouvillian obtaining the right eigenvectors $\mathcal{L}\hat{\eta}_j = \lambda_j\hat{\eta}_j$. We then construct the matrix $\mathcal{U}$ whose columns are the $\hat{\eta}_j$. We invert $\mathcal{U}$ obtaining $\mathcal{U}^{-1}$ such that $\mathcal{U}^{-1}\mathcal{L}\mathcal{U} = \mathcal{D}$ where $\mathcal{D}$ is the diagonal matrix containing the Liouvillian eigenvalues. The rows of $\mathcal{U}^{-1}$ are finally $\hat{\sigma}_j^{\dagger}$.}. Any density matrix $\hat{\rho}(t) $ can thus be decomposed as \cite{minganti_quantum_2019}
\begin{equation}\label{eq:spectral}
\hat{\rho}(t) = \sum_{j} c_{j} (t) \, \hat{\eta}_j =  \sum_{j} c_{j} (t=0) \, e^{\lambda_j t} \, \hat{\eta}_j,
\end{equation}
with spectral weights $c_{j} (t) = \operatorname{Tr}[\hat{\sigma}_j^{\dagger}\hat{\rho}(t)]$. 
The steady state $\hat{\rho}_{\rm ss}$ is defined as the state which does not evolve under the action of the Liouvillian, $\LL \hat{\rho}_{\rm ss} =0$.
As such, $\hat{\rho}_{\rm ss}$ is the right eigenoperator of the Liouvillian associated with the eigenvalue $\lambda_0=0$.
For the models we consider in this paper, the steady state is unique and $\hat{\rho}_{\rm ss} = \lim_{t\to \infty} \hat{\rho}(t)$.

We are interested in describing quantum chaos in this class of systems.
Before illustrating the original results, we review the basic concepts of quantum chaos and motivate our method.

\subsection{Spectral approaches to quantum chaos}

According to the quantum chaos conjecture, a Hermitian system whose classical limit is characterized by an integrable dynamics exhibits a spectrum with spacings distributed according to Poissonian statistics, typical of uncorrelated random variables \cite{berry_level_1977}. Conversely, the spectrum of quantum systems with a chaotic classical dynamics presents a spacing distribution described by random matrix theory \cite{casati_connection_1980, bohigas_characterization_1984}. 
These differences in spacing distributions arise from the repulsion between eigenvalues which is absent in integrable systems due to conserved quantities. 
If the system does not admit a meaningful classical limit, a similar spectral distinction can be nevertheless observed.
For example, integrable spin chains display Poissonian statistics, while ergodic phases of matter exhibit spectra that conform to the predictions of random matrix theory \cite{ponte_manybody_2015, serbyn_spectral_2016, bordia_periodically_2017, abanin_colloquium_2019}.
Due to this phenomenological success, a random matrix structure in the Hamiltonian is often used as the sole criterion to define quantum chaos \cite{kos_manybody2018}.

The spectral approach to quantum chaos can be extended to the Lindblad master equation. Here, the statistical distribution of the spacings of the Liouvillian eigenvalues $\lambda_j \in \mathbb C$ is used to characterize DQC [c.f. Eq.\eqref{Eq:Liouvillian_eigenvalues}]. 
In analogy to the Hamiltonian case (see Appendix~\ref{sec:appendix}),
we consider the distribution of nearest-neighbor eigenvalue spacings
\begin{equation}\label{eqs:distribution}
    p(s) = \sum_{j} \delta(s_{j} -s),
\end{equation}
where $s_j=|\lambda_j - \lambda_j^{\rm NN}|$, with $\lambda_j^{\rm NN}$ the eigenvalue closest to $\lambda_j$ in the complex plane. 
In integrable systems, $s$ follows the 2D Poisson distribution
\begin{equation}
    p_{\rm 2D}(s) = \frac{\pi}{2}se^{-\frac{\pi}{4}s^2}.
\end{equation}
In chaotic systems, $s$ follows the Ginibre distribution of Gaussian non-Hermitian random matrices ensembles
\begin{equation}\label{eqs:ginibre}
    p_{\textrm{GinUE}}(s) = \left(\prod_{k=1}^{+\infty}\frac{\Gamma(1+k, s^2)}{k!}\right)\sum_{j=1}^{+\infty}\frac{2s^{2j+1}e^{-s^2}}{\Gamma(1+j, s^2)}.
\end{equation}
A crossover from integrability to chaos is captured by a distribution intermediate between the two \cite{akemann_universal_2019}.
We remark that an unfolding procedure, in which the uncorrelated part is removed from $p(s)$ in Eq.~\eqref{eqs:distribution}, is required to match the predictions of random matrix theory \cite{markum_non-hermitian_1999}.
Throughout this work, we adopt the unfolding procedure detailed in Ref.~\cite{akemann_universal_2019} and reviewed in Appendix~\ref{sec:liouvillian_unfolding}.
Furthermore, as noted in Ref.~\cite{akemann_universal_2019}, more reliable indicators of chaos can be obtained by considering the bulk statistics, i.e., the spacing statistics of the spectrum from which the eigenvalues close to the real axis have been removed.
Here, we will adopt this practice.

A recently introduced quantity that does not require the unfolding procedure, is the complex spacing ratio \cite{sa_complex_2020}
\begin{equation}
    z_j  = \frac{\lambda_j^{\rm NN} - \lambda_j}{\lambda_j^{\rm NNN} - \lambda_j}= r_j e^{i \theta_j},
\end{equation}
where $\lambda_j^{\rm NNN}$  is the second-nearest neighbor to $\lambda_j$ in the complex plane.
The average values $\langle r \rangle$ of $r_j$  and $\langle \cos(\theta) \rangle$ of $\cos(\theta_j)$, can be used as indicators of chaos in the Liouvillian case.
For a 2D Poisson distribution $\langle r \rangle=0.66$, and  $-\langle \cos(\theta) \rangle=0$.
For the Ginibre distribution $\langle r \rangle=0.74$, and  $-\langle \cos(\theta) \rangle=0.24$.

\begin{figure*}[t!]
\includegraphics[width=\textwidth]{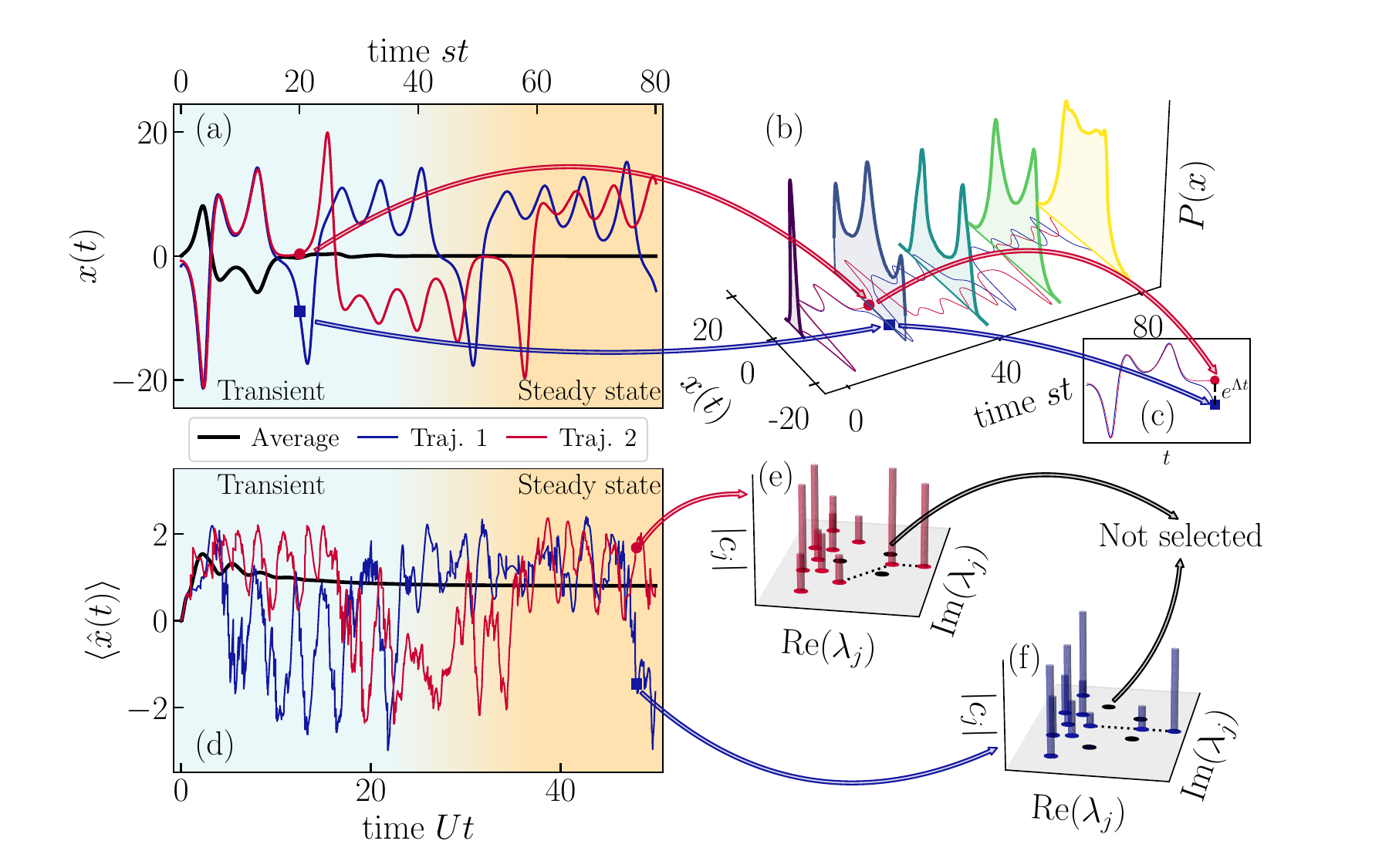}\vspace{1.8em}
\caption{Chaos in classical and quantum probabilistic systems.
(a) The classical Lorenz system in Eq.~\eqref{Eq:Lorenz} with $r/s = 2.8$ and and $b/s = 0.2667$.  
The red and blue curves represent the dynamics obtained from two different initial conditions, sampled from a unitary Gaussian distribution centered around $\vec{r} = (0, 1, 1)$.
The black solid line represents the result obtained by averaging over $5 \times 10^5$ initial conditions.
The average dynamics can be divided into a transient, where both the average and single trajectories evolve, and a steady-state, where the average is stationary but single trajectories still evolve.
(b) As a function of time, the dynamics of the probability distribution $P(x)$ compared to the results of single trajectories.
(c) Depiction of the Lyapunov exponent: despite starting from almost identical initial conditions, the trajectories associated with the red and blue curves eventually diverge.
(d) As in panel (a), the quantum dynamics for the Bose-Hubbard dimer, introduced in Eq.~\eqref{eqs:lindblad}, for the same parameters as in Fig.~\ref{fig:steady_vs_transient_chaos} and $F/U = 3$. The blue and red curves represent the results of single quantum trajectories, while the black curve is the result of the master equation (average over infinitely many quantum trajectories).
Also here, the dynamics has a transient before the onset of the steady state.
(e-f) According to the SSQT criterion, introduced in this article, the definition of chaos or integrability relies on the behavior of single quantum trajectories in conjunction with the statistical analysis of the Liovillian spectrum. To classify steady-state dynamics as integrable or chaotic, the spectral weights $c_j(t)$ defined in Eq.~\eqref{eq:spectralrho} determine the relevance of Liouvillian eigenvalues, and are illustratively represented on the vertical axis. Only some of the Liouvillian eigenvalues have a sizeable weight $|c_j|$ and are relevant in determining the steady-state dynamics.
The other eigenvalues are not selected by the dynamics of single trajectories, and should not be considered when performing a spectral analysis.
On this selected set of Liouvillian eigenvalues, the universal predictions of non-Hermitian random matrix theory distinguish regular and chaotic steady-state motion.}
\label{fig:artistic_picture}
\end{figure*}

\subsection{Chaos in average?}

An open quantum system inevitably converges towards its steady state $\hat\rho_{\rm ss}$ \cite{breuer_theory_2007}. 
Throughout the paper, we always assume that $\hat\rho_{\rm ss}$ is unique \cite{yoshida_uniqueness_2024}.
Two remarks are necessary.
First, a description of open quantum systems in terms of the density matrix is intrinsically probabilistic.
Second, once the steady state has been reached, any memory of the initial condition is lost, and the steady-state properties are only determined by the structure of the Lindblad master equation \eqref{eqs:lindblad_general}.
A definition of DQC relying only on the spectral properties of the Liouvillian will inevitably describe \textit{chaos as a transient and probabilistic phenomenon}.

To understand why such a description of chaos may be ambiguous, consider the Lorenz system, one of the most celebrated models exhibiting classical chaos. It is described by
\begin{equation}\label{Eq:Lorenz}
\begin{split}
    &\frac{\partial x}{\partial t} =\sigma(y-x), \quad \frac{\partial y}{\partial t} =x(\rho - z) - y,\\
    &\frac{\partial z}{\partial t} = xy - \beta z
\end{split}
\end{equation}
where $\sigma$, $\rho$, and $\beta$ are parameters, while $x$, $y$, and $z$ are the spatial coordinates.

Given an initial condition $\vec{r}(0) = \{x(0), \, y(0), \, z(0) \}$, the solution $\vec{r}(t)$ to these equations is \textit{chaotic} because: (i) $\vec{r}(t)$ is bounded [there exists $M$ for which $|\vec{r}(t)| < M$]; (ii) $\vec{r}(t)$ is not periodic [there is no $T$ for which $\vec{r}(t) = \vec{r}(t+T)$]; (iii) given two sufficiently close initial conditions, $\vec{r}(0)$ and $\vec{s}(0)$, the corresponding solutions, $\vec{r}(t)$ and $\vec{s}(t)$, exhibit exponential sensitivity to the initial conditions [$| \vec{r}(t) - \vec{s}(t) | \simeq e^{\Lambda t}$ for short enough times, where $\Lambda$ is called a Lyapunov exponent] \cite{strogatz_nonlinear_2018}. We show the time evolution of the Lorenz system for two slightly different initial conditions in Fig.~\ref{fig:artistic_picture} (a). We characterize their divergence in Fig.~\ref{fig:artistic_picture} (c).

Consider now a probability distribution $P(\vec{r}, t=0)$, representing a probabilistic initial condition, where each point of the distribution $\vec{r}$ evolves according to Eq.~\eqref{Eq:Lorenz}.
The definition of chaos using the previous criteria is now ambiguous when applied to the time-evolved distribution $P(\vec{r}, t)$.
Indeed, if the system is chaotic, any continuous initial probability distribution will converge towards the stationary distribution $P_{\rm ss}(\vec{r})$ and the averages associated with this distribution will converge to their stationary value [see Figs.~\ref{fig:artistic_picture} (a) and (b)].
Therefore, considering only 
$P(\vec{r}, t)$ reduces chaos to a transient phenomenon.
We conclude that rigorously ascertaining the chaotic nature of a classical probabilistic process requires looking at single trajectories.

In the dissipative quantum case, however, the largest majority of works define chaos through the spectral properties of the generators of the dynamics.
This definition has the same limitation as its classical probabilistic counterpart.
Motivated by this classical analogy, we propose to define an open quantum system as chaotic if single instances of the dynamics are chaotic.
To probe DQC, we apply random matrix theory tools to the Liouvillian eigenvalues involved in the dynamics of each of those instances, commonly called quantum trajectories.
The method we introduce below is the mathematical formulation of this intuition.

\section{Spectral statistics of quantum trajectories} \label{sec:SSQT}

In this section, we provide our operational definition of dissipative quantum chaos through the SSQT criterion and we clarify the notions of steady-state and transient quantum chaos.

We \textit{define an open quantum system as chaotic if its spectrum, selected according to the SSQT, follows the predictions of random matrix theory, independently of the behavior of its classical limit}.
We thus conjecture that, as in the classical case, the passage from integrable to chaotic dynamics is signaled by the behavior of single trajectories.

\subsection{Operational definition}

\begin{figure*}
    \centering
    \includegraphics[width=\textwidth]{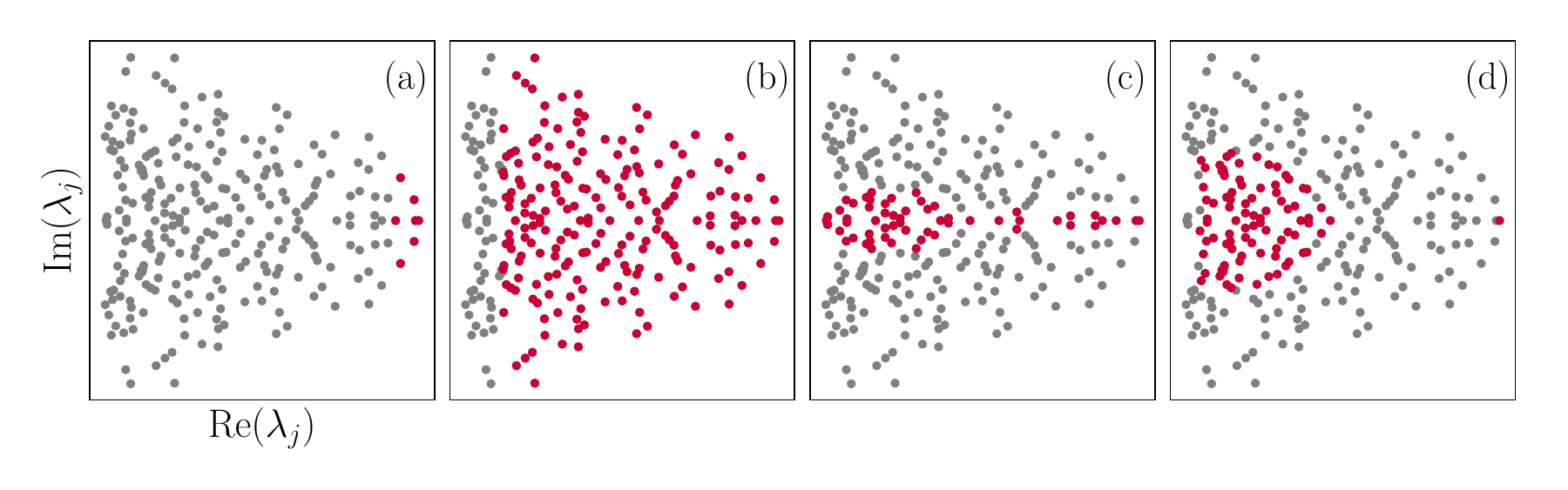}
    \caption{Depiction of how the SSQT selects a portion of the eigenvalues (in red) of the whole Liouvillian spectrum (gray). The examples reported here are those observed within this article. The SSQT may select:
    (a) Few eigenvalues around the steady state. In this case, the steady state is integrable because it is described by just a few oscillation frequencies and decay rates.
    (b) A large portion of eigenvalues, spreading from the steady state towards the depth of the spectrum.
    (c) A ``stripe'' of eigenvalues, distributed deep within the spectrum, but close to the purely real ax.
    (d) A set of eigenvalues deep in the spectrum, completely disconnected from the steady state.
    }
    \label{fig:spectra_example}
\end{figure*}

The master equation \eqref{eqs:lindblad_general} admits a stochastic unraveling in terms of quantum trajectories $\ket{\psi(t)}$, combining the Hamiltonian dynamics with continuous monitoring of the environment \cite{wiseman_quantum_2009,jacobs_2014}.
For a counting trajectory \footnote{Different unravelings, correspond to different types of measurements. Among the most common unravelings are the homodyne measurement \cite{carmichael_open_1993}, resulting in a Wiener process for the system's state $\ket{\psi(t)}$, and the photon-counting measurement \cite{molmer_monte_1993,wiseman_quantum_2009}, giving rise to the celebrated quantum jumps.}, within each time step $dt$ a quantum jump occurs with probability $dp = \gamma \sum_\mu \bra{\psi(t)}\hat{L}_\mu^{\dagger}\hat{L}_\mu\ket{\psi(t)}dt$ and $\ket{\psi(t)}$ evolves into
\begin{equation}\label{eqs:jump_operators}
    \ket{\psi(t + dt)}  \propto \hat{L}_\mu\ket{\psi(t)},
\end{equation}
where the jump operator $\hat{L}_\mu$ is sampled from the probability distribution 
\begin{equation}
    p_\mu = \frac{ \bra{\psi(t)}\hat{L}_\mu^{\dagger}\hat{L}_\mu\ket{\psi(t)}}{\sum_\mu\bra{\psi(t)}\hat{L}_\mu^{\dagger}\hat{L}_\mu\ket{\psi(t)}} .
\end{equation}
With probability $1-dp$ no quantum jump occurs, and $\ket{\psi(t)}$ evolves into
\begin{equation}\label{eqs:non_hermitian_hamiltonian}
    \ket{\psi(t + dt)} \propto (\mathds{1} - idt\hat{H}_{\rm nh})\ket{\psi(t)},
\end{equation}
where $\hat{H}_{\rm nh} = \hat{H} -i \sum_\mu \gamma_\mu \hat{L}_\mu^{\dagger}\hat{L}_\mu/2$ is the non-Hermitian Hamiltonian. 
After each time step, the state $\ket{\psi(t+dt)}$ is renormalized. 
The density operator $\hat{\rho}$ is the average over the statistical ensemble of stochastic quantum trajectories \cite{breuer_theory_2007, molmer_monte_1993, carmichael_open_1993, gardiner_zoller}. 
We argue that it is the dynamics of each trajectory that defines chaos and integrability.

To define DQC, we take $N_{\rm traj}$ independent quantum trajectories $\ket{\psi_m(t)}$ with the same initial state  $\ket{\psi_m(0)} = \ket{\psi_0}$  and evolve them according to Eqs.~\eqref{eqs:jump_operators}~and~\eqref{eqs:non_hermitian_hamiltonian}.
The density matrix of a single trajectory and its spectral decomposition according to Eq.~\eqref{eq:spectral} read
\begin{equation}
\begin{split}
    \label{eq:spectralrho}
    \hat{\rho}_m(t) =& \ketbra{\psi_m(t)} 
    = \sum_{j} c_{m,j} (t)  \, \, \hat{\eta}_{j}.
\end{split}
\end{equation}
Through this expansion we associate to each eigenvalue $\lambda_j$ the relative spectral weight $c_{m,j}(t)$. 
While DQC is usually characterized only by the statistical properties of $\{\lambda_j\}$ \cite{grobe_quantum_1988, akemann_universal_2019, sa_complex_2020}, we rather argue that DQC is encapsulated in \textit{both} $\lambda_j$ and $c_{m,j}(t)$.
The idea is illustrated in Figs.~\ref{fig:artistic_picture} (d-f).

To determine the presence of DQC from $\lambda_j$ and $c_{m, j}(t)$, we select the Liouvillian eigenvalues $\lambda_j$ for which the trajectory $\ket{\psi_m(t)}$ has a sizeable spectral component.
We do this by introducing a cutoff $c_{\rm min}$ that selects the eigenvalues and eigenoperators that are physically relevant at time $t$, i.e., for which $|c_{m,j}(t)|>c_{\rm min}$ [see Figs.~\ref{fig:artistic_picture}(e) and (f)].
On these, we perform the analysis of the statistical distribution of the Liouvillian eigenvalues, e.g., the complex spacing ratio $\langle \cos(\theta) \rangle_{m} (t)$. 
The results are then averaged over multiple independent quantum trajectories to obtain 
\begin{equation}\label{eqs:cos_theta}
    \langle \cos(\theta) \rangle (t) = \sum_{m=1}^{N_{\rm traj}} \frac{\langle \cos(\theta) \rangle_{m} (t)}{N_{\rm traj}}.
\end{equation}
Notice that this definition of $\langle \cos(\theta) \rangle (t)$ does not correspond to the evaluation on the averaged density matrix.
If the number of selected eigenvalues is small, the system's dynamics is assumed to be integrable as only few rates describe its dynamics.
It is worth noting that, due to the absence of global order in the complex plane, such a set of eigenvalues can not be chosen, e.g., by imposing a minimal and maximal energy cutoff as in closed quantum systems.

To set $c_{\rm min}$, we introduce an heuristic procedure based on the center of mass $C_m$ of the spectral coefficients $|c_{m, j}|$, defined as
\begin{equation}\label{eqs:cutoff}
    C_m = \frac{\sum_j|c_{m, j}|^2}{\sum_j|c_{m, j}|}.
\end{equation}
We fix $c_{\textrm{min}} = \bar{C}\times 10^{-k}$ with $k \in [2, 4]$, and $\bar{C} = \sum_m C_m/N_{\rm traj}$.
In computing $C_m$ we discard the $|c_{m, j}|>1$ \footnote{This choice is motivated by the following reasoning.
    If the steady state is pure, $|c_{m, 0}|=1$.
    However,  $|c_{m, 0}|<1$ for non-pure states.
    Furthermore, due to the non-orthogonal nature of the eigenoperators of Liouvillian (that is non-Hermitian), $|c_{m, j}|$ may be significantly larger than $1$.
    In this regard, discarding coefficients that are larger than 1 in computing $C_m$ ensures that  the tails of the distribution of eigenvalues are taken into account and that the steady state is always a relevant eigenstate.}.
When comparing $\bar{C}$ to the distribution of the spectral coefficients $p(|c|)$, obtained by binning the set $\{ |c_{m, j}| \,\forall\, m, j\}$, we notice that for a chaotic state, $p(|c|)$ peaks around $\bar{C}$ (and $c_{\rm min}$) so that many eigenvalues are selected as expected for DQC.
If instead the peak of $p(|c|)$ falls much before $\bar{C}$ (and $c_{\rm min}$), as is the case for regular, almost pure states, then only a few eigenvalues are selected and the SSQT criterion correctly predicts the system's integrability.
Physical examples are shown in the Appendix~\ref{App:Choice_cmin}.
This way of setting $c_{\rm min}$ may be model-dependent, however, the predictions we draw from the SSQT remain universal.

We observe that when the system is invariant under weak or strong symmetries \cite{albert_symmetries_2014, buca_note_2012}, the portion of the Hilbert space spanned by a quantum trajectory may depend on the specific unraveling \cite{bartolo_homodyne_2017, sanchez_munoz_symmetries_2019}.
For models that are not invariant under any symmetry, such as the ones considered below, we expect the specifics of the unraveling to play a negligible role.

In Fig.~\ref{fig:spectra_example} we schematically report a case history of
how the SSQT selects the Liouvillian eigenvalues according to the simulation of the physical models discussed below.
We remark that in the absence of the criterion provided by the SSQT there is no likely way to select in a meaningful way the eigenvalues involved in the system's dynamics.

\subsection{Steady-state and transient quantum chaos}

The choice of time $t$ at which the above analysis is carried out allows introducing the concepts of transient and steady-state quantum chaos. 

If $t$ is chosen within the early stages of the dynamics, our criterion characterizes \textit{transient chaos}. 
In this case, the chaotic or integrable nature of the system will depend on the initial condition $\hat{\rho}(0)$. 
The SSQT criterion thus supersedes other spectral criteria and, upon an appropriate choice of initial state, its transient chaos definition recovers the results of the spectral criterion established by the GHS conjecture \cite{grobe_quantum_1988}.

In the steady-state limit $t \to \infty$, the density matrix is stationary, but single quantum trajectories may still be non-stationary.  
Then, the SSQT analysis will single out the properties wich are independent of the initial condition, defining a criterion for \textit{steady-state chaos} that depends solely on the structure of the Lindblad master equation \eqref{eqs:lindblad_general} \footnote{In this case, and since the quantum trajectories of an open quantum system with a single steady state are ergodic \cite{beaulieu_observation2023}, the SSQT criterion at the steady state can be obtained by averaging in \textit{time} rather than in \textit{number} of trajectories.}.

We report an extensive comparison between our criterion and other ones commonly used in the literature, showing their possible ambiguities, in the Appendix \ref{Sec:Comparison}.

\subsection{Chaos in bosonic systems}
\label{Sec:Chaos_in_bosons}

An important feature of the SSQT criterion is that it allows studying chaos in non-particle-number conserving bosonic systems that are of great interest both for quantum simulation \cite{fink_signatures_2018, chen_quantum_2023, beaulieu_observation2023} and quantum computation \cite{mirrahimi_dynamically_2014, devoret_confining2015, touzard_2018, chamberland_building2022, lescanne_exponential_2020, gautier_2022, di_candia_critical_2023, gravina_critical2023}. 

Let us briefly clarify the challenge in discussing chaos in the absence of $U(1)$ symmetry (see also appendix \ref{sec:Ham}). 
Quantum chaos in number-conserving [i.e., $U(1)$-symmetric] bosonic systems has been extensively studied both in the classical and quantum regime \cite{buchleitner_2003, kolovsky_quantum_2004, tomadin_2007, kollath_statistical_2010, rautenberg_classical_2020, pausch_2021, goto_chaos_2021, pausch_optimal_2022}. 
So far, however, no theoretical characterization of quantum chaos with the spectral approach in a driven-dissipative system where the $U(1)$ symmetry is lifted has yet been provided. 
What makes the study of driven systems in the absence of $U(1)$-symmetry challenging, and sets these models apart from other driven or tilted cases, is that now a cutoff in the otherwise infinite-dimensional bosonic Hilbert space has to be provided. 
Failing to do that results in spectral statistics that are dominated by states with large particle numbers, which are asymptotically integrable thereby hiding possible chaotic dynamics. 
For isolated systems one can perform the level statistics on a given energy window, corresponding to, e.g., the energy provided by a quench.
A similar choice would be arbitrary for the 2D Liouvillian spectrum, as concepts such as energy conservation break down in a Liouvillian framework.
Furthermore, from a mathematical point of view, it makes little sense to order the Liouvillian eigenvalues, as they are complex [see also Fig.~\ref{fig:spectra_example}].
The SSQT criterion, instead, automatically determines the relevant eigenvalues along the driven-dissipative dynamics that need to be analyzed.

\section{Application I: The driven-dissipative Bose-Hubbard model} \label{sec:quantum_simulation}

Quantum fluids of light involve coupled bosonic modes with significant nonlinearity (anharmonicity). 
They are a very active research area, with these systems being proposed as quantum simulators for various phenomena including topological states, disordered lattices, flat bands, turbulence, skyrmions, spin-Hall and quantum Hall effects, quantized vortices, and universal fluctuations \cite{carusotto_quantum_2013, Nori_quantum2014, Carusotto2020}. 
Within this panorama, the Bose-Hubbard model, both driven and undriven, closed or open, has a central role as a paradigmatic case study of the emergent features of these systems.

We now apply our theoretical framework to study DQC in the driven-dissipative Bose-Hubbard dimer, the building block of larger lattice architectures. 
We show the presence of both transient and steady-state quantum chaos. 
We also discuss the breakdown of the quantum-to-classical conjecture \cite{grobe_quantum_1988}, showing the presence of steady-state quantum chaos despite a regular classical ($\hbar\rightarrow 0$) and semiclassical (first order in $\hbar$) dynamics.

\begin{figure*}[t!]
\includegraphics[width=\textwidth]{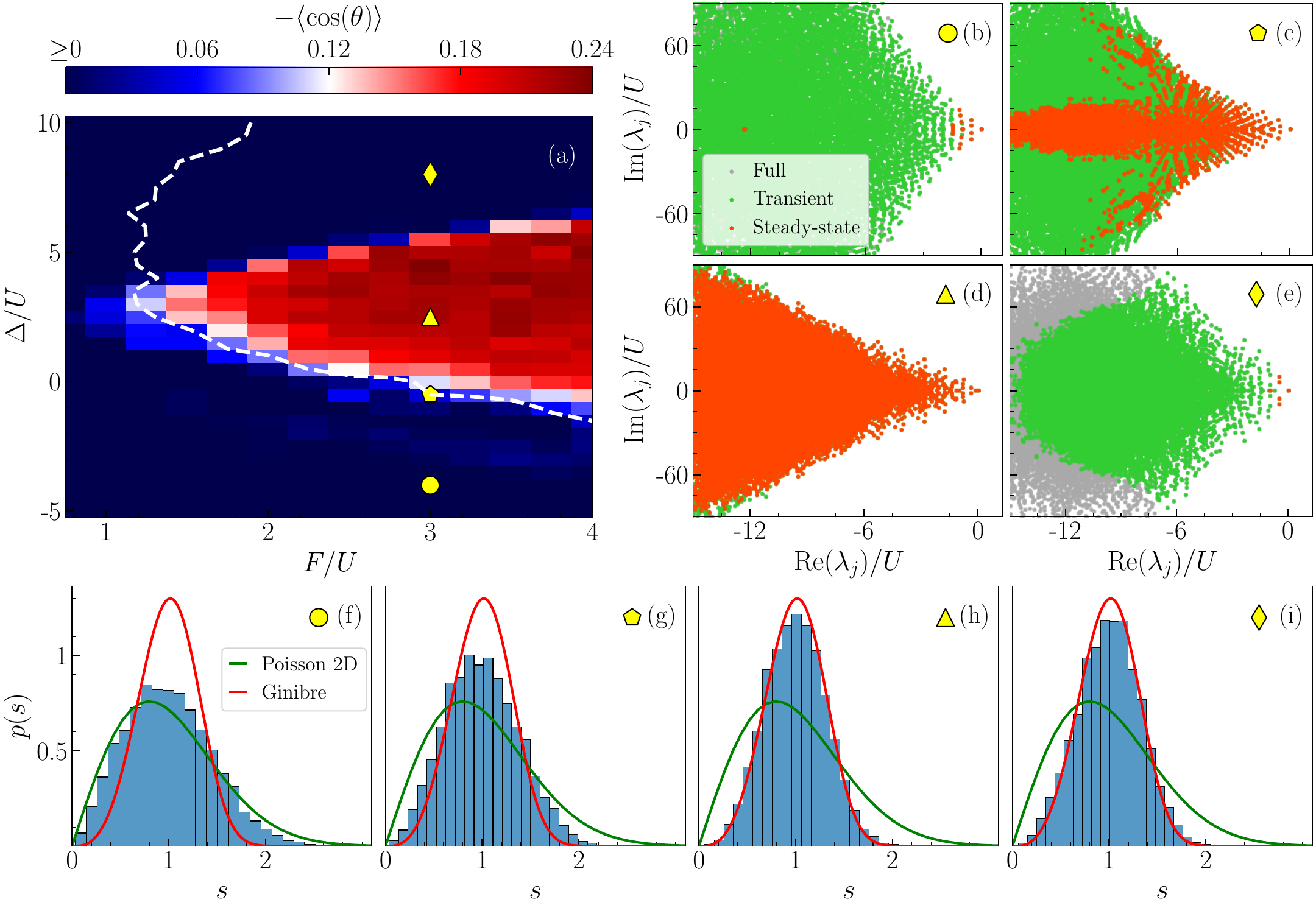}
\caption{
Steady-state and transient dissipative quantum chaos in the Bose-Hubbard dimer using the SSQT criterion described in Sec.~\ref{sec:SSQT}.
(a) Phase diagram of the single number indicator $\langle \cos(\theta) \rangle$ defined in Eq.~\eqref{eqs:cos_theta} as a function of the detuning $\Delta/U$ and of the drive $F/U$.
The colormap refers to steady-state chaos (results are obtained upon averaging over $100$ independent quantum trajectories), while the dashed white line bounds the region of transient chaos. 
When the number of selected eigenvalues is $N_{\lambda}<500$, a statistically significant analysis cannot be carried out, and we set $-\langle \cos(\theta) \rangle =0$.
(b-e) Liouvillian spectra for $F/U=3$, and increasing values of $\Delta$. The full spectrum is denoted by gray dots.
The relevant eigenvalues for the transient (green dots) and the steady-state (red dots) cases are superimposed, and they corresponds to a single quantum trajectory.
All data have  been obtained by numerically diagonalizing the Liouvillian, and selecting the relevant eigenvalues  with $c_{\rm min}=\bar{C}\times 10^{-4}$ (see Appendix \ref{App:Choice_cmin}).
The initial state is the coherent state $\ket{\psi_m(0)} = \ket{\alpha}\otimes\ket{\alpha}$ for $\alpha = \sqrt{3F/(\Delta - i \gamma)}$.
(f-i) Histograms of the spacing distribution $p(s)$ calculated over the whole Liouvillian spectrum for the same four points selected in panels (b-e). The green (red) curve is a guideline indicating an ideal 2D-Poisson (Ginibre) distribution.
The values of the detuning are: (b) $\Delta/U=-4$, (c) $\Delta/U = -0.5$, (d) $\Delta/U = 2.5$, (e) and $\Delta/U = 8$.
Other parameters are set to $g/U=2$ and $\gamma/U=1$. The cavity cutoff has been fixed to $N_c=12$.
}
\label{fig:steady_vs_transient_chaos}
\end{figure*}

\begin{figure*}[t!]
\includegraphics[width=\textwidth]{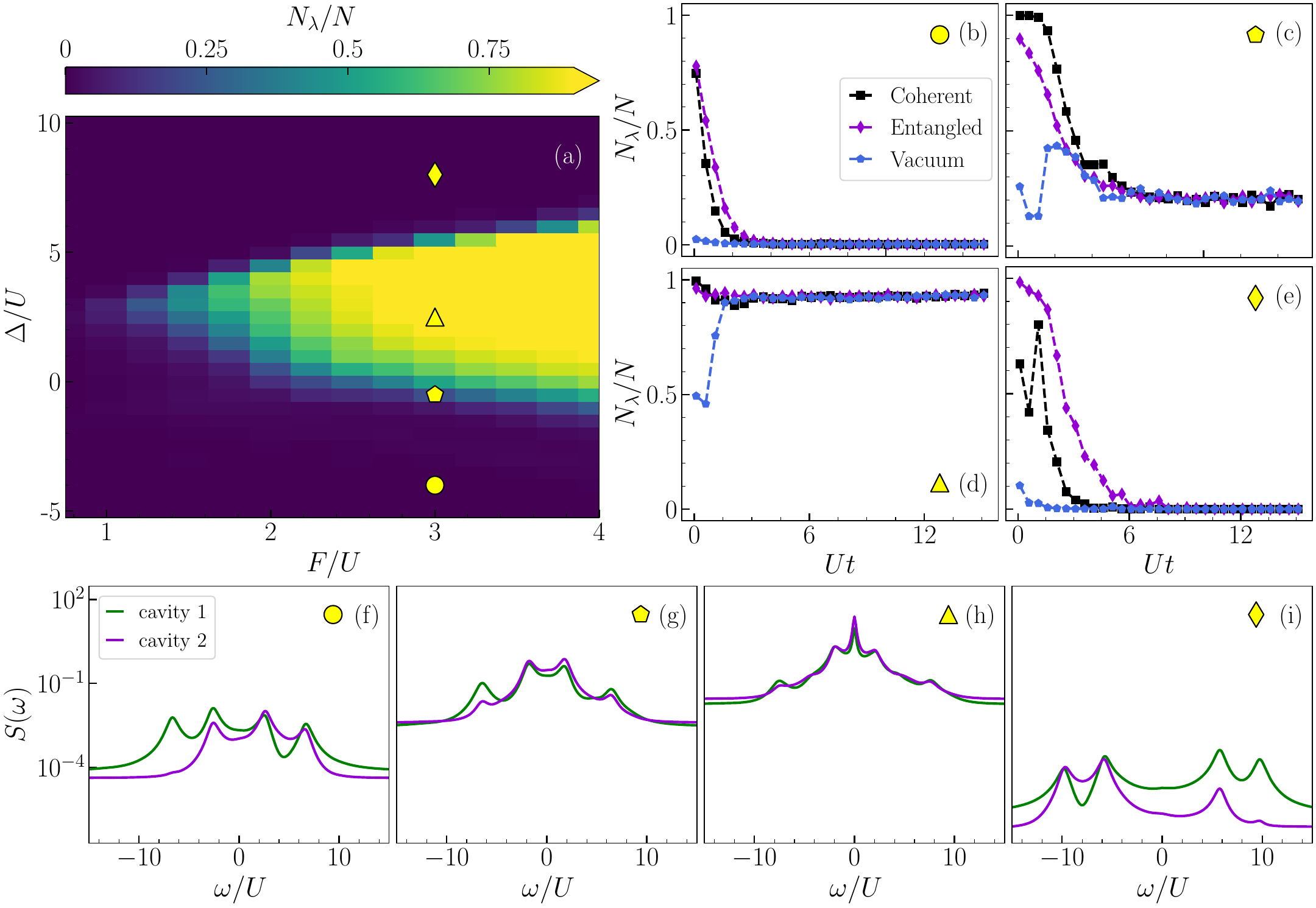}
\caption{
Analysis of the number of relevant eigenvalues $N_\lambda(t)$ obtained with the SSQT criterion in Sec.~\ref{sec:SSQT} for the Bose-Hubbard dimer. 
(a) $N_\lambda/N$ (where $N$ is the total number of Liouvillian eigenvalues) at the steady state as a function of the detuning $\Delta/U$ and of the drive $F/U$.  
(a). The phase diagram is obtained upon averaging over $100$ independent quantum trajectories.
(b-e) As a function of time, $N_\lambda(t)/N$ for $F/U=3$, and increasing values of $\Delta$ as in Figs.~\ref{fig:steady_vs_transient_chaos} (b-e) for three different types of initial states: the coherent state chosen for Figs.~\ref{fig:steady_vs_transient_chaos} (b-e), the vacuum state $\ket{\psi_m(0)}=\ket{0}\otimes\ket{0}$, and the entangled states in Eq.~\eqref{eqs:mixed_states}. While transient dynamics exhibits different values of $N_{\lambda}/N$, depending on the type of initial state $\ket{\psi_m(0)}$, the steady-state dynamics does not depend on this choice. Data are computed upon averaging over $500$ independent quantum trajectories. (f-i) Emission spectra $S(\omega)$ in the steady state for the mode $\hat{a}_1$ (green) and the mode $\hat{a}_2$ (purple) computed over the same four points selected in panels (b-e).
}
\label{fig:N_lamb}
\end{figure*}

\subsection{The model} 
\label{sec:model}

The model we consider here is a dissipative Bose-Hubbard dimer, with a drive applied only on one resonator.
In the frame rotating at the frequency of the driving (pumping) field, the Hamiltonian of the system reads
\begin{equation}\label{eqs:hamiltonian}
\begin{split}
    \hat{H} &= \sum_{j=1}^2\left(-\Delta \hat{a}_j^{\dagger}\hat{a}_j + \frac{1}{2}U\hat{a}_j^{\dagger}\hat{a}_j^{\dagger}\hat{a}_j\hat{a}_j\right)\\ &- g\left(\hat{a}_{2}^{\dagger}\hat{a}_1 + \hat{a}_1^{\dagger}\hat{a}_{2}\right) + F(\hat{a}_1^{\dagger} + \hat{a}_1),
\end{split}
\end{equation}
where $\hat{a}_j^\dagger$ and $\hat{a}_j$ are the bosonic creation and annihilation operators of the $j$-th cavity mode. 
Here, $\Delta = \omega_p-\omega_c$ is the pump-to-cavity detuning, $U$ is the strength of the Kerr nonlinearity, $g$ is the hopping strength, and $F$ is the driving field amplitude.
Losses are assumed to act homogeneously on both cavities with photon loss being the dominant dissipation process.
The dynamics of the system is described by the master equation 
\begin{equation}\label{eqs:lindblad}
\frac{\partial\hat{\rho}}{\partial t} = -i[\hat{H}, \hat{\rho}] + \gamma \sum_{j=1}^2\left(\hat{a}_j\hat{\rho}\hat{a}_j^{\dagger} - \frac{1}{2}\acomm{\hat{a}_j^{\dagger}\hat{a}_j}{\hat{\rho}}\right),
\end{equation}
where $\gamma$ is the loss rate, and $\hat{H}$ is given by Eq.~\eqref{eqs:hamiltonian}. 
As discussed above, the master equation \eqref{eqs:lindblad} admits a stochastic unraveling in terms of quantum trajectories. 
As the model we consider is not invariant under any weak or strong Liouvillian symmetry \cite{buca_note_2012,albert_symmetries_2014}, 
we expect the results to be independent of the unraveling 
\cite{bartolo_homodyne_2017, sanchez_munoz_symmetries_2019}.
Throughout the paper, we employ a photon-counting unraveling because of its numerical efficiency.

As discussed in Sec.~\ref{Sec:Chaos_in_bosons} and detailed in Appendix~\ref{sec:Ham},
in the presence of a driving field, the $U(1)$ invariance of the bare model is lifted, particle number is no longer conserved, and states evolve in time to potentially cover the entirety of the Hilbert space.
The local interaction term, scaling as $U\langle \hat{a}^{\dagger \, 2}\hat{a}^2 \rangle \sim U n^2$, dominates over that of all other states, scaling at most linearly in the cavity occupation $n=\langle \hat{a}^\dagger\hat{a}\rangle$ 
\footnote{One can easily show that, in the large occupation limit, the Liouvillian eigenvalue $\lambda \simeq -\kappa/2 (n+m) -i U/2 (n^2-m^2) $ with $n, m \in \mathbb{N}$ are associated with the Fock (number) states.
Taking a large Liouvillian will lead to a spectral distance dominated by these regular eigenvalues.
Each quantum trajectory, instead, will only be characterized by a limited number of Fock states, and the eigenvalues associated with the high-Fock number states do not participate in the dynamics.
}.
This is an explicit example of how the SSQT criterion provides a natural way to select the relevant portion of the spectrum, ensuring that the selected eigenvalues do not incur in this asymptotic decoupling issue.

In what follows all calculations have been performed using the QuTiP open source library \cite{Johansson2012,Johansson2013}.

\subsection{Phase diagram}
\label{sec:phase_diagram}

Figure~\ref{fig:steady_vs_transient_chaos} (a) shows the indicator $\langle\cos(\theta)\rangle$ defined in Eq.~\eqref{eqs:cos_theta}, computed as a function of $\Delta/U$ and $F/U$, for the eigenvalues relevant to the steady-state \footnote{the choice of a cavity cutoff $N_c=12$ ensures the convergence of the spectral statistics as proposed in \cite{sa_complex_2020} and \cite{prasad_dissipative_2022}. We however note that with the proposed cutoff for some of the points in the phase diagram (in particular with $F/U \ge 3.5$), expectation values have not yet reached convergence. This however does not change the spectral distinction between chaos and integrability.}. 
The result clearly identifies a steady-state chaotic phase emerging for sufficiently large values of $F$ and an appropriate choice of $\Delta$.
The dashed line in Fig.~\ref{fig:steady_vs_transient_chaos} (a) highlights the phase boundary between the integrable and chaotic regions, as obtained by applying the criterion for transient chaos at $t=0$. 
The phase boundary coincides with that of steady-state chaos in the region of negative detuning. 
For large positive values of $\Delta$ instead, the diagram shows a region where transient chaos occurs in spite of an integrable steady state. 
Liouvillian spectra with the relevant eigenvalues, for the points highlighted in Fig.~\ref{fig:steady_vs_transient_chaos} (a), are displayed in Figs.~\ref{fig:steady_vs_transient_chaos} (b-e). 
The spacing distributions \eqref{eqs:distribution} associated to each of these points, computed over the whole Liouvillian spectrum, are plotted in Figs.~\ref{fig:steady_vs_transient_chaos} (f-i). 
We observe that while our criterion clearly classifies the steady-state dynamics in Fig.~\ref{fig:steady_vs_transient_chaos} (e) as completely regular (only a few eigenvalues are involved in the dynamics), a spectral analysis on the full Liouvillian spectrum [Fig.~\ref{fig:steady_vs_transient_chaos} (i)] does not predict the nature of the steady state.

To study the passage from the transient to the steady state dynamics, we now consider three different types of initial states: coherent states, the vacuum state, and entangled states of the form
\begin{equation}\label{eqs:mixed_states}
    \ket{\psi_m(0)} = \sum_{j=0}^5\frac{\ket{j}\otimes\ket{j}}{\sqrt{6}}.
\end{equation}
We denote the average number of relevant eigenvalues at time $t$ with $N_\lambda(t)$. 
We show $N_{\lambda}(t)/N$ (where $N$ is the total number of Liouvillian eigenvalues) in Fig.~\ref{fig:N_lamb}. 
In the steady state, we note a direct correspondence between the $\langle\cos(\theta)\rangle$ and $N_{\lambda}$ [c.f Figs.~\ref{fig:steady_vs_transient_chaos} (a)~and~\ref{fig:N_lamb} (a)].
In the transient dynamics, instead, we find that $\langle\cos(\theta)\rangle(t) \simeq \langle\cos(\theta)\rangle(0) \, N_{\lambda}(t)/N_{\lambda}(0)$. 
We plot $N_{\lambda}(t)$ in Fig.~\ref{fig:N_lamb} (b-e).
As expected, transient dynamics depends on the type of initial states, while the steady-state features do not depend on this choice. We remark that this feature is in stark contrast with the unitary dynamics of an isolated quantum system, where the selected eigenvalues depend solely on the initial state (once a cutoff value $c_{\rm min}$ has been set). 
Indeed the spectral weights $c_j(t)$ in Eq.~\eqref{eq:spectral} reduce to $|c_j(t)| = |\bra{\phi_j}\ket{\psi(t)}|^2 = |e^{-iE_jt}\bra{\phi_j}\ket{\psi(0)}|^2 = |\bra{\phi_j}\ket{\psi(0)}|^2$ where $E_j, \ket{\phi_j}$ are the eigenvalues and eigenvectors of the Hamiltonian and $\ket{\psi(t)}$ solves the Schr\"odinger equation.

We finally show that the presence of DQC in the Bose-Hubbard dimer has distinctive features in the emission spectrum, an accessible quantity in quantum optics experiments. It is defined as
\begin{equation}\label{eqs:emission_spectrum}
       S(\omega) = 2 \textrm{Re}\int_{0}^{+\infty}d\tau e^{-i\omega\tau}\lim_{t\rightarrow+\infty}\langle \hat{a}^{\dagger}(t+\tau)\hat{a}(t)\rangle.
\end{equation}
Following Ref.~\cite{sanchez_munoz_symmetries_2019}, and using the quantum regression theorem \cite{breuer_theory_2007}, we express \footnote{It is sufficient to re-write Eq.~\eqref{eqs:emission_spectrum} as 
\begin{equation}
    S(\omega) = 2 \textrm{Re}\int_{0}^{+\infty}d\tau e^{-i\omega\tau}\operatorname{Tr}[\hat{a}^{\dagger}e^{\mathcal{L}\tau}(\hat{a}\hat{\rho}_{\textrm{ss}})].
\end{equation}
At this point one uses the eigenmatrix decomposition \eqref{eq:spectral} over $(\hat{a}\hat{\rho}_{\textrm{ss}})$. Finally, since $e^{\mathcal{L}\tau}\hat{\eta}_j = e^{\lambda_j\tau}\hat{\eta}_j$, one arrives at Eq.~\eqref{eqs:emission_spectrum-eigenvalues}.}
\begin{equation}\label{eqs:emission_spectrum-eigenvalues}
\begin{split}
    S(\omega) &= \sum_{j, \gamma_j=0}c_{\sigma_j}c_{\eta_j}2\pi\delta(\omega-\omega_j)\\ &+ \sum_{j, \gamma_j\ne 0}c_{\sigma_j}c_{\eta_j}\frac{i(\omega-\omega_j) - \gamma_j}{(\omega-\omega_j)^2 + \gamma_j^2},
\end{split}
\end{equation}
where $\lambda_j = \gamma_j + i\omega_j$ and $c_{\sigma_j} = \operatorname{Tr}(\hat{\sigma}_j^{\dagger}\hat{a}\hat{\rho}_{\textrm{ss}})$, $c_{\eta_j} = \operatorname{Tr}(\hat{a}^\dagger\hat{\eta}_j)$.
In Fig.~\ref{fig:N_lamb} (f-i) we plot $S(\omega)$ for both modes $\hat{a}_1$ and $\hat{a}_2$, for the same highlighted points as in panel Fig.~\ref{fig:N_lamb} (a).
We see that when the model is integrable [Fig.~\ref{fig:N_lamb} (f)], a sequence of distinct spectral peaks is visible. 
As we enter the chaotic phase [Figs.~\ref{fig:N_lamb} (g-h)], $S(\omega)$ increases by several orders of magnitude, becoming broader and without a well-defined structure. 
This feature reflects the sharp transition from a few to thousands of eigenvalues involved in steady-state dynamics of quantum trajectories as illustrated in Fig.~\ref{fig:N_lamb} (a).
Finally, for larger detuning the system becomes integrable in the steady state, as signaled by the behavior of $S(\omega)$ [Fig.~\ref{fig:N_lamb} (i)].
This feature is completely missed by a straightforward analysis of the Liouvillian level statistics, as evident from Figs.~\ref{fig:steady_vs_transient_chaos} (h-i).

\allowdisplaybreaks

\subsection{Breakdown of the classical-to-quantum correspondence}
\label{sec:classical}

\begin{figure*}[t!]
\includegraphics[width=\textwidth]{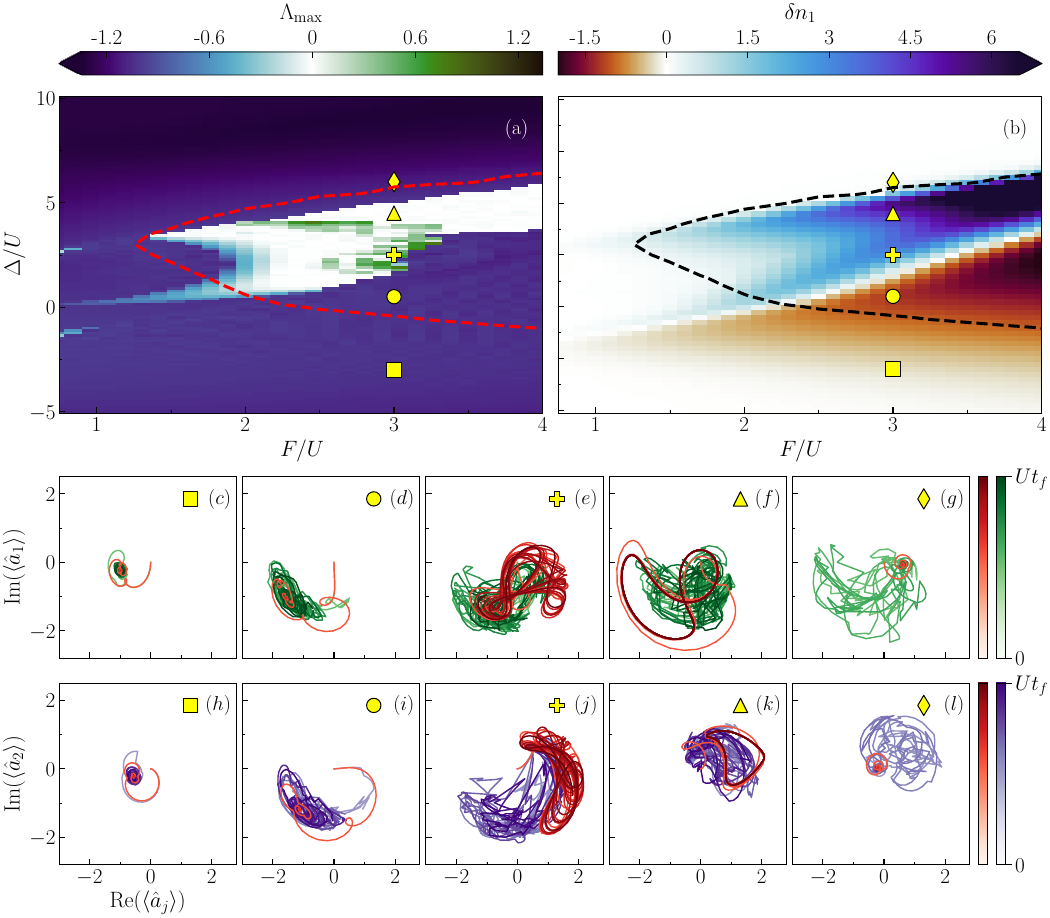}
\caption{Classical vs full quantum analysis. (a) The largest Lyapunov exponent $\Lambda_{\rm max}$, calculated according to the method in Appendix~\ref{sec:Lyapunov}. 
The system displays a stripe of limit cycles $\Lambda_{\rm max}=0$, and within this region we find classical chaos $\Lambda_{\rm max}>0$. 
The dashed red line bounds the region of steady-state quantum chaos.
(b) Deviation from Poissonian distribution $\delta n_1$, defined in Eq.~\eqref{eqs:delta_n}, for the first cavity. 
The region with super-Poissonian light ($\delta n_1 > 0$) coincides with the classical region hosting limit cycles and strange attractors.
The steady-state quantum chaotic region (dashed line) extends over both the super-Poissonian and sub-Poissonian regions.
(c-g) Single classical (in red) and quantum trajectory (in green) with $Ut_f=30$ for the first cavity; (h-l) the same but for the second cavity (quantum trajectories are in purple). The five cases correspond to selected parameter values highlighted with different symbols in panel (a). 
(c) and (h) Both the classical and the quantum solutions approach a fixed point. 
(d) and (i) The classical case approaches a fixed point, while the quantum trajectory displays a chaotic attractor. 
(e) and (j) Both the classical and quantum trajectories show chaotic attractors.
(f) and (k) The classical trajectory displays a limit cycle while the quantum trajectory shows a chaotic attractor.
(g) and (l) The classical trajectory approaches a fixed point, while the quantum trajectory displays transient chaotic behavior before approaching the fixed point. 
The values of detunings are: (c) and (h) $\Delta/U=-3$, (d) and (i) $\Delta/U=0.5$, (e) and (j) $\Delta/U=2.5$, (f) and (k) $\Delta/U=4.5$, (g) and (l) $\Delta/U=6$.
Other parameters as in Fig.~\ref{fig:steady_vs_transient_chaos}.}
\label{fig:Q&C_quantum}
\end{figure*}

We investigate here the classical limit of the system described by Eqs.~\eqref{eqs:hamiltonian} and \eqref{eqs:lindblad}.
We employ the Gross-Pitaevskii mean-field (i.e., coherent state) approximation \footnote{A rigorous derivation of the semiclassical limit of the Lindblad master equation indicates that the correct classical limit, i.e. $\hbar\to0$ should include second-order correlations between fields \cite{Dubois_2021}. Here we will identify the classical limit as the the mean-field approximation, which assumes that second-order correlators factor into field products.}.
For each resonator in the Bose-Hubbard chain, we assume that $\hat{\rho}=\bigotimes_j\hat{\rho}_j$ at all times, with $\hat{\rho}_j = \ketbra{\alpha_j}$ and $\hat{a}_j\ket{\alpha_j} = \alpha_j \ket{\alpha_j}$ \cite{Casteels_2017,debnath_nonequilibrium_2017, giraldo_driven-dissipative_2020}.
Doing so for the master equation \eqref{eqs:lindblad} leads to

\allowdisplaybreaks
\begin{equation}\label{eqs:GP}
\begin{aligned}
    \frac{\partial}{\partial t} \alpha_1 &= \left(i\Delta -\frac{1}{2}\gamma\right)\alpha_1 -iU|\alpha_1|^2\alpha_1 + ig\alpha_2 -iF\,,\\
    \frac{\partial}{\partial t} \alpha_2 &= \left(i\Delta -\frac{1}{2}\gamma\right)\alpha_2 - iU|\alpha_2|^2\alpha_2 + ig\alpha_{1}\,. 
\end{aligned}
\end{equation}

Classical chaos for dynamical systems can be characterized through the Lyapunov exponents $\Lambda$ \cite{strogatz_nonlinear_2018}. The system is chaotic if the largest Lyapunov exponent $\Lambda_{\rm max}$ is strictly positive. 
Lyapunov exponents can also distinguish bounded attractors hosting limit cycles from dense regions in phase space where classical chaos arises. 
We evaluate $\Lambda_{\rm max}$ as a function of $\Delta/U$ and $F/U$. Details on the numerical estimation of the largest Lyapunov exponent are reported in the Appendix~\ref{sec:Lyapunov}. The results are shown in Fig.~\ref{fig:Q&C_quantum} 
(a).
We observe a large region characterized by a vanishing Lyapunov exponent, where the system dynamics follows a limit cycle.
Within this region, a smaller chaotic region is present, where the dynamics are characterized by a strange attractor with a positive Lyapunov exponent. 
The comparison of the classical phase diagram with that for steady-state DQC shows that the quantum system displays chaos in a larger region of the parameter space than its classical limit. Looking at the transient chaotic region [white dashed line in Fig.~\ref{fig:steady_vs_transient_chaos} (a)] the difference is even more marked. 
This observation suggests the lack of correspondence between classical and quantum chaos in both the transient \textit{and} the steady-state of a driven-dissipative system
\cite{grobe_quantum_1988}.

To gain more insight into the emergence of dissipative quantum chaos for these parameters where the classical limit is not chaotic, we study the fluctuations in the photon number. We define the deviation from the Poissonian distribution 
\begin{equation}\label{eqs:delta_n}
\begin{aligned}
\delta n_j &=\langle \hat{a}_j^\dagger\hat{a}_j^\dagger \hat{a}_j\hat{a}_j\rangle_{\rm ss} - \langle \hat{a}_j^\dagger \hat{a}_j\rangle_{\rm ss}^2 \\
&= (\Delta{n}_j)^2 - \langle\hat{n}_j\rangle_{\rm ss}
\end{aligned}
\end{equation}
where $\langle \hat{O} \rangle_{\rm ss} \equiv \operatorname{Tr}(\hat{\rho}_{\rm ss} \hat{O})$ and $(\Delta{n}_j)^2=\langle\hat{n}_j^2\rangle_{\rm ss}-\langle\hat{n}_j\rangle_{\rm ss}^2$. $\delta n_j=0$ for Poissonian photon-number statistics, and takes positive (negative) values for a super (sub)-Poissonian distribution. 
We plot $\delta n_1$ in Fig.~\ref{fig:Q&C_quantum} (b). 
$\delta n_1$ divides the parameter region in which steady-state chaos occurs in two distinct zones.
In one, the system displays super-Poissonian statistics, and comparison with Fig.~\ref{fig:Q&C_quantum} (a) indicates that the classical limit displays either chaos or limit cycles. 
The other zone is characterized by sub-Poissonian statistics, and the classical limit is correspondingly characterized by fixed points. 
While Poissonian or super-Poissonian light admits a classical description, sub-Poissonian light arises from genuine quantum mechanical effects \cite{scully_quantum_1997}.

The lack of correspondence can also be visualized by considering
the expectation value of the fields for the single classical and quantum trajectories, as shown in  Figs.~\ref{fig:Q&C_quantum} (c-l).
Panels (c) and (h) correspond to a steady-state integrable system. 
In this case, both the classical and quantum trajectories approach fixed points, with the quantum trajectories displaying few quantum jumps along their dynamics. 
Panels (d) and (i) correspond to steady-state chaos with sub-Poissonian photon-number statistics. 
Here, the classical trajectory approaches a fixed point, but quantum jumps prevent the quantum trajectory from reaching the same fixed point. 
Hence, fluctuations induced by the environment are necessary to trigger chaos. 
Each quantum jump resets the quantum trajectory to a random state and prevents the system from reaching the corresponding classical fixed point, in a way that is reminiscent of the quantum Zeno effect \cite{misra_zenos_1977}. 
The sub-Poissonian statistics is therefore explained, as the variations in the photon number along the dynamics are due to the quantum jumps, which here are random events with a sub-Poissonian distribution. 
Panels (e) and (j) correspond to parameters for which both the classical and quantum systems are chaotic. 
Here, the classical trajectories display a chaotic attractor, and the quantum trajectories are similarly characterized by a chaotic attractor randomly perturbed by quantum jumps. 
Panels (f) and (k) correspond to parameters where the classical system displays a limit cycle, while quantum jumps randomly perturb the trajectories leading to a chaotic attractor. 
The two latter cases are characterized by super-Poissonian statistics typical of classical strongly fluctuating processes.
Indeed, along the classical dynamics, the field intensity varies from small to large values along a continuous curve \cite{walls_quantum_2008}. 
While quantum jumps do not alter the photon-number statistics, their fluctuations transform classical limit cycles into a chaotic dynamics. 
Finally, panels (g) and (l) display a case where transient chaos but integrable steady-state occur. 
Here again, the quantum trajectory is characterized by a chaotic behavior solely induced by fluctuations, while the classical trajectory regularly approaches a fixed point.

The overall analysis provides a clear explanation of the nature of steady-state chaos. In a driven-dissipative system, steady-state chaos can be caused by environment-induced fluctuations that reset the quantum trajectories at random times, effectively preventing the system from following its classical path in phase space. This picture explains why DQC occurs on a significantly larger region of parameters than that characterized by classical chaos, where the classical limit only displays fixed points or limit cycles. 
The analysis of other quantities, such as the emission spectra or photon number within the resonator, does not show any qualitative difference in the description of the whole quantum chaotic region, despite the discrepancy with the classical limit.
These findings ultimately justify our definition of DQC solely in terms of the quantum properties of the system and highlight how a quantum-to-classical correspondence may break down in driven-dissipative quantum systems.

\subsection{Beyond first order in $\hbar$: truncated Wigner analysis}
\label{sec:semi_classical}

We now investigate if the DQC observed in the full quantum description can be already captured perturbatively, or rather is an intrinsic quantum mechanical feature. 
Semiclassical approaches, in the sense that they account for quantum effects only perturbatively, include cluster mean-field approaches \cite{biella_cluster2016, huybrechts_cluster2019}, quantum cumulant expansions \cite{kubo_generalized_1962, plankensteiner_quantumcumulantsjl_2022, huybrechts_cumulant2023} and the truncated Wigner approximation (TWA) \cite{hillery_distribution_1984, carmichael_statistical_1999, polkovnikov_phase_2010}.  
In particular, the TWA can be formally shown to capture fluctuations at order $\hbar$, and proved successful in the characterization of open and closed bosonic systems (see for example Refs.~\cite{wimberger_fragmentation2013, vicentini_critical2018, schlagheck_enhancement2019}). 
Within the Wigner formalism, the full Lindblad equation \eqref{eqs:lindblad} is mapped onto a partial differential equation for the Wigner function. 
This mapping is exact, but the many-body interactions lead to terms of high order whose numerical treatment is highly challenging. 
The TWA consists of discarding these high-order terms and keeping only terms of second order. 
Then, the equation for the Wigner function reduces to a Fokker-Plank equation which can be mapped onto the set of stochastic differential equations for the phase-space variables.
\begin{equation}\label{eqs:TWA}
\begin{aligned}
    \frac{\partial}{\partial t} \alpha_1 &= \left(i\Delta -\frac{1}{2}\gamma\right)\alpha_1 -iU(|\alpha_1|^2 - 1)\alpha_1 \\&+ ig\alpha_2 -iF + \xi_1(t)\sqrt{\gamma/2}\,,\\
    \frac{\partial}{\partial t} \alpha_2 &= \left(i\Delta -\frac{1}{2}\gamma\right)\alpha_2 - iU(|\alpha_2|^2 - 1)\alpha_2 \\&+ ig\alpha_{1}+ \xi_2(t)\sqrt{\gamma/2}\,.
\end{aligned}
\end{equation}
Here, $\xi_j$ is a stochastic Wiener process such that $\langle\xi_j(t)\xi_k^*(t')\rangle=\delta_{jk}\delta(t-t')$.

\begin{figure}[t!]
\includegraphics[width=0.46\textwidth]{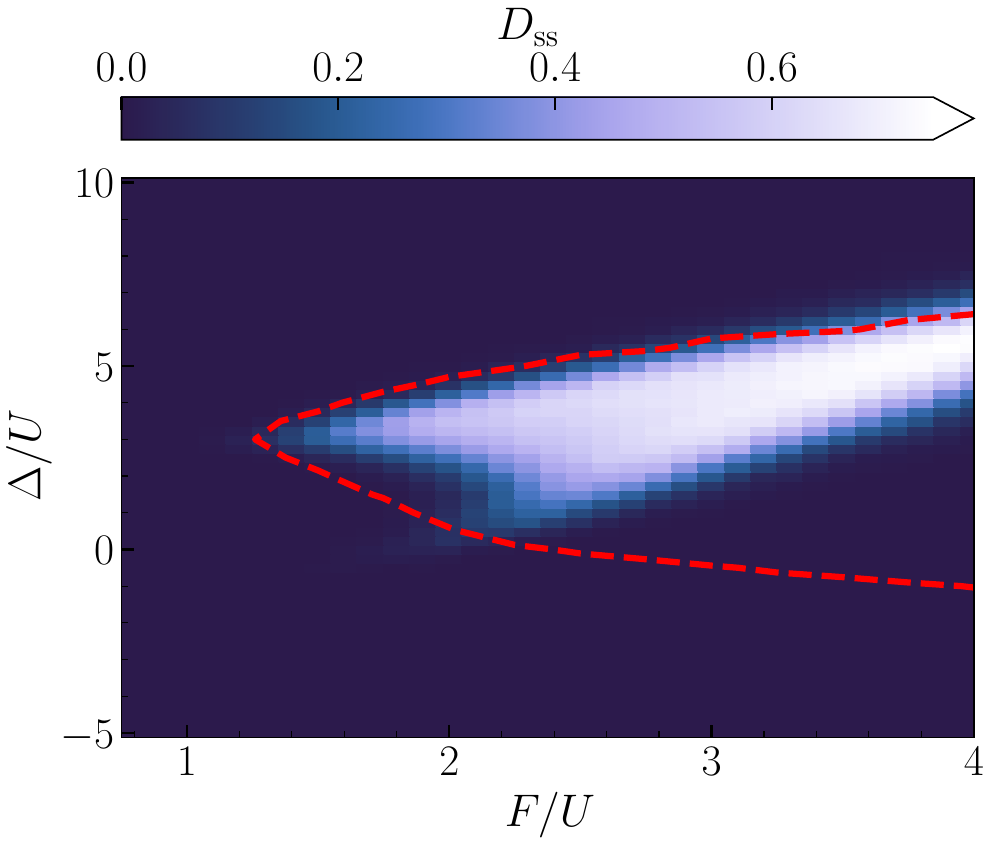}
\caption{Semiclassical decorrelator for the second cavity $D_{\rm ss}(\tau)$ computed within the truncated Wigner approximation (TWA) averaging over $500$ independent Wigner stochastic trajectories and a steady-state time of $U\tau = 100$. 
The region displaying semi-classical chaos [$D_{\rm ss}(\tau)\simeq 1$] matches the quantum region characterized by super-Poissonian fluctuations in Fig.~\ref{fig:Q&C_quantum} (b) and extends only slightly beyond the classical region characterized by $\Lambda_{\textrm{max}} \ge 0$ in Fig.~\ref{fig:Q&C_quantum} (a). The dashed line delimits the region of steady-state quantum chaos. 
The quantum chaotic region characterized by sub-Poissonian fluctuations is not captured by the quantum corrections provided by TWA. 
We also fix the initial change in the initial condition as $\varepsilon= 0.01(1 + i)/\sqrt{2}$.
Physical parameters chosen as in Fig.~\ref{fig:steady_vs_transient_chaos} (a).
}
\label{fig:TWA}
\end{figure}

To study chaos within the TWA, we employ a variant of the semiclassical out-of-time-order correlator (OTOC) $D(t)$ introduced in Ref.~\cite{das_spreading2018} for spin systems and successfully applied to many-body classical and semiclassical systems \cite{bilitewski_temperature2018, schuckert_thermal2019, bilitewski_classical2021, sibaram_thermal2021, deger_arresting2022, mondal2023emergence}. 
The computation of the semiclassical OTOC (or decorrelator) involves two replicas of the system, say $\alpha$, $\tilde{\alpha}$, that differ by $|\varepsilon| \ll 1$ in the initial condition \cite{das_spreading2018}.
The semiclassical decorrelator measures how those two replicas de-correlate in time. For an integrable system, the two replicas are expected to remain correlated over a long time, so that $D(t)=0$ (no de-correlation). 
If the dynamics is instead chaotic, the two replicas will rapidly lose any correlation, as in a classically chaotic system, leading to $D(t)=1$ (maximum de-correlation). 
For a dissipative system evolving towards a steady state, we can define also the steady-state semiclassical decorrelator as $\lim_{t\rightarrow+\infty}D(t) = D_{\rm ss}$.

We extend this criterion to our semiclassical state, and we define the semi-classical decorrelator as 
\begin{equation}
    D_{\rm ss}(\tau) = 1 - \left\langle e^{-d_{L^2}(\alpha(\tau), \tilde{\alpha}(\tau))}\right\rangle.
\end{equation}
Each $\alpha(0)$ is the solution of Eqs.~\eqref{eqs:TWA} having been obtained by evolving Eqs.~\eqref{eqs:TWA} for a time long enough for the average to have reached the steady state.
$\alpha(\tau)$ is then computed integrating Eqs.~\eqref{eqs:TWA} for a time $\tau$.
Each replica $\tilde{\alpha}(\tau)$ is obtained by initializing $\tilde{\alpha}(0) =  \alpha(0) + \varepsilon$ and evolving for the same time $\tau$ Eqs.~\eqref{eqs:TWA} using the \textit{same} noise extracted to evolve $\alpha(\tau)$.
$\langle\,\cdot\,\rangle$ denotes the average over many independent Wigner trajectories.  
We then compute the $L^2$ distance between the solutions $\alpha$ and $\tilde{\alpha}$, defined as
\begin{equation}\label{eqs:TWA_OTOC}
\begin{split}
    d_{L^2}(\alpha, \beta) &= \sqrt{[\textrm{Re}(\alpha) - \textrm{Re}(\beta)]^2 + [\textrm{Im}(\alpha) - \textrm{Im}(\beta)]^2}.
\end{split}
\end{equation}

Results for $D_{\rm ss}(\tau)$  are reported in Fig.~\ref{fig:TWA} for the second cavity. 
The region characterized by $D_{\rm ss}(\tau) \simeq 1$ matches the quantum chaotic region characterized by super-Poissonian fluctuations and extends slightly beyond the classical region exhibiting $\Lambda_{\rm max} \ge 0$. However, similarly to the mean-field analysis, also the semi-classical approach fails to predict chaos in the quantum chaotic region characterized by sub-Poissonian fluctuations. 

This finding provides further evidence that steady-state quantum chaos can be an emergent feature of driven-dissipative bosonic systems, lacking classical or semi-classical counterparts. 
It demonstrates the unique nature of DQC, and highlights its fundamental difference from quantum chaos in Hamiltonian systems.

\section{Application II: Readout of a transmon qubit} \label{sec:quantum_information}

In this section, we study the dispersive readout of a transmon qubit in circuit quantum electrodynamics architectures.
While recent years witnessed tremendous progress towards fast and high-fidelity readout of qubits \cite{jeffrey_fast_2014, walter_rapid2017, heinsoo_rapid_2018, sunada_fast_2022, sunada_photon-noise-tolerant_2024, spring_fast_2024}, understanding the limitations of the building blocks of currently employed readout protocols is paramount to achieve the extremely high fidelities required for, e.g., parity checks in quantum error correction protocols.
As discussed in Refs.~\cite{berke_transmon_2022, cohen_reminescence2023, chavez-carlos_driving_2025, dumas_measurement-induced_2024} (and the references therein), also in this context quantum chaos can emerge, impairing quantum information processing by degrading the stored quantum information and reducing the performance of gates and readouts.
While these pioneering studies mainly focused on energy-based arguments, this section examines how the chaotic features of the system are qualitatively and quantitatively modified when accounting for dissipation in the modeling of noise-intermediate scale quantum devices.

\label{sec:model_transmon}

A transmon qubit is a bosonic mode characterized by a strong anharmonicity.
The resulting uneven energy spacing allows selective control of the ground and excited states, thereby approximating an idealized two-level system.
Transmons are obtained by adding a large shunt capacitance to a cooper-pair box (see, e.g., Ref.~\cite{blais_circuit_2021}).
This design lowers the charging energy $E_C$ without significantly modifying the Josephson energy $E_J$. 
In the limit $E_C \ll E_J$, the circuit Hamiltonian is
\begin{equation}\label{eqs:transmon_full}
    \hat{H}_t = 4E_C\hat{n}^2 - E_J\cos\hat{\phi},
\end{equation}
where $\hat{n} = \hat{Q}/2e$ is the charge number operator, $\hat{\phi} = 2e\hat{\Phi}$ is the phase operator, and $[\hat{Q}, \hat{\Phi}] = i\hbar\mathds{1}$.  
The charge and phase operators can be rewritten as
\begin{equation}\label{eqs:bosonic_fields}
\begin{split}
    &\hat{n} = \frac{i}{2}\left(\frac{E_J}{2E_C}\right)^{1/4}(\hat{b}^{\dagger}-\hat{b}),\\ &\hat{\phi} = \left(\frac{2E_C}{E_J}\right)^{1/4}(\hat{b}^{\dagger}+\hat{b}),
\end{split}
\end{equation}
where $[\hat{b}, \hat{b}^{\dagger}] = \mathds{1}$. 
Substituting Eq.~\eqref{eqs:bosonic_fields} in Eq.~\eqref{eqs:transmon_full} and discarding counter-rotating terms results in 
(see Appendix~\ref{sec:Transmon_appendix})
\begin{equation}\label{eqs:transmon}
    \hat{H}_t = \omega_t\hat{b}^{\dagger}\hat{b} - \left( \frac{E_C}{2} - \chi_{\rm eff}^{(2)} \right) \hat b^{\dagger 2} \hat b^2 + \sum_{k=3}^{5} \chi_{\rm eff}^{(k)} \hat b^{\dagger k} \hat b^k,
\end{equation}
where $\omega_t$ is the transmon resonance frequency, $\chi_{\rm eff}^{(k)}$ are polynomial functions of $E_C/E_J$, and terms beyond the fifth order have been neglected. 

Reading out a qubit can be accomplished by considering a driven linear cavity mode $\hat a$ with Hamiltonian 
\begin{equation}\label{eqs:resonator}
\begin{split}
    \hat H_r &= \omega_r\hat a^\dagger \hat a + F\cos(\omega_pt)(\hat{a} + \hat{a}^\dagger)\\&\simeq\omega_r\hat a^\dagger \hat a + F(\hat{a} e^{i \omega_p t} + \hat{a}^{\dagger} e^{-i \omega_p t}),
\end{split}
\end{equation}
where $\omega_r$ is the readout resonance frequency, $F$ is the drive amplitude, and $\omega_p$ is the drive frequency.
The latter part of Eq.~\eqref{eqs:resonator} follows from discarding the counter-rotating terms in the pump.
Throughout the section we always assume $\omega_p = \omega_r$ (cavity driven on resonance).
The cavity and the transmon are then coupled through
\begin{equation}\label{eqs:coupling}
    \hat H_c = - g(\hat{a}^{\dagger}\hat{b} + \hat{b}^{\dagger}\hat{a}),
\end{equation}
where we have dropped counter-rotating terms coming from a dipole-dipole coupling.
The feedline controlling the readout cavity is used both to drive it and to collect its output. 
Therefore, in the frame rotating at the pump frequency, the system dynamics is described by the Lindblad master equation 
\begin{equation}\label{eqs:lindblad_circuitQED}
    \frac{\partial \hat{\rho}}{\partial t} = -i[\hat{H}, \hat{\rho}] + \gamma_r\left(\hat{a}\hat{\rho}\hat{a}^{\dagger} - \frac{1}{2}\acomm{\hat{a}^{\dagger}\hat{a}}{\hat{\rho}}\right),
\end{equation}
with 
\begin{equation}\label{eqs:hamiltonian_circuitQED}
\begin{split}
    \hat{H} &=  \hat{H}_t + \hat{H}_r + \hat{H}_c \\ &= -\Delta\hat{b}^{\dagger}\hat{b} + F(\hat{a}^{\dagger} + \hat{a})  - \left(\frac{E_C}{2} - \chi_{\textrm{eff}}^{(2)}\right)\hat{b}^{\dagger 2}\hat{b}^{2} \\ & \quad + \sum_{k=3}^{5} \chi_{\rm eff}^{(k)} \hat b^{\dagger k} \hat b^k - g(\hat{a}^{\dagger}\hat{b} + \hat{b}^{\dagger}\hat{a}).
\end{split}
\end{equation}
Here, $\Delta = \omega_p-\omega_t =  \omega_r-\omega_t$ is the resonator-to-transmon detuning, and $\gamma_r$ is the resonator's loss rate. For this proof-of-concept, we assume the transmon not to be subject to any dissipative or decoherence process.

In all the numerical results presented in this section, unless otherwise stated, we use the full quantum model in Eqs.~\eqref{eqs:lindblad_circuitQED} and \eqref{eqs:hamiltonian_circuitQED}. 
Differently from Refs.~\cite{cohen_reminescence2023, dumas_measurement-induced_2024}, we are discarding the counter-rotating terms appearing from the cosine potential and the coupling with the resonator, and the gate charge in Eq.~\eqref{eqs:transmon_full}.
Such terms are relevant in describing chaos in the system, yet, the model we consider here is able to capture the emergence of DQC, and allows us to easily apply the framework developed in the previous sections.
We assume the following parameters as fixed

\begin{center}
\begin{tabular}{|c|c|c|c|c|}
    \hline
    $\quad g/2\pi\quad$ & $\quad\gamma_r/2\pi\quad$ &  $\quad\Delta/2\pi\quad$ & $\quad E_J/E_C\quad$\\
    \hline
    100MHz & 50MHz  & -750MHz & 50\\
    \hline
\end{tabular}
\end{center}

The above parameters have similar magnitudes of those contained in Refs.~\cite{walter_rapid2017, heinsoo_rapid_2018, sunada_fast_2022}, except a slightly larger resonator's decay rate.
We focus here on the minimal setup for dispersive readout, without considering Purcell filters or multiplexing.
We consider the transmon charging energy $E_C$ and the drive amplitude $F$, as the only varying parameters.
In particular, we consider charging energies in the interval $E_C/2\pi = 300-700\textrm{MHz}$. 
This leads to a transmon frequency range
\begin{equation}
    \omega_t/2\pi \simeq \left(\sqrt{8 E_CE_J} - E_C\right)/2\pi = 5.7-13.3\textrm{GHz}.
\end{equation}
We notice that since we are keeping fixed the detuning $\Delta = \omega_r-\omega_t$, the resonator frequency is varying accordingly.

\subsubsection{Jaynes-Cumming limit}
 
We now describe the dispersive readout protocol.
For illustrative purposes, let us consider a simplification of Eqs.~\eqref{eqs:lindblad_circuitQED} and \eqref{eqs:hamiltonian_circuitQED}, which is obtained by truncating all the transmon's energy levels beyond the ground and first excited state (the so-called two-level system approximation).
The Hamiltonian in Eq.~\eqref{eqs:hamiltonian_circuitQED} reduces to the driven-dissipative Jaynes-Cumming model
\begin{equation}\label{eqs:JC}
    \hat{H}_{\rm JC} = -\Delta\frac{\hat{\sigma}^z}{2} +F(\hat{a}^{\dagger} + \hat{a}) - g(\hat{a}^\dagger\hat{\sigma}^{-} + \hat{a}\hat{\sigma}^{+}).
\end{equation} 
This truncation is essentially exact in the limit of infinite anharmonicity, i.e., if $E_C\to\infty$.

The readout protocol is typically performed in the dispersive limit, where $g \ll |\Delta|$. A Schrieffer-Wolff transformation on the undriven Jaynes-Cumming Hamiltonian (see Ref.~\cite{blais_circuit_2021} for a full derivation) brings Eq.~\eqref{eqs:JC} to the well known dispersive Hamiltonian 
\begin{equation}\label{eqs:dispersive_jaynes_cumming}
    \hat{H}_{\rm disp} = \left(-\Delta + \frac{g^2}{\Delta}\right)\frac{\hat{\sigma}^z}{2} + F(\hat{a} + \hat{a}^\dagger) + \frac{g^2}{\Delta}\hat{a}^\dagger\hat{a}\hat{\sigma}^z.
\end{equation}
The advantage of considering Eq.~\eqref{eqs:dispersive_jaynes_cumming} is that now the equations of motion for the cavity and the qubit can be straightforwardly solved. 
If we assume that $\langle\hat{a}^\dagger\hat{a}\hat{\sigma}_z\rangle\simeq\langle\hat{a}^\dagger\hat{a}\rangle\langle\hat{\sigma}_z\rangle$, we obtain the following equation for the cavity field's coherence 
\begin{equation}
    \frac{\partial}{\partial t}\langle\hat{a}\rangle = \left(\frac{ig^2}{\Delta}\langle\hat{\sigma}_z\rangle - \frac{1}{2}\gamma_r\right)\langle\hat{a}\rangle - iF.
\end{equation}
The steady-state field in the cavity is given by
\begin{equation}\label{eqs:quasi_steady_states}
    \langle\hat{a}\rangle_{\rm ss} = \frac{F}{g^2\langle\hat{\sigma}_z\rangle/\Delta + i\gamma_r/2},
\end{equation}
thus, it depends on how the system has been initialized.
This dependence on $\langle\hat{\sigma}^z\rangle = \pm 1$ allows to deduce the initial state of the qubit by monitoring, e.g., the field quadratures $x$ and $p$ \cite{walter_rapid2017, heinsoo_rapid_2018, sunada_fast_2022}. Notice that, within this simple picture, the cavity photon number does not depend on the initial state of the transmon.
Furthermore, as $\hat{\sigma}^z$ commutes with $\hat{H}_{\rm disp}$, the dispersive readout is a quantum nondemolition measurement in this limit.
Finally, dispersive approximation requires a resonator's photon number not exceeding the critical photon number $n_{\rm crit} = (\Delta/2g)^2$. 
If $n_{\rm crit}$ is exceeded, a large transfer of photons between the qubit and the cavity occurs, and the Hamiltonian in Eq.~\eqref{eqs:dispersive_jaynes_cumming} is no longer a valid description of the system. 

Even in the absence of a qubit photon-loss mechanism, the qubit inherits a small decay rate $\gamma_t = (g/\Delta)^2\gamma_r$ from the readout cavity, due to the Purcell effect \cite{blais_circuit_2021}. 
As such, the (unique) steady state of the circuit QED setup is $\hat{\rho}_{\rm ss} = \ketbra{\alpha}\otimes\ketbra{0}$, where $\alpha = F/(-g^2/\Delta + i\gamma_r/2)$, and it is reached on a time scale dictated by $\gamma_t$.
The full point of dispersive readout is to be faster than this time scale, such that the system remains in the $\langle\hat{\sigma}^z\rangle$-dependent state in Eq.~\eqref{eqs:quasi_steady_states} for the measurement time.

To derive Eqs.~\eqref{eqs:dispersive_jaynes_cumming} and \eqref{eqs:quasi_steady_states} we made the fundamental assumption that the transmon qubit can be approximated by a two-level system.
This is not necessary for the dispersive approximation, which instead only relies on the conditions $g/\Delta \ll 1$ and $\langle\hat{a}^\dagger\hat{a}\rangle<n_{\rm crit}$.
It is possible to derive a more accurate dispersive Hamiltonian that takes into account the multi-level structure of the transmon (see Ref.~\cite{blais_circuit_2021} for a full derivation), but the equations of motion can not be solved analytically anymore.

\subsection{Phase diagram}
\label{sec:transmon_phasediagram}

\begin{figure}[t!]
\includegraphics[width=0.47\textwidth]{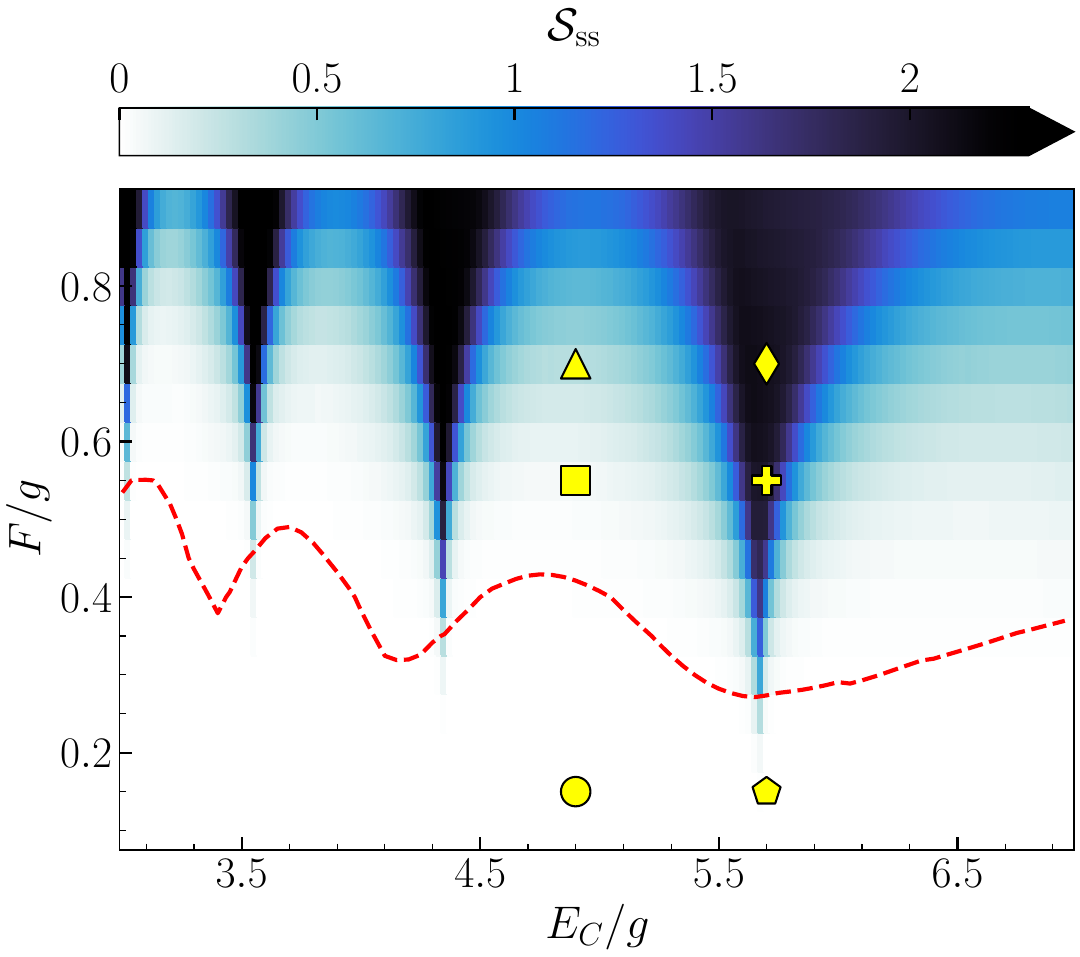}
\caption{
Von Neumann entropy $\mathcal{S}_{\rm ss}$ of the steady state of the circuit QED model described by Eqs.~\eqref{eqs:lindblad_circuitQED} and \eqref{eqs:hamiltonian_circuitQED} as a function of the resonator drive amplitude $F/g$ and the transmon charging energy $E_C/g$. 
The red dashed line indicates the boundary of the region where $\mathcal{S}>1$ at $gt=100$ when the transmon is initialized in $\ket{1}$.
Other parameters are set to $E_J = 50E_C$, $\Delta/g= -7.5$, $\gamma_r/g = 0.5$.
}
\label{fig:transmon_phasediagram}
\end{figure}

\begin{figure*}[t]
\includegraphics[width=\textwidth]{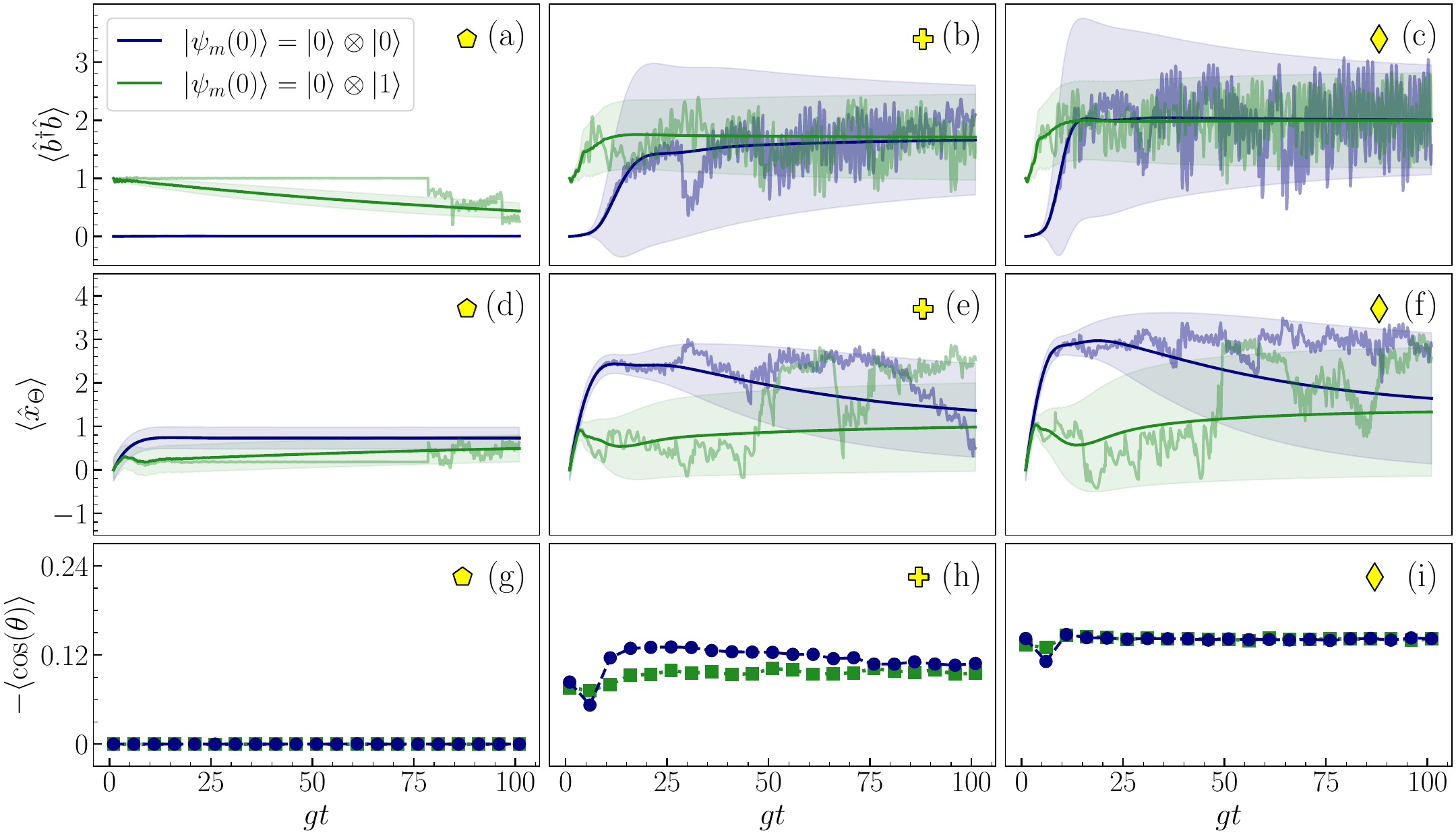}
\caption{Circuit QED architecture described by Eqs.~\eqref{eqs:lindblad_circuitQED} and \eqref{eqs:hamiltonian_circuitQED} at $E_C/g=5.7$ and for the three highlighted points in Fig.~\ref{fig:transmon_phasediagram} corresponding to $F/g=0.15$, $F/g=0.55$, $F/g=0.7$.
(a-c) Transmon's dynamics. 
Blue curves indicate that the initial state of the transmon is $\ket{0}$. 
Green curves indicate that the initial state of the transmon is $\ket{1}$.
The initial state of the resonator is always $\ket{0}$.
Dark-colored curves represent the transmon photon number $\langle\hat{b}^{\dagger}\hat{b}\rangle$ computed with the master equation \eqref{eqs:lindblad_circuitQED}.
Colored shadings respresent the photon-number variance $\Delta n_b^2 = \langle(\hat{b}^{\dagger}\hat{b})^2\rangle - \langle\hat{b}^{\dagger}\hat{b}\rangle^2$ always computed with the master equation.
Light-colored curves represent the transmon photon number along a single quantum trajectory.
(d-f) Readout's dynamics.
Following the same color scheme of panels (a-c) we plot the rotated resonator position $\langle\hat{x}_{\Theta}\rangle$, with $\Theta$ defined in \cite{Note10}, both with the master equation and along a single trajectory, and the variance $\Delta x_{\Theta}^2$.
(g-i) Application of the SSQT on the system's dynamics.
We plot the time evolution of  $\langle\cos(\theta)\rangle$ averaging over $100$ independent quantum trajectories initialized in $\ket{\psi_m(0)} = \ket{0}\otimes\ket{1}$ (green dotted lines with square markers) and $\ket{\psi_m(0)} = \ket{0}\otimes\ket{0}$ (blue dashed lines with circular markers). 
When the number of selected eigenvalues is $N_{\lambda}<500$, we directly set $\langle\cos(\theta)\rangle=0$.
The cutoff in the Hilbert space has been fixed at $N_b = 7$ for the transmon and $N_a = 19$ for the resonator. $c_{\rm min}$ has been fixed to $\bar{C}\times10^{-2}$ (see Appendix~\ref{App:Choice_cmin}).}
\label{fig:transmon_steadystate}
\end{figure*}

\begin{figure*}[t]
\includegraphics[width=\textwidth]{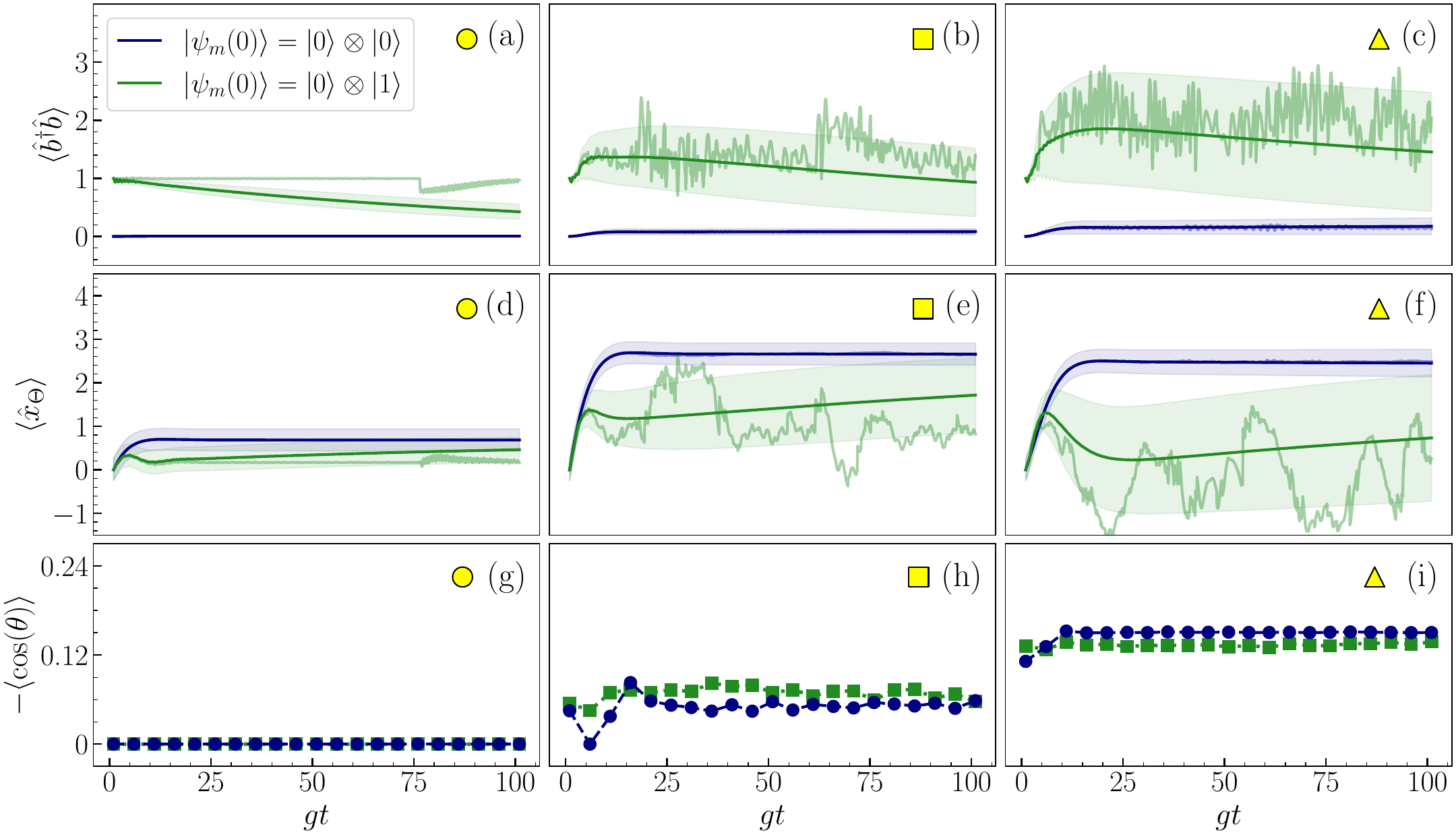}
\caption{As in Fig.~\ref{fig:transmon_steadystate}, but with $E_C/g=4.9$.}
\label{fig:transmon_transient}
\end{figure*}

\begin{figure}[h!]
    \centering
    \includegraphics[width=0.45 \textwidth]{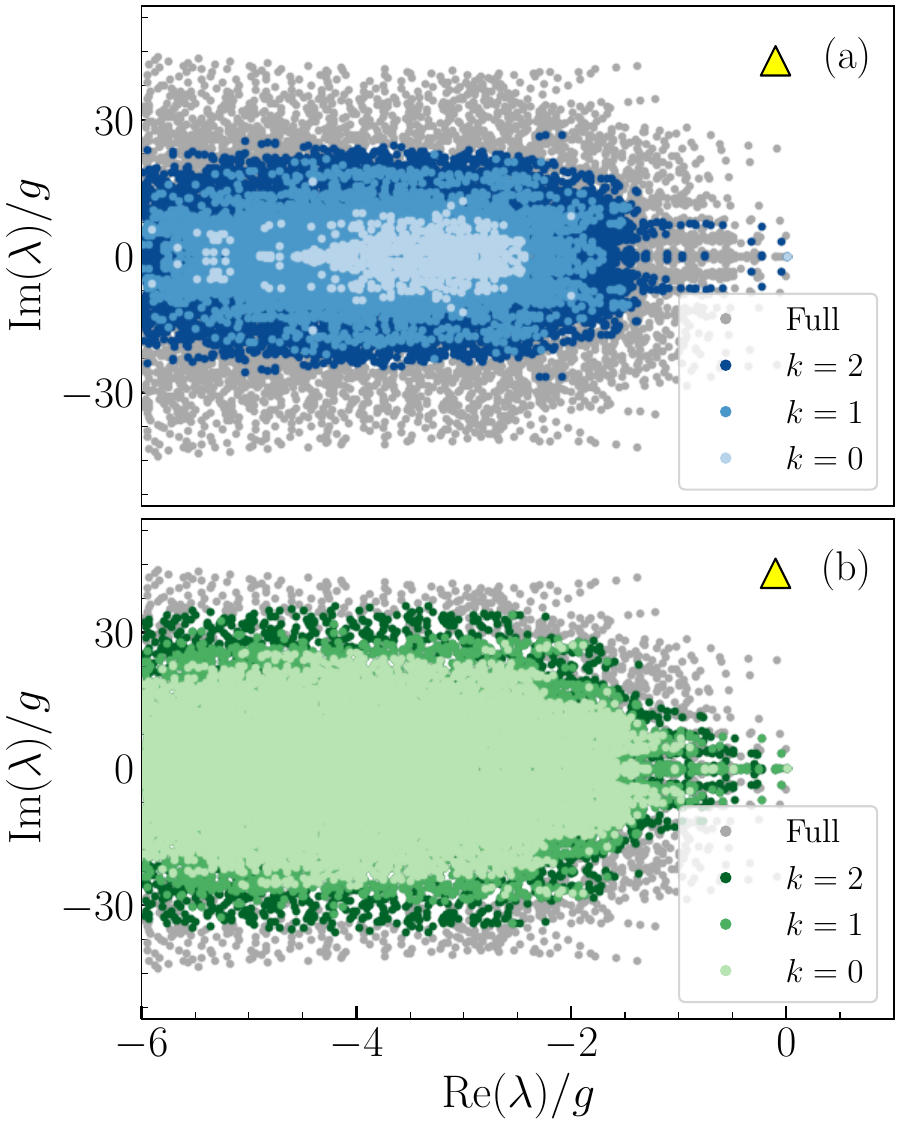}\vspace{0.5em}
    \caption{Liouvillian spectra for the circuit QED architecture for $E_C/g=4.9$ in the strongly driven regime corresponding to $F/g=0.7$. 
    At $gt=50$, the activated spectrum of single trajectories initialized in:
    (a) $\ket{\psi_m(0)}=\ket{0}\otimes\ket{0}$; (b) $\ket{\psi_m(0)}=\ket{0}\otimes\ket{1}$.
    In both plots, black dots indicate the full spectrum, while the eigenvalues selected by the SSQT are shown three values of $c_{\rm min} = \bar{C}\times10^{-k}$: $k=0, 1, 2$ (light-to-dark blue/green). 
    While in (a) the set of relevant eigenvalues is deep in the spectrum and disconnected from the steady state [c.f. the scheme in Fig.~\ref{fig:spectra_example} (d)], in (b) the selected eigenvalues spread from the steady state towards the depth of the spectrum, [c.f.  Fig.~\ref{fig:spectra_example} (b) and  Fig.~\ref{fig:steady_vs_transient_chaos} (d)].
    Other parameters are set as in Fig.~\ref{fig:transmon_phasediagram}.
    }
    \label{fig:Spectra_transmon}
\end{figure}

We first study the system's phase diagram at the steady state $\hat{\rho}_{\textrm{ss}}$. 
Fig.~\ref{fig:transmon_phasediagram} shows the Von Neumann entropy $\mathcal{S}_{\rm ss} = -\operatorname{Tr}(\hat{\rho}_{\textrm{ss}}\log\hat{\rho}_{\textrm{ss}})$ as a function of the drive amplitude $F/g$ and the transmon charging energy $E_C/g$, while keeping the ratio $g/\Delta$ and $E_C/E_J$ fixed.
The result brings evidence of a complete departure from the two-level system and dispersive approximations which, even within a driven-dissipative picture, predict $\mathcal{S}_{\rm ss}=0$, i.e., a pure state. 
We stress that, for the parameters we are considering, the critical photon number is $n_{\textrm{crit}} = (\Delta / 2 g)^2 \simeq 14$, while the largest steady-state photon number in Fig.~\ref{fig:transmon_phasediagram} is $n_{\textrm{ss}} = \langle \hat{a}^\dagger \hat{a} \rangle_{\textrm{ss}} \simeq 9.3$.
Interestingly, the regions where a high-entropy steady state emerges correspond to a choice of the parameter where multiphoton resonances between the vacuum and the $n$-th excited state of the transmon can occur, as detailed in Appendix \ref{sec:Transmon_appendix}, where we show that the breakdown of the dispersive approximation can be understood through a mean-field analysis, where the cavity effectively drives the transmon.
Thus, the multilevel structure of the transmon can play a crucial role in the dynamics, deviating from the predictions of the idealized models even when $n<n_{\rm crit}$, as also pointed out in Refs.~\cite{cohen_reminescence2023, dumas_measurement-induced_2024}.

Although the phase diagram refers to the steady-state entropy, very similar values of $\mathcal{S}$ can be observed along the transient dynamics when the system is initialized in $\hat{\rho}(0) = \ketbra{0}\otimes\ketbra{0}$ (not shown). A completely different entropy phase diagram is found when considering the transient dynamics following the initial state $\hat{\rho}(0) = \ketbra{0}\otimes\ketbra{1}$. 
In this case the region where the density matrix is highly entropic is significantly larger than before and occurs also in between two multiphoton resonances, as illustrated in  
Fig.~\ref{fig:transmon_phasediagram}.

\subsection{Emergence of dissipative quantum chaos}
\label{sec:chaos_transmon}

We now study the emergence of transient and steady-state chaos in the dispersive readout.

\subsubsection{DQC at the multiphoton resonance}

We consider the dispersive readout across a multiphoton resonance, studying the three points highlighted in Fig.~\ref{fig:transmon_phasediagram} at $E_C/g=5.7$, for initial states $\hat{\rho}(0) = \ketbra{0}\otimes\ketbra{0}$ and $\hat{\rho}(0) = \ketbra{0}\otimes\ketbra{1}$.
In Figs.~\ref{fig:transmon_steadystate} (a-c) we show the dynamics of the average transmon photon number $n_b = \langle\hat{b}^{\dagger}\hat{b}\rangle$ and its variance $\Delta n_b^2 = \langle(\hat{b}^{\dagger}\hat{b})^2\rangle - \langle\hat{b}^{\dagger}\hat{b}\rangle^2$, both computed by solving the master equation \eqref{eqs:lindblad_circuitQED}. We superimpose the results of a single quantum trajectory $\langle \psi_m(t)|\hat{b}^{\dagger}\hat{b}|\psi_m(t)\rangle$.
At low drive amplitude $F/g$ the two-level approximation remains valid.
The qubit's dynamics is confined to the 2D manifold spanned by $\ket{0}$ and $\ket{1}$ and the  behavior of the quantum trajectories is very regular (until a jump in the qubit occurs).
When the drive amplitude is increased, instead, the system's dynamics and steady-state become remarkably different from the previous case. 
Regardless of the qubit's initial state, $n_b$ rapidly converges to $n_b>1$, the variance $\Delta n_b^2$ becomes significantly large, and the quantum trajectory exhibits a highly fluctuating behavior. 

The transmon's dynamics is mirrored by the readout.
In Figs.~\ref{fig:transmon_steadystate} (d-f) we plot the rotated position $\langle\hat{x}_{\Theta}\rangle$ of the field within the resonator, where $\Theta$ is chosen to maximize the distance between the averaged resonator positions in phase space when the qubit is initialized in $\ket{0}$ and in $\ket{1}$ \footnote{To compute $\hat{x}_{\Theta}$ it is sufficient to perform a rotation in the $xp$-plane of an angle $\Theta$ so that
\begin{equation}
    \hat{x}_{\Theta} = \hat{x}\cos({\Theta}) + \hat{p}\sin({\Theta})
\end{equation}
and $\Theta = \arctan\frac{p_1-p_0}{x_1-x_0}$ where $x_{0, 1}$ and $p_{0, 1}$ are the position and the momentum of the resonator when the transmon is initialized in $\ket{0}$ or in $\ket{1}$. In the numerical simulations we computed $x_{0, 1}$ and $p_{0, 1}$ at $gt=50$, and we set the corresponding $\Theta$ for all the times.}. In experiments, 
$\langle\hat{x}_{\Theta}\rangle$ is normally reconstructed through homodyne detection \cite{blais_circuit_2021}.
We also show the evolution of $\Delta x_\Theta^2 = \langle\hat{x}_{\Theta}^2\rangle - \langle\hat{x}_{\Theta}\rangle^2$ (again computed with the master equation) and of $\expval{\hat{x}_{\Theta}}$ along a single quantum trajectory.
At low $F/g$, the cavity field behaves accordingly to the dispersive readout protocol, but the two cavity states are not well distinguishable because of the small number of photons in the cavity.
For larger $F/g$ single quantum trajectories are highly fluctuating regardless of the qubit's initialization.
As the distribution of the field quadratures of the two states overlap, the dispersive readout protocol is not applicable.

We now demonstrate that this behavior originates from the emergence of transient and steady-state quantum chaos.
In Figs.~\ref{fig:transmon_steadystate} (g-i) we plot $\langle\cos(\theta)\rangle$  as a function of time for the two initial states.
This indicator signals the transition from a regular state at $F/g=0.15$ to a strongly nonintegrable state characterized by $-\langle\cos(\theta)\rangle \simeq 0.13$ for higher drive amplitudes, as the system enters in the high entropy phase.
Such a transition takes place regardless of the initial state.
These observations indicate that the two-level system approximation of Eq.~\eqref{eqs:transmon} is no longer valid and that the multiphoton resonances do not simply induce a transition out of the qubit manifold, but they also lead to DQC.

\subsubsection{DQC outside the multiphoton resonance}

We consider the dispersive readout outside a multiphoton resonance.
In particular, we examine the three points highlighted in Fig.~\ref{fig:transmon_phasediagram} at $E_C/g=4.9$.
Repeating the previous analysis for this choice of parameters, in Figs.~\ref{fig:transmon_transient} (a-c) we study the dynamics of the transmon photon number $n_b = \langle\hat{b}^{\dagger}\hat{b}\rangle$. 
At low drive amplitude $F/g$, we see no difference with the previous case, as the two-level approximation is valid in both.
Increasing the drive amplitude leads to a different behavior in the dynamics with respect to Figs.~\ref{fig:transmon_steadystate} (b) and (c).
Initializing the qubit in $\ket{0}$, the photon number in the transmon slightly deviates from zero, showing rapid and small fluctuations not predicted by the two-level approximation.
Surprisingly, when the qubit is initialized in $\ket{1}$, $\langle\hat{b}^\dagger\hat{b}\rangle$ immediately and significantly deviates from $1$ due to large fluctuations.

Also in this case, the dynamics of the readout cavity closely follows the behavior of the qubit, as shown in Figs.~\ref{fig:transmon_transient} (d-f).
While for $\hat{\rho}(0) = \ketbra{0}\otimes\ketbra{0}$, the cavity field evolves towards a coherent-like state (with rapid small fluctuations in the quantum trajectory), if $\hat{\rho}(0) = \ketbra{0}\otimes\ketbra{1}$, the trajectory becomes highly fluctuating and $\Delta x_\Theta^2$ is much larger.
In this case, the cavity states are profoundly different according to the qubit's initialization, in a fashion that is reminiscent of the definition of transient chaos provided in the previous sections.
As discussed in the Appendix~\ref{sec:Transmon_appendix}, the transient dynamics for these parameters is not captured by the dispersive readout approximation, as the state of the cavity significantly differs from a coherent state.

To link the observed behavior with the emergence of DQC, we apply again the SSQT criterion.
Results are presented in Figs.~\ref{fig:transmon_transient} (g-i).
At $F/g = 0.15$ the system is integrable.
At $F/g = 0.55$, both the transient dynamics of $\ket{\psi_m(0)}=\ket{0}\otimes\ket{1}$ and $\ket{\psi_m(0)}=\ket{0}\otimes\ket{0}$ exhibit nonintegrable behavior. 
Finally, at $F/g = 0.7$ the system enters a chaotic phase and $-\langle\cos(\theta)\rangle \simeq 0.13$, similar to the one in Fig.~\ref{fig:transmon_steadystate} (i).
We note that, while the emergence of chaos greatly affect the dynamics of the system, an analysis based on purely spectral criteria fails to describe the actual readout behavior.

To further support this analysis, , we study  the Liouvillian eigenvalues activation according to the SSQT criterion for
$E_C/g=4.9$ and $F/g=0.7$.
According to Fig. \ref{fig:transmon_transient} (i),
random matrix theory predicts non-integrability for both initial states, while  single quantum trajectories show different behavior.
We first consider the initial state $\ket{\psi_m(0)}=\ket{0}\otimes\ket{0}$.
In Fig.~\ref{fig:Spectra_transmon} (a) we plot the full Liouvillian spectrum and the Liouvillian eigenvalues $\{\lambda_j\}$ selected by the SSQT with cutoffs $c_{\rm min} = \bar{C}\times 10^{-k}$: $k=0, 1, 2$, along the dynamics of a single quantum trajectory. 
The most relevant eigenvalues, corresponding to smaller values of $k$, are placed \textit{deep} in the spectrum and are disconnected from $\lambda=0$.
Thus, they describe \textit{fast} physical processes that lead to a negligible deviation of the observable along a single trajectory from the value predicted in the steady state.

We then consider the initial state $\ket{\psi_m(0)}=\ket{0}\otimes\ket{1}$.
In Fig.~\ref{fig:Spectra_transmon} (b) we plot again the full Liouvillian spectrum and the Liouvillian eigenvalues $\{\lambda_j\}$ selected by the SSQT criterion, for the same choices of $k$. 
Contrarily to the previous case, $\{\lambda_j\}$ spread from $\lambda=0$ to the depth of the spectrum, describing now also \textit{slow} physical processes.
These result in large oscillations, bringing the transmon to a chaotic manifold that completely breaks down the two-level system approximation.
This configuration resembles the chaotic Bose-Hubbard dimer studied in Sec.~\ref{sec:quantum_simulation} and illustrated in Figs.~\ref{fig:steady_vs_transient_chaos} (c-d).

\subsection{Readout performance and DQC}
\label{sec:readout_transmon}

\begin{figure*}[t]
\includegraphics[width=0.9\textwidth]{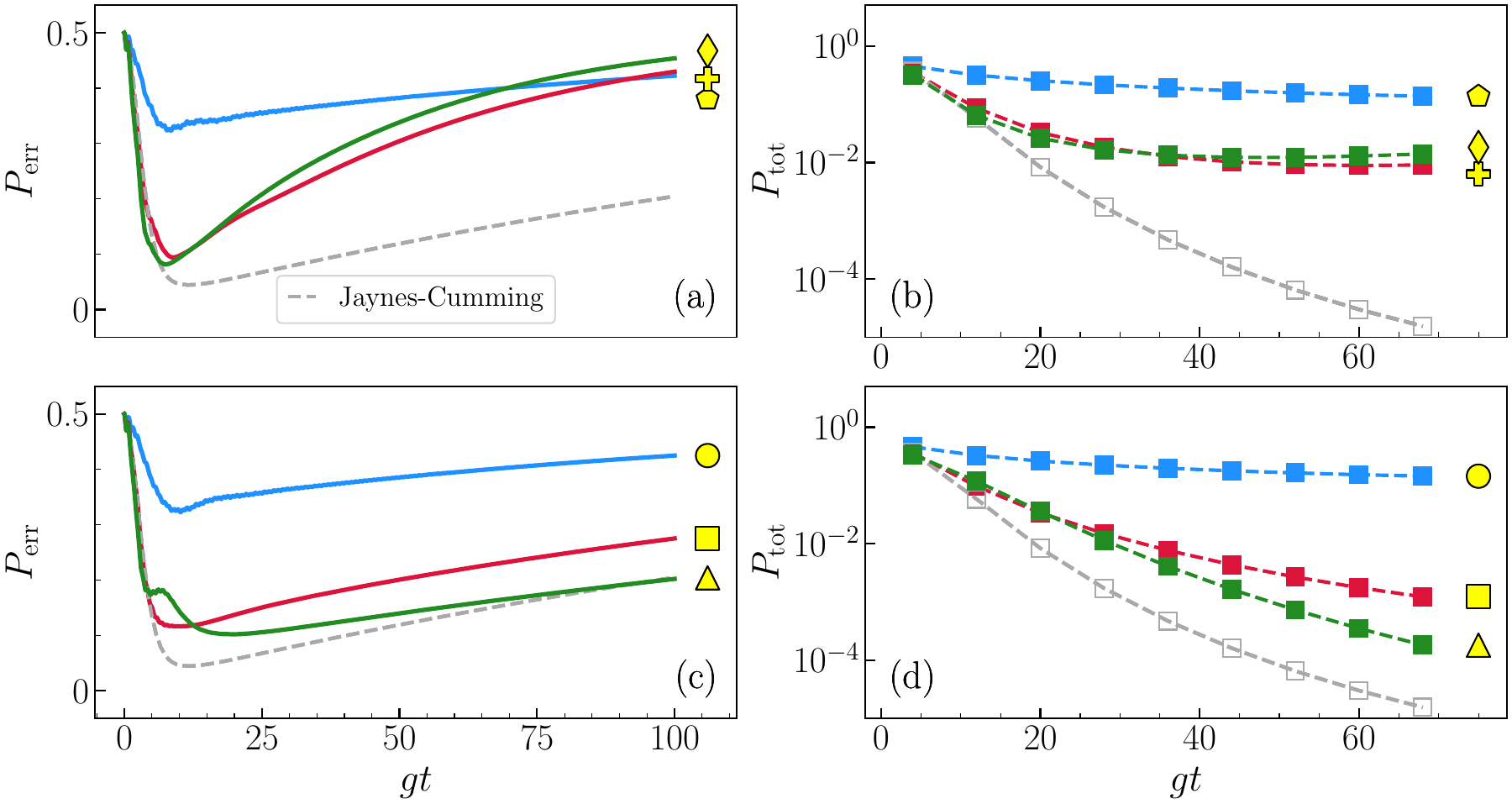}
\caption{Dynamics of the instantaneous and total error probability computed according to Eqs.~\eqref{eqs:error_probability} and \eqref{eqs:total_error_probability}, respectively.
(a) $P_{\rm err}$ as a function of time for $E_C/g=5.7$ (at the multiphoton resonance) for the same three drive amplitudes considered in Fig.~\ref{fig:transmon_steadystate}.
The gray-dashed lines corresponds to the error probability given by the driven-dissipative Jaynes-Cumming model in Eq.~\eqref{eqs:JC}, with parameters $\Delta/g$ and $\gamma_r/g$ chosen from the table in Sec.~\ref{sec:model_transmon} and $F/g=0.7$.
(b) $P_{\rm tot}$ as a function of time, using the same parameters and color scheme of panel (a).
The points over which $P_{\rm tot}$ has been computed corresponds to an odd length of the lists $\mathcal{P}$ and $\mathcal{Q}$ appearing in Eq.~\eqref{eqs:total_error_probability}.
(c) and (d) same of panels (a) and (b) but for $E_C/g=4.9$ (outside the multiphoton resonance).
Other parameters are set as in Fig.~\ref{fig:transmon_phasediagram}.}
\label{fig:transmon_readout}
\end{figure*}

We have established the emergence of DQC in Eqs.~\eqref{eqs:lindblad_circuitQED} and \eqref{eqs:hamiltonian_circuitQED}.
We now characterize how the presence of chaos affects the readout performance from a quantum information perspective.
For the readout to work, a necessary condition is that the reduced state of the resonator must depend on the qubit's initial state.
Namely, the two states given by $s=0$ and $s=1$ in the the reduced-density matrix
\begin{equation}
    \hat{\rho}_s(t) = \operatorname{Tr}_t\left[e^{\mathcal{L}t}\left(\ketbra{0}\otimes\ketbra{s}\right)\right],
\end{equation}
need to be distinguishable.
We define the istantaneous error probability $P_{\rm err}$ as \cite{di_candia_critical_2023}
\begin{equation}\label{eqs:error_probability}
    P_{\textrm{err}}(t) = \frac{1}{2}\left(1-\frac{1}{2}||\hat{\rho}_0(t)-\hat{\rho}_1(t)||\right),
\end{equation}
where $||\hat{O}|| = \operatorname{Tr}\left(\sqrt{\hat{O}^{\dagger}\hat{O}}\right)$ is the trace norm and $||\hat{\rho}_0(t)-\hat{\rho}_1(t)||$ is the distinguishability between the two states. 
For $P_{\rm err}(t) = 0$ the two cavity states are orthogonal, while for 
$P_{\rm err}(t) = 0.5$, they coincide.
Importantly, $P_{\rm err}$ is meaningful within an infinitesimal time interval $dt$: it provides an upper bound to the amount of information we can obtain, about the state of the system, from a measurement carried out within the interval $dt$.

In experiments, the readout is not performed over an infinitesimal time interval $dt$, but rather over an time interval whose typical duration is 50-100ns \cite{walter_rapid2017, heinsoo_rapid_2018, sunada_fast_2022} -- shorter than the cavity-induced transmon decay time.
Collecting the signal over an extended time interval allows to significantly lower the instantaneous error probability.
To estimate the maximally achievable readout fidelity from Eq. \eqref{eqs:error_probability}, we adopt the following procedure.
First, we suppose to be able to reconstruct with a quantum state tomography the resonator's reduced density matrix at time $t$.
We then collect $N$ measurements sampling $P_{\rm err}$ each $2/\gamma_r = 4g$ (so that the system ``decorrelates" again after a measurement). 
Each error probability is integrated over the time step separating two samples.
We obtain a list of sampled error probabilities $\{P_{\rm err}^{(1)}, P_{\rm err}^{(2)}, ..., P_{\rm err}^{(N)}\}$ where
\begin{equation}
    P_{\rm err}^{(k)} = \frac{1}{t_{k}-t_{k-1}}\int_{t_{k-1}}^{t_k}dtP_{\rm err}(t).
\end{equation}
If we have done less than $k/2$ errors, we assume that the measurement gives us the correct result.
The total error probability of having done an error along the whole measurement can be therefore constructed with the binomial expansion
\begin{equation}\label{eqs:total_error_probability}
    P_{\rm tot}^{(k)} = \sum_{j=1}^{N_k}\left(\prod_{\ell\in \mathcal{P}_j}P_{\rm err}^{(\ell)}\right)\left[\prod_{\ell\in \mathcal{Q}_j}\left(1 - P_{\rm err}^{(\ell)}\right)\right].
\end{equation}
Here, $\mathcal{P}$ is the ensemble of lists containing: the list with $k$ error probabilities, all the lists with $k-1$ error probabilities, all the lists with $k-2$ error probabilities and so on until $k/2$. $\mathcal{Q}$ is instead the ensemble of lists containing: no elements, all the lists with $1$ error probabilities, all the lists with $2$ error probabilities and so on until $k/2$. Clearly, $\mathcal{P}_j\cup\mathcal{Q}_j = \{P_{\textrm{err}}^{(1)}, ..., P_{\textrm{err}}^{(k)}\}$ and $\mathcal{P}_j\cap\mathcal{Q}_j = \varnothing$. $N_k$ is the total length of $\mathcal{P}$ and $\mathcal{Q}$.

In Fig.~\ref{fig:transmon_readout} we plot $P_{\rm err}$ and $P_{\rm tot}$ as a function of time for the driven-dissipative circuit QED setup of Eqs.~\eqref{eqs:lindblad_circuitQED} and \eqref{eqs:hamiltonian_circuitQED} for the same parameters as in Figs.~\ref{fig:transmon_steadystate} and \ref{fig:transmon_transient}.
Notice that, since Eq.~\eqref{eqs:total_error_probability} is sensitive to the parity of $k$, we plot the $P_{\rm tot}$ corresponding to odd $k$.
In Fig.~\ref{fig:transmon_readout} (a) we plot $P_{\rm err}$ for $E_C/g=5.7$ at the multiphoton resonance.
At small $F/g$, $P_{\rm err}$ is large, due to the small number of photons in the readout cavity.
When increasing the drive amplitude, $P_{\rm err}$ decreases for short times but, at longer times, it rapidly increases.
For comparison, Fig.~\ref{fig:transmon_readout} (a) shows also $P_{\rm err}$ as obtained from the driven-dissipative Jaynes-Cumming model in Eq.~\eqref{eqs:JC}, with $F/g=0.7$. In this case, the error probability is significantly lower than in the full model, again pointing out at the importance of modeling the transmon beyond the two-level approximation.
The readout fidelity is assessed through the total error probability $P_{\rm tot}$ given by Eq.~\eqref{eqs:total_error_probability} and the results are plotted in Fig.~\ref{fig:transmon_readout} (b).
At small $F/g$, almost no information can be extracted from the readout process.
As we increase the drive amplitude, the system is brought in the chaotic manifold and $P_{\rm tot}$, after an initial decrease, reaches a plateau at values $P_{\rm tot}>10^{-2}$.
The plateau coincides with the rapid increase of $P_{\rm err}$, and with the emergence of dissipative quantum chaos, as shown in Fig.~\ref{fig:transmon_steadystate}.
In the two-level limit instead, $P_{\rm tot}$ decreases as the measurement duration is increased, eventually leading to an almost ideal measurement.
These findings suggest that, due to the transmon energy spectrum and to the chaotic dynamics of the readout cavity, the information about the initial state is degraded during the readout process, and a longer measurement does not necessarily improve the readout fidelity. 

In Fig.~\ref{fig:transmon_readout} (c) we plot $P_{\rm err}$ for $E_C/g=4.9$, i.e. outside the multiphoton resonance.
Despite the presence of DQC, the dispersive readout protocol is more accurate, and the instantaneous error probability decreases as $F/g$ is increased. In this case, the prediction of the two-level limit is close to the one of the full model.
A similar conclusion is drawn from the behavior of $P_{\rm tot}$ in Fig.~\ref{fig:transmon_readout} (d). 
In this case, the readout fidelity increases systematically with the driving field, leading to values of total error probability as small as $10^{-4}$.

Overall, our analysis illustrates how DQC, under the lenses of the SSQT criterion, could hinder the performance of superconducting-based noisy intermediate scale quantum devices. It highlights the importance of the SSQT for the correct assessment of the performance of the readout protocol.  

\section{Conclusions and outlook} \label{sec:conclusion}

We develop a general understanding of chaos and integrability in the dynamics of open quantum systems. 
We define the distinction between chaos and integrability through the features of single dynamical instances commonly referred to as quantum trajectories.
We extract the set of Liouvillian eigenvalues involved in the dynamics of individual trajectories, over which universal predictions of non-Hermitian random matrix theory apply.
This allows us to introduce the notions of steady-state and transient quantum chaos as separated and general phenomena in open systems.
While transient chaos is determined by the initial state, steady-state chaos, for the models we studied, depends only on the structure of the Lindblad master equation.
To characterize open quantum dynamics, we develop a model-independent criterion, the spectral statistics of quantum trajectories. 
We demonstrate that the SSQT correctly identifies chaos in several dissipative systems, where other criteria provide ambiguous results.

We apply our theoretical framework to two examples of driven-dissipative bosonic systems.
First, we study a driven-dissipative Bose-Hubbard dimer, explaining a rich phase diagram where integrability, steady-state, and transient chaos coexist. 
We further show that chaotic quantum trajectories can have effects on experimentally measurable quantities such as 
the emission spectrum.
This is an important step towards the understanding of realistic platforms for quantum simulation and a concrete example of how DQC manifests in quantum systems.
Second, we focus on the textbook example of the dispersive readout of a superconducting transmon qubit and we
demonstrate how the presence of DQC affect it.
We explain how the mechanism of eigenvalues activation across the Liouvillian spectrum, described by the SSQT, plays a fundamental role in determining the performance of the readout protocol. 
Our work paves the way for the investigation of chaos in realistic quantum computing and sensing devices, where the effects of both classical and quantum chaotic dynamics have been recently predicted \cite{berke_transmon_2022, cohen_reminescence2023, chavez-carlos_driving_2025, dumas_measurement-induced_2024}. 

Finally, we provide a physical example where the quantum-to-classical correspondence proposed in Ref.~\cite{grobe_quantum_1988} manifestly breaks down.
We identify a regime of parameters where steady-state quantum chaos persists in both the classical (where all quantum correlations are neglected) and semiclassical (considering quantum correlations to order $\hbar$) limits. We also find parameters where an emergent steady-state quantum chaos does not admit any classical or semiclassical counterpart. 
While these regimes exhibit different features at the classical level, several observables remain unchanged in the quantum regime.
We demonstrate how, in this region, quantum fluctuations are extremely relevant and are induced by the quantum noise effect of dissipation, which causes random deviations from regular phase-space trajectories, effectively turning classical integrable regions into chaotic attractors.

In the future, we plan to extend our results to the study of 
driven-dissipative bosonic systems designed for quantum computation \cite{gravina_critical2023} and sensing \cite{di_candia_critical_2023}.
These often display weak or strong Liouvillian symmetries. An interesting question then concerns the study of DQC in systems invariant under Liouvillian symmetries, as in this case the results are expected to depend significantly on the specific unraveling of the master equation \cite{bartolo_homodyne_2017}.
Looking instead at the quantum-to-classical correspondence, further study is warranted to systematically understand which conditions can lead to the departure from the GHS conjecture in open quantum systems.

\begin{acknowledgments}
We acknowledge enlightening discussions with Alberto Biella, Fabio Mauceri, Alberto Mercurio, Lorenzo Fioroni, and Léo Paul Peyruchat. 
This work was supported by the Swiss National Science Foundation through Projects No. 200020\_185015, 200020\_215172, and 20QU-1\_215928, and was conducted with the financial support of the EPFL Science Seed Fund 2021.

V.S. and F.M. contributed equally to this work.
\end{acknowledgments}

\appendix

\section{Details on methods}\label{sec:appendix}

\subsection{Liouvillian unfolding}\label{sec:liouvillian_unfolding}

The unfolding is a common procedure to characterize integrable or chaotic systems.
As explained in Ref.~\cite{guhr_random-matrix_1998}, in Hamiltonian systems it allows distinguishing the fluctuations, which are universal, from the averaged spectral density, which is system-specific.

Unfolding procedures are similarly achieved in open quantum systems by separating average and fluctuations in the spectral density distribution as
\begin{equation}
    \rho(\lambda) = \sum_{j=1}^N\delta^{(2)}(\lambda-\lambda_j) = \rho_{\rm av}(\lambda) + \rho_{\rm fl}(\lambda),
\end{equation}
where $\lambda_j$ are the complex-valued Liouvillian eigenvalues.
The method proposed by Ref.~\cite{akemann_universal_2019} consists of approximating the sum of delta functions in the previous equation with the smooth sum of Gaussians
\begin{equation}
    \tilde{\rho}_{av}(\lambda) \simeq \frac{1}{2\pi\sigma^2N}\sum_{j=1}^N\exp\left(-\frac{1}{2\sigma^2}|\lambda - \lambda_j|^2\right).
\end{equation}
Then the $j$-th spacing $s_j = |\lambda_j - \lambda_j^{\rm NN}|$ is mapped onto $s_j' = s_j\sqrt{\tilde{\rho}_{av}(\lambda_j)}/\bar{s}$, where $\bar{s}$ is chosen such that $\sum_{j=1}^Ns_j'/N = 1$, and the statistical distribution of $s_j'$ is evaluated. 
The parameter $\sigma$ should be chosen to make the unfolding procedure effective. It was found that setting $\sigma = 4.5\times\sum_{j=1}^Ns_j/N$ achieves such an optimal unfolding \cite{akemann_universal_2019}.

\subsection{Choice of $c_{\rm min}$}
\label{App:Choice_cmin}

\begin{figure}[h!]
    \centering
    \includegraphics[width=0.435 \textwidth]{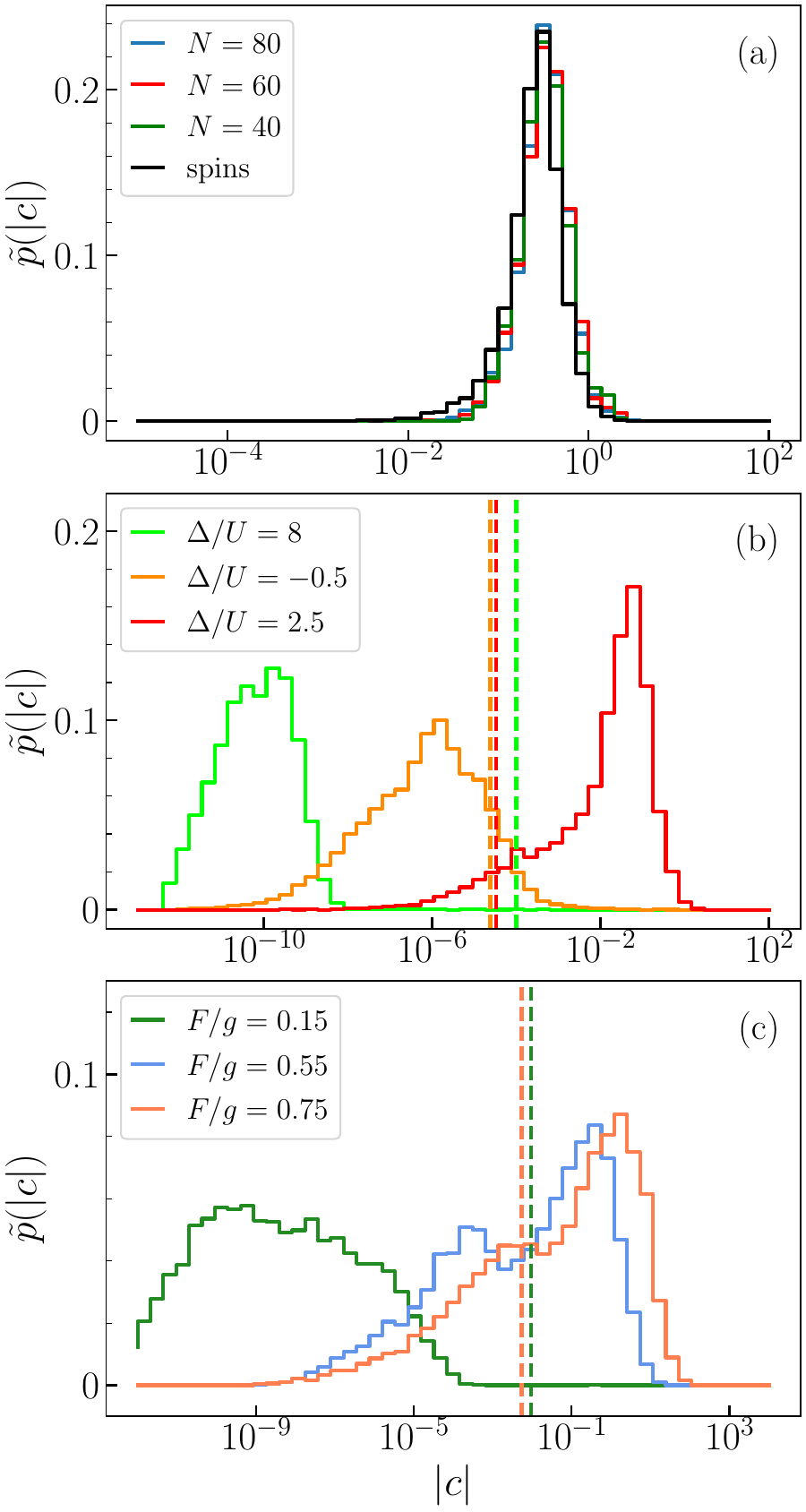}\vspace{0.5em}
    \caption{Distribution $\tilde{p}(|c|) = p(|c|)/\int d|c|p(|c|)$ obtained by binning the weights $\{ |c_{m, j}| \, \forall m, j\}$ as described in the text. 
    (a) $\tilde{p}(|c|)$ for the random chaotic Liouvillian described in Sec.~\ref{sec:rand_liouv}, and different sizes of the Hilbert space. In black, we also superimpose the distribution for the spin model in Sec.~\ref{sec:open_spin} in the deeply chaotic configuration $F/J =1$.
    (b) $\tilde{p}(|c|)$ for the Bose-Hubbard model described in Sec.~\ref{sec:quantum_simulation} and different detunings, having fixed $F/U = 3$, $J/U=2$ and $\gamma/U = 1$.
    (c) $\tilde{p}(|c|)$ for the circuit QED architecture described in Sec.~\ref{sec:quantum_information} and different drive amplitudes, having fixed $E_C/J=5.7$, $E_J=50E_C$, $\Delta_r/J = 0$, $\Delta_t/J=-7.5$.
    The dashed lines in panels (b) and (c) represent the $c_{\rm min}$ cutoff chosen as described in Sec.~\ref{App:Choice_cmin}.
    For panel (b) we set $c_{\rm min} = C_m\times10^{-4}$, for panel (c) we set $c_{\rm min} = C_m\times10^{-2}$. 
    Results have been obtained over one single quantum trajectory.
    }
    \label{fig:c_distribution}
\end{figure}

Consider a set of quantum trajectories and the corresponding $(\lambda_j, \vec{c}_j)$ as defined in Sec.~\ref{sec:SSQT}. 
The distribution of the weights $p(|c|)$ is obtained by binning the set $\{ |c_{m, j}| \,\forall\, m, j\}$, where $j$ labels the eigenvalue and $m$ each quantum trajectory.
As we show in Fig.~\ref{fig:c_distribution}, for all the chaotic models considered in the article, the distribution is very regular, and shows a clear peak $C_{\rm max}$.

We then fix $c_{\rm min}$ as the center of mass of the distribution $\bar{C}$ of the spectral coefficients as proposed in Sec.~\ref{sec:SSQT} [c.f. Eq.~\eqref{eqs:cutoff}], discarding $|c_{m, j}|>1$ \cite{Note4}.
We finally fix $c_{\textrm{min}} = \bar{C}\times 10^{-k}$, with $k\in[2,4]$. 
This is the case of the models studied in the main text.
For the models studied in the appendix, instead, given the almost equal distribution of $p(|c|)$ in Fig.~\ref{fig:c_distribution} (a), for the sake of simplicity we fixed $c_{\rm min} = 10^{-3}$.

As shown in Fig.~\ref{fig:c_distribution}, given a Liouvillian system, this way to set $c_{\rm min}$ produces cutoffs almost independent of the physical parameters of the model.
We observe that if $c_{\textrm{min}}\simeq C_{\rm max}$, many eigenoperators contribute to the spectral decomposition of each quantum trajectory. 
The present choice correctly captures most of these components, neglecting the few outliers.
If, on the other hand, $c_{\textrm{min}}\gg C_{\rm max}$, each quantum trajectory has a nonvanishing projection only on a few key eigenoperators.
The present choice of $c_{\rm min}$ correctly neglects all irrelevant components. 

\subsection{Classical analysis}\label{sec:Lyapunov}

The numerical estimation of the largest Lyapunov exponent $\Lambda_{\rm max}$ provides a clean signature of chaos in classical dynamical systems described by a set of ordinary differential equations.
Here we adopt the orbit separation method described in Ref.~\cite{sprott_chaos_2003}. 

We consider two trajectories $\vec{y}_1(t)$ and $\vec{y}_2(t)$ with initial conditions $\vec{y}_1(0)=\vec{y}_0$ and $\vec{y}_2(0) = \vec{y}_0 + \vec{\varepsilon}$ and with discretized time $[0, \Delta t, ..., n\Delta t]$.
We assume an uniform perturbation $\vec{\varepsilon} = [\varepsilon, ..., \varepsilon]$.
While the first trajectory $\vec{y}_1$ evolves according to the given ordinary differential equations system and the initial condition, the second trajectory (the perturbed trajectory) $\vec{y}_2$ is re-adjusted at each time step to be always at a distance $\varepsilon$ from $\vec{y}_1$.
In particular, at the $k$-th time step we measure the distance $d_k = |\vec{y}_1 - \vec{y}_2|$. 
Then the perturbed orbit is re-initialized according to
\begin{equation}
    \vec{y}_2 = \vec{y}_1 + \frac{\varepsilon}{d_k}(\vec{y}_1 - \vec{y}_2).
\end{equation}
After a transient dynamics, the logarithm of the relative separation $\ell_j = \log(|d_j/\varepsilon|)$ starts to be stored. After $n$ iterations the numerical largest Lyapunov exponent is computed as
\begin{equation}
    \Lambda_{\rm max} = \frac{1}{\Delta t}\sum_{j=1}^{n}\frac{\ell_j}{n}.
\end{equation}

\section{Comparison of criteria for steady-state chaos}\label{Sec:Comparison}

\subsection{Statistics of Liouvillian eigenvalues}\label{sec:rand_liouv}

We show here that the bare statistics of Liouvillian eigenvalues does not provide information about steady-state dynamics. 

To do this, we show that, starting from a Liouvillian $\LL$ with chaotic spectral signatures, a second Liouvillian $\LL^\prime$ with the \emph{same} bulk spectral statistics, and a pure steady state $\hat{\rho}_{\rm ss} = \ketbra{\Psi}$ can always be constructed. 
As for any quantum trajectory $\ket{\psi_m(t)}$ associated with $\LL'$ there exists $t_m$ such that $\ket{\psi_m(t\geq t_m)} = \ket{\Psi}$,
steady-state chaos can never occur in $\LL^\prime$.

\begin{figure}[t!]
\includegraphics[width=0.5 \textwidth]{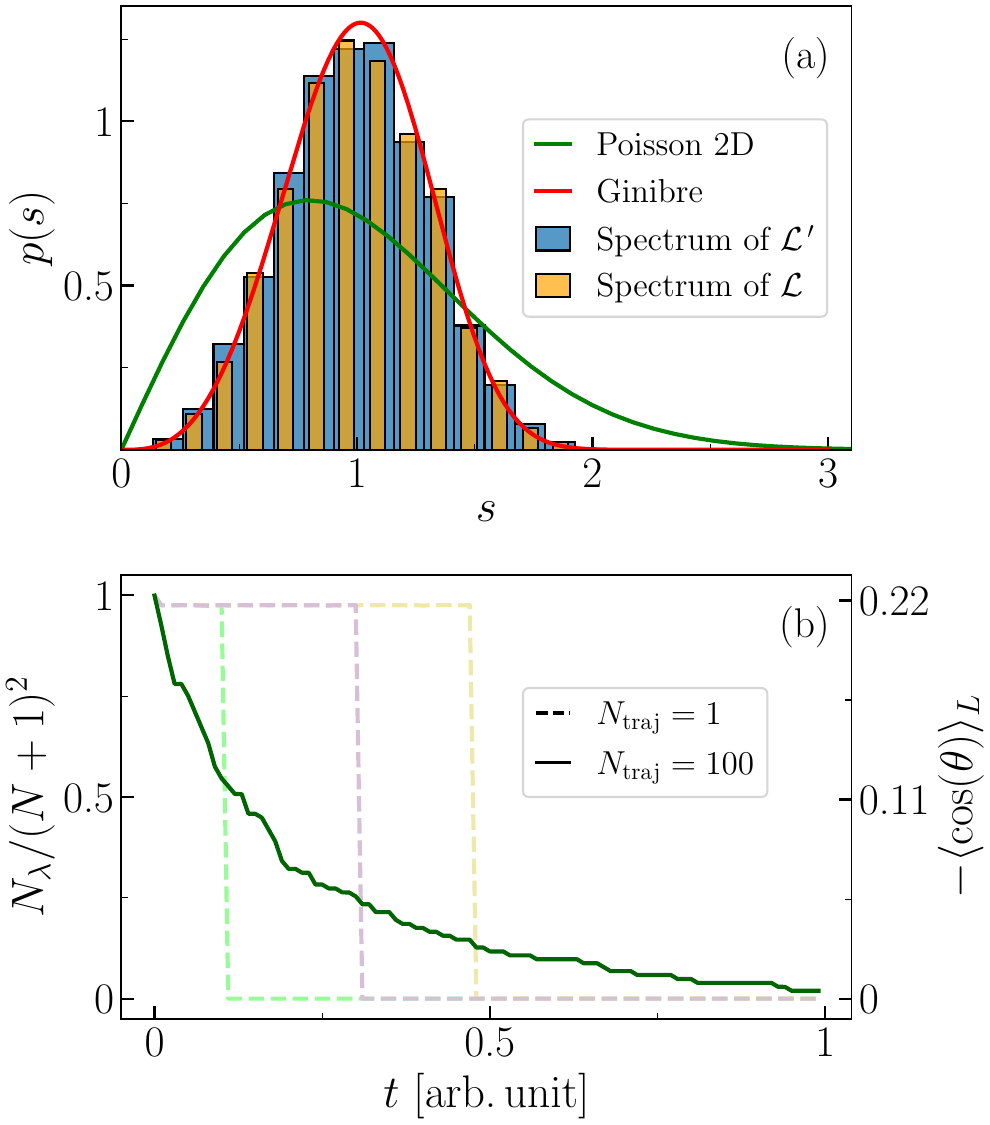}
\caption{Limitations of the full level statistics approach in determining steady-state chaos.
(a) Raw spectral statistics for the Liouvillian $\LL$ and $\LL^{\prime}$ defined in the main text. The green (red) curve is a guideline indicating an ideal 2D-Poisson (Ginibre) distribution.
(b) As a function of time, the number of relevant eigenvalues $N_{\lambda}(t)$. Initially, $N_{\lambda} \simeq (N+1)^2$ for any quantum trajectory (dashed lines), i.e., all the Liouvillian eigenvalues are relevant. At random times, a quantum jump leads the system towards the stationary state $\ket{N}$ so that $N_{\lambda}=1$. This behaviour is captured by the average dynamics (solid line), and exactly matched by the indicator $\langle\cos(\theta)\rangle$ [we set $\langle\cos(\theta)\rangle = 0$ when $N_\lambda=1$]. Parameters: $\beta=2$, $r=2$, $N=80$, and $g_{\rm eff}=\SI{100}{\rm{arb. unit}}$ (defined with respect to the variance of the Gaussian ensemble from which the Hamiltonian is drawn), and $\forall m$ $\ket{\psi_{m}(0)}$ is chosen randomly within the Hilbert space.}
\label{fig:enlarged_random_liouvillian}
\end{figure}

As a general example, we consider random Liouvillians. 
Random Liouvillians are the generators of a class of completely-positive and trace-preserving maps, whose spectra match the universal predictions of non-Hermitian random matrix theory \cite{denisov_universal_2019, can_random_2019, Sa_2020, sa_spectral_2020, costa_spectral2023, yang2024decoherence}.
The procedure we adopt is the following: 
\begin{itemize}
    \item We construct a random Liouvillian $\LL$ on a Hilbert space spanned by the orthornormal basis $\{\ket{0}, \ket{1}, \dots \ket{N-1} \}$.
    Following the procedure in Ref.~\cite{Sa_2020}, we define a basis for the operators in the Hilbert space  $\{\hat G_i\}$ with $i=0, \ldots, N^2-1$ such that $\operatorname{Tr}[\hat G_i^{\dagger} \hat G_j]=\delta_{i j}$, with $\hat G_0= \hat{\mathbb{1}} / \sqrt{N}$. We construct $r$ jump operators as $\hat L_{\mu}= g \sum_{j=1}^{N^2-1} \hat G_j w_{j, \mu}$, with $\hat L_{\mu}$ traceless and $w$ a matrix sampled from a Ginibre ensemble.
    We then draw one instance of the Hamiltonian $\hat{H}$ from a Gaussian unitary ensemble, and define $\LL $ according to Eq.~\eqref{eqs:lindblad_general}. The coupling parameter is given in terms of the rescaled parameter $g_{\rm eff} = (2r\beta N)^{1/4}g$.
    \item We determine the steady state $\hat{\rho}_{\rm ss}$ of this random Liouvillian and diagonalize it as $\hat{\rho}_{\rm ss} = \sum_j p_j \ketbra{\Psi_j}$.
    \item We extend the Hilbert space to include a new state $\ket{N}$.
    \item We embed $\LL$ into the larger Hilbert space, and construct the new Liouvillian $\LL' = \LL + \mathcal{D}[\ketbra{N}{\Psi_0}]$, where $\ket{\Psi_0}$ is the most probable eigenstate of $\hat{\rho}_{\rm ss}$. The new steady state is $\hat{\rho}_{\rm ss}' = \ketbra{N}$ by construction.
\end{itemize}
The Liouvillian $\LL'$ defines a completely-positive and trace-preserving map.
In Figure~\ref{fig:enlarged_random_liouvillian} (a) we display its bulk statistics alongside that of $\LL$. Both $\LL$ and $\LL^\prime$ follow a Ginibre distribution with only the former allowing for steady-state chaos. While the spectral statistics is unable to capture the steady-state properties of $\LL^\prime$, we show in Fig.~\ref{fig:enlarged_random_liouvillian} (b) that the SSQT criterion manages to correctly describe its transient chaotic dynamics giving way to an integrable steady state.

Notice that this procedure is general, can be extended to non-random Liouvillians, and that the Hilbert space extension can include more than one state.
From a physical perspective, this mathematical construction is a particular instance of dissipative quantum state engineering \cite{Verstraete_quantum_2009}, where the dynamics of an open quantum system is confined within a small portion of the Hilbert space.

\begin{figure}[t]
\hspace*{-3em}
\includegraphics[width=0.45\textwidth]{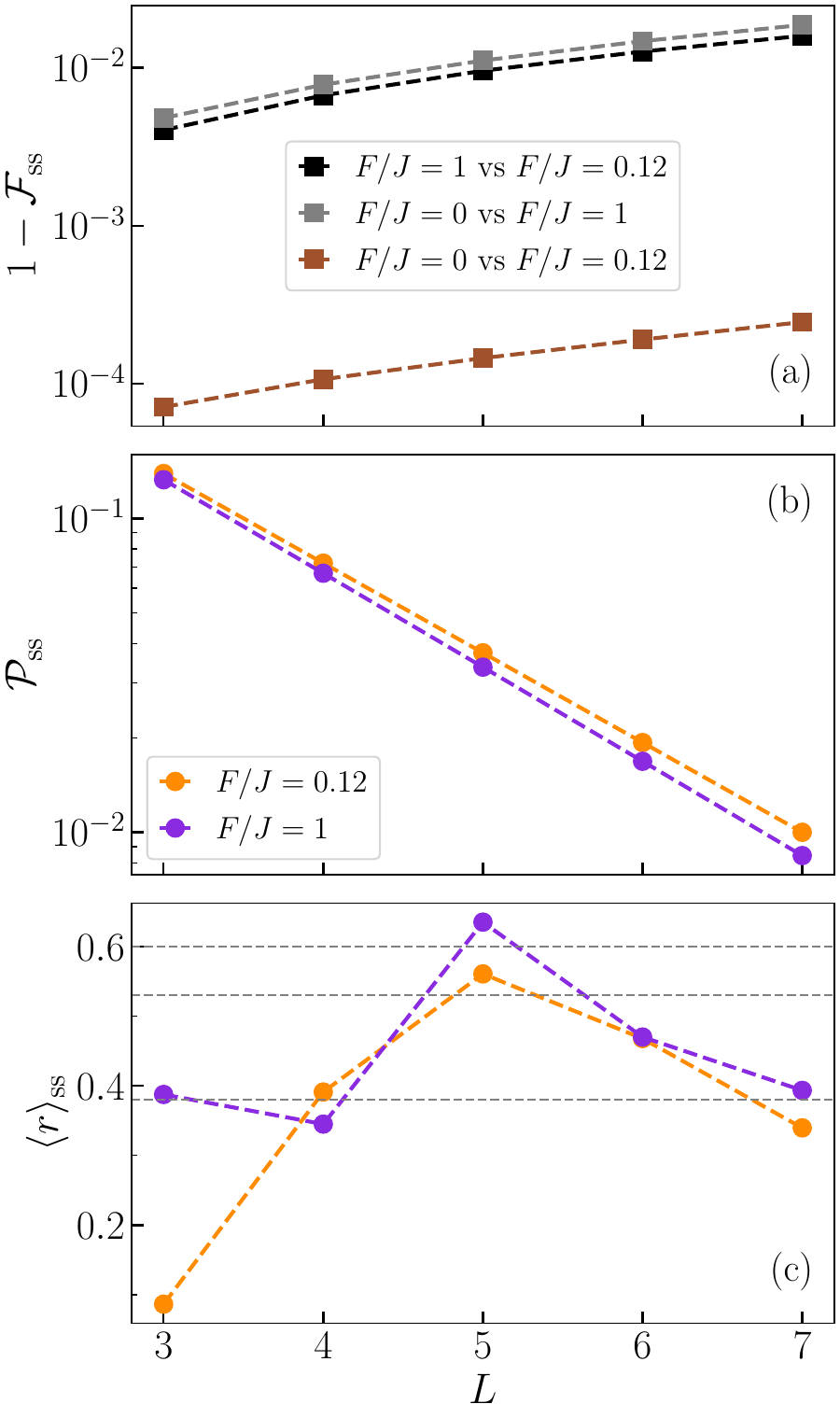}
\caption{Limitations of a steady-state density matrix analysis in a coherently-driven open spin chain. (a) Fidelity $\mathcal{F}(\rho, \sigma) = \left(\operatorname{Tr} \sqrt{\sqrt\rho \sigma\sqrt\rho}\right)^2$  between the steady states of the three configurations $F/J =0$, $F/J =0.12$, and $F/J =1$ for increasing system size $L$.
(b) As a function of $L$, the purity $\mathcal{P}_{\rm ss} = \operatorname{Tr}(\hat{\rho}_{\rm ss}^2)$ for the three configurations (the cases $F/J =0$ and $F/J =0.12$ overlap within this graphical resolution). (c) As a function of $L$, the quantifier  $ \langle r \rangle_{\rm ss}$.
The three dashed lines represent the Hamiltonian ratio for (from the bottom to the top) uncorrelated random variables, the Gaussian Orthogonal Ensemble, the Gaussian Unitary Ensemble.
Parameters: $\gamma/J=1$, $\gamma_1^{+}/J = 0.5$, $\gamma_1^{-}/J = 1.2$, $\gamma_L^{+}/J = 1$, $\gamma_L^{-}/J = 0.8$.
}
\label{fig:steady_state_analysis}
\end{figure}

\begin{figure}[t]
\includegraphics[width=0.45\textwidth]{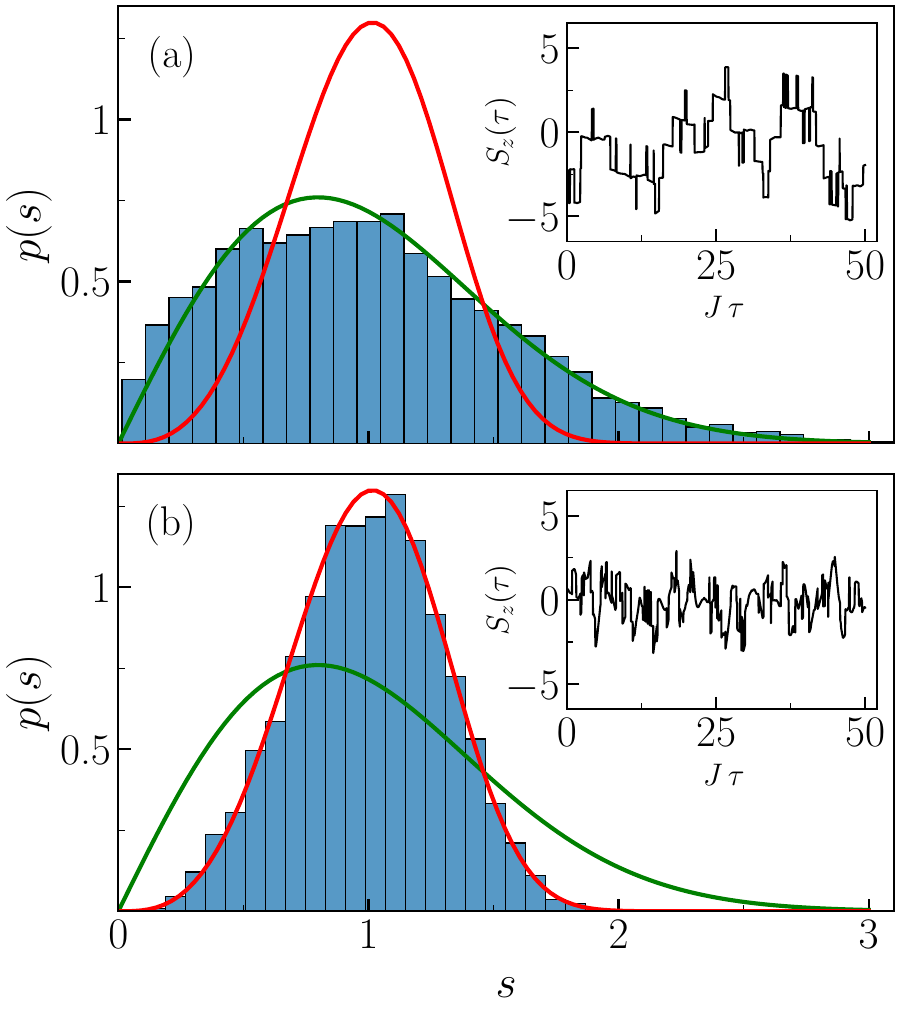}
\caption{Relevant spectral statistics obtained applying the SSQT criterion over a single quantum trajectory on the coherently-driven open spin chain. (a) $F/J=0.12$. (b) $F/J=1$. 
The insets show the total spin $S_z = \sum_{j=1}^L \langle \hat{\sigma}_j^z \rangle$ along single quantum trajectories. 
The green (red) curve is a guideline indicating an ideal 2D-Poisson (Ginibre) distribution.
Parameters: same as in Fig.~\ref{fig:steady_state_analysis} and $L=7$.}
\label{fig:open-spin_chain}
\end{figure}

\subsection{Entanglement spectrum and analysis of the steady state}\label{sec:open_spin}

We argue here that the structure of the steady-state density matrix $\hat{\rho}_{\rm ss}$ does not provide enough information about steady-state dynamics. 

A possible diagnostic of regularity and chaos in the steady state, applied instead of the spectral statistics of the Liouvillian, has been discussed in Refs.~\cite{Sa_2020, sa_spectral_2020, costa_spectral2023} for the study of random Liouvillians and random Kraus maps.
One formally defines the effective Hamiltonian
\begin{equation}\label{Eq:effective_Hamiltonian_entanglement}
    \hat{H}_{\rm ss} = -\log\left( \hat{\rho}_{\rm ss} \right),
\end{equation}
and studies the statistical distribution of its eigenvalue spacing.
In particular, an indicator is the Hamiltonian ratio $\langle r \rangle_{\rm ss}$ of $\hat{H}_{\rm ss}$ [c.f. Eq.~\eqref{eqs:ratio}].
It ranges from $\langle r \rangle_{\rm ss}=0.38$, the ratio of a 1D Poisson distribution corresponding to an integrable Hamiltonian system, to $\langle r \rangle_{\rm ss}=0.60$, the ratio of a Gaussian Unitary Ensemble distribution characterizing a chaotic Hamiltonian system.

The model we consider is a 1D chain of $L$ two-level systems, whose Hamiltonian reads 
\begin{equation}
    \hat{H}=J \sum_{j=1}^{L-1}\left(\hat{\sigma}_{j}^x \hat{\sigma}_{j+1}^x+\hat{\sigma}_{j}^y \hat{\sigma}_{j+1}^y\right) + F \sum_{j=1}^{L-1} \hat{\sigma}_j^{x}
\end{equation}
where $\hat{\sigma}_j^{x, \, y, \, z}$ are the Pauli matrices.
The presence of a drive $F$ generalizes the model studied in Refs. \cite{akemann_universal_2019, sa_complex_2020}.
The system is subject to bulk dephasing of all spins and amplitude damping and gain at the boundaries. 
The Liouvillian reads
\begin{equation}
\begin{split}
    \LL \hat{\rho} =& -i [\hat{H}, \hat{\rho}] + \gamma\sum_{j=1}^{L} \mathcal{D}[\hat{\sigma}_{j}^z] \hat{\rho}
     \\ 
     & \quad
     + \left(\gamma_1^{+} \mathcal{D}[\hat{\sigma}_{1}^+] +\gamma_1^{-} \mathcal{D}[\hat{\sigma}_{1}^-] \right) \hat{\rho}     \\
    & \quad 
    + \left(\gamma_L^{+} \mathcal{D}[\hat{\sigma}_{L}^+]
    + \gamma_L^{-} \mathcal{D}[\hat{\sigma}_{L}^-]\right) \hat{\rho}    
\end{split}    
\end{equation}

We investigate the model at different intensities of the drive amplitude: $F/J=0$ (no drive), $F/J=0.12$ (weak drive), $F/J=1$  (strong drive).
As shown in Fig.~\ref{fig:steady_state_analysis} (a), the steady state changes only marginally in the three regimes.
Furthermore, as displayed in Fig.~\ref{fig:steady_state_analysis} (b), the steady state becomes only slightly more mixed upon increasing the drive amplitude.
As such we conclude that methods based on the steady state alone should not observe significant differences between the three cases.
This is indeed confirmed in Fig.~\ref{fig:steady_state_analysis} (c), where the parameter $\langle r \rangle_{\rm ss}$ is displayed as a function of the system size [c.f. Eq.~\eqref{Eq:effective_Hamiltonian_entanglement} and the discussion below].

From Fig.~\ref{fig:steady_state_analysis} (c) we also deduce the unreliability of $\langle r \rangle_{\rm ss}$ as a model-independent predictor of DQC. 
In particular, for $F=0$ the model is known to be Bethe-ansatz integrable for all $L$ \cite{akemann_universal_2019, sa_complex_2020} contrary to the predictions of $\langle r \rangle_{\rm ss}$ which varies with $L$ between integrability and chaos.
The same dependence of $\langle r \rangle_{\rm ss}$ on $L$ also occurs for $F/J=0.12$ ($F/J=1$), whereas the spectral statistics criterion always predicts integrability (chaos).

We show in Fig.~\ref{fig:open-spin_chain} that the SSQT criterion consistently characterizes the weak and strong drive configurations.
For $F/J=0.12$ (and also $F/J=0$) the relevant level statistics is 2D Poissonian [c.f. Fig.~\ref{fig:open-spin_chain} (a)], while for $F/J=1$ it follows the Ginibre distribution [c.f. Fig.~\ref{fig:open-spin_chain} (b)]. 
For this study, we chose the largest considered system size, $L=7$ and we applied the SSQT over a single Monte Carlo trajectory.
We also notice the profound difference between single quantum trajectories for $F/J=0.12$ and $F/J=1$, shown as insets to Fig.~\ref{fig:open-spin_chain}.
This analysis demonstrates the effectiveness of our criterion in characterizing the dynamical properties of single trajectories, which ultimately determine the chaotic nature of the system.

\subsection{Out-of-time-order correlators}

\begin{figure}[t!]
\includegraphics[width=0.46\textwidth]{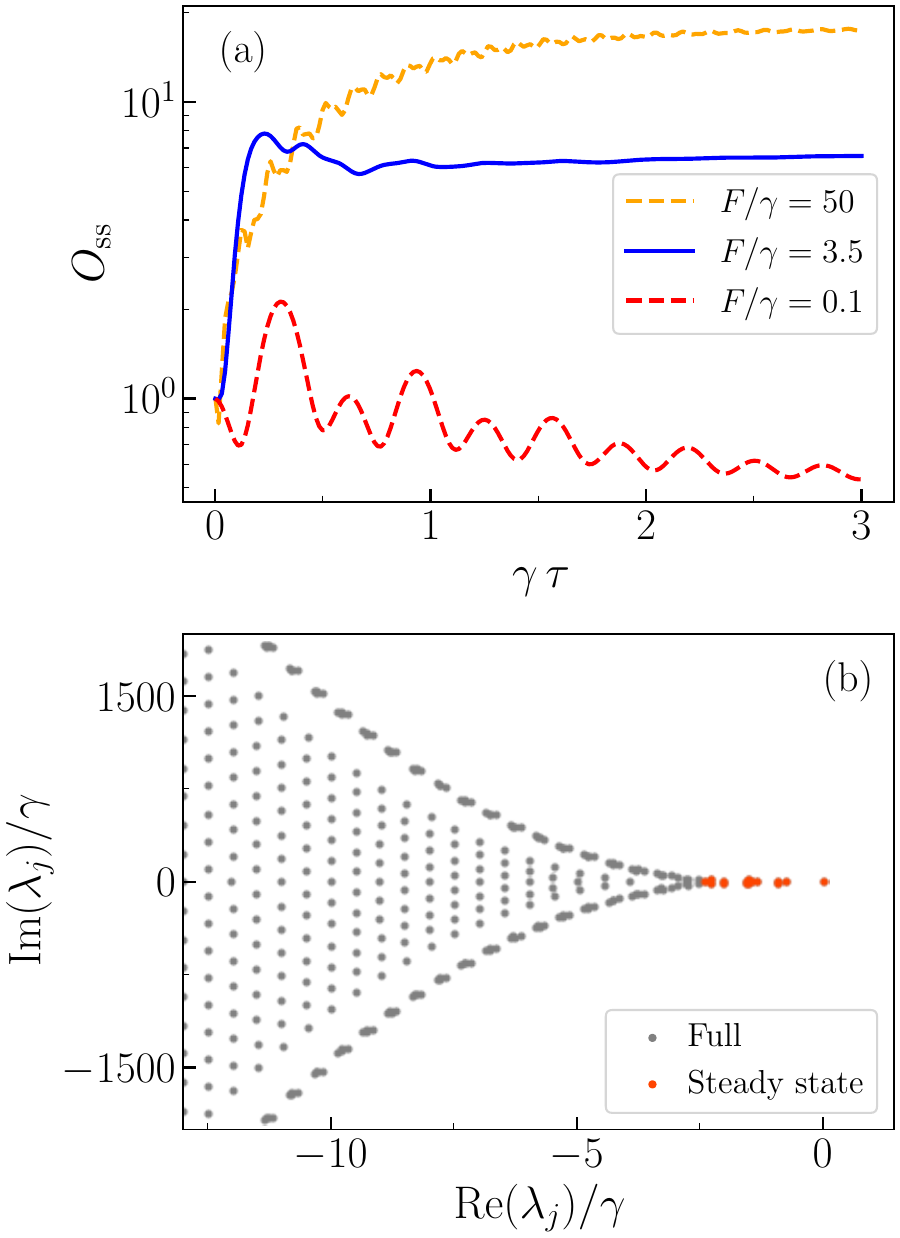}
\caption{Out-of-time order correlators for the one-photon driven-dissipative Kerr resonator. (a) The OTOC defined in Eq.~\eqref{Eq:OTOC_def} calculated for the steady state $\hat{\rho}_{\rm ss}$ in the low photon regime ($F/\gamma=0.1$), close to the critical point ($F/\gamma=3.5$) and in the high photon regime ($F/\gamma=50$). While away from the critical point $O_{\rm ss}(\tau)$ correctly signals the integrability of the model, close to criticality, OTOC's behavior becomes ambiguous. (b) For $F/\gamma=3.5$, the full Liouvillian spectrum is denoted by grey dots, and selected eigenvalues with the SSQT in the steady state are denoted by red dots. The SSQT analysis correctly predicts the model's integrability. Other parameters are set to be $\Delta/\gamma=10$, $U/\gamma=10$.
}
\label{fig:OTOCs}
\end{figure}

Out-of-time order correlators (OTOCs) are a popular probe of chaotic dynamics in quantum systems \cite{maldacena_bound_2016, hashimoto_out_time-order_2017, garcia-mata_out_time-order_2022}.
Indeed, in the presence of chaotic behavior, OTOCs diverge and thus signal the presence of chaos \cite{richter_manybody2018}. Rapid exponential growth is associated with a quantum Lyapunov exponent, while the reached bound is often called the ``bound on chaos" \cite{maldacena_bound_2016}. In non-chaotic systems, OTOCs usually display a linear or sub-linear initial growth. 

Despite being a ``necessary'' condition for quantum chaos, OTOCs may be also diverging in non-chaotic settings.
OTOCs measure scrambling in the quantum system, i.e., the spread of local information.
Although scrambling and chaos are often concomitant, recently the idea that scrambling implies chaos has been questioned \cite{gopalakrishnan_scrambling2018, gopalakrishnan_floquet2018, xu_scrambling2020, rozenbaum_scrambling2020}. 
Examples of exponentially-growing OTOCs even in the absence of quantum chaos have been shown to be caused by features such as isolated saddle points \cite{xu_scrambling2020}.
As such, the presence of scrambling and exponentially diverging OTOCs thus does not equate to chaos.

OTOCs have been extended to open quantum systems \cite{zanardi_information_2021}. Whether decoherence can lead to similar mechanisms for which OTOCs can grow exponentially in integrable systems remains unclear.

\begin{figure}[t!]
\includegraphics[width=0.46\textwidth]{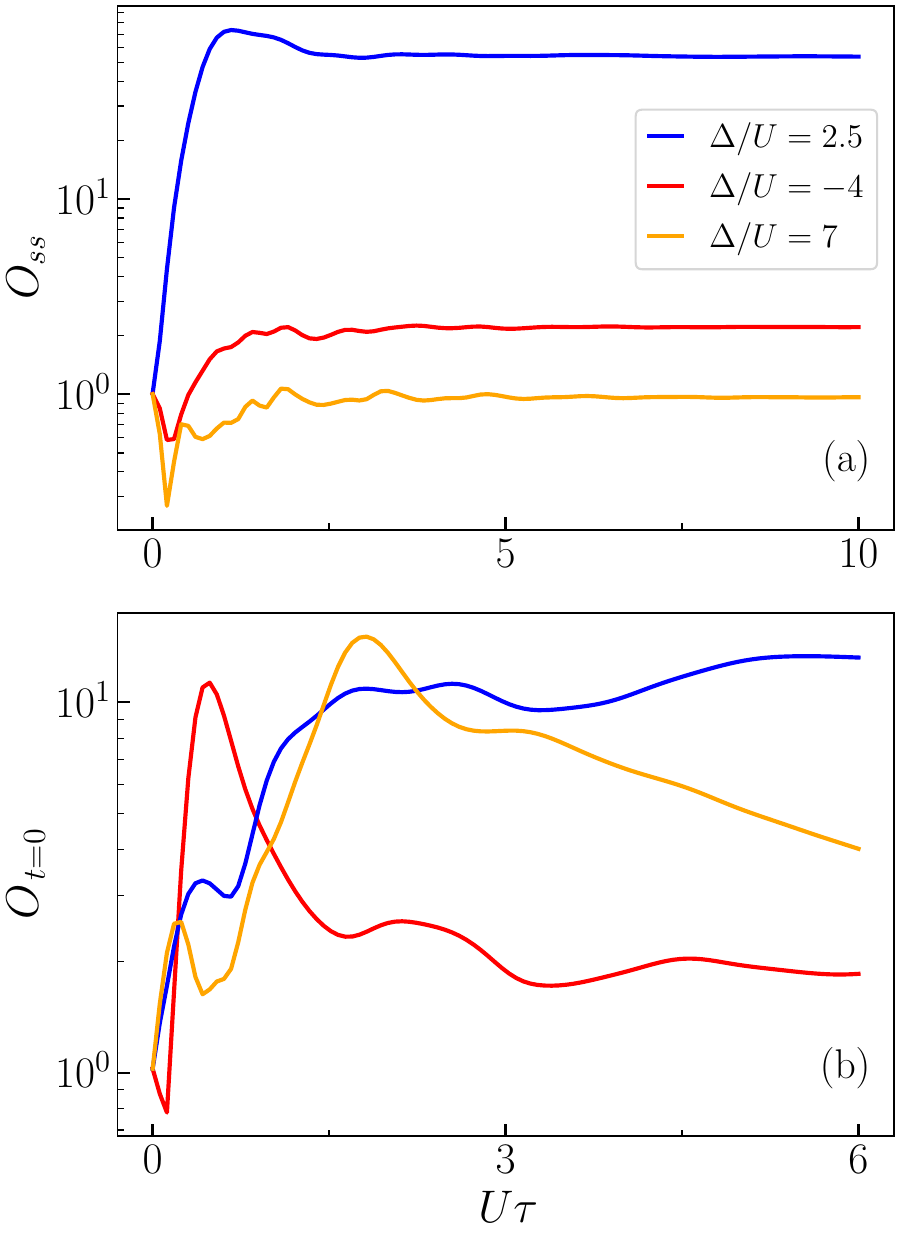}
\caption{The OTOC defined in Eq.~\eqref{Eq:OTOC_def} for the mode $\hat{a}_1$ calculated (a) at the steady state and (b) at $t=0$, for the driven-dissipative Bose-Hubbard model discussed in Sec.~\ref{sec:quantum_simulation}. The three selected configurations correspond to integrability ($\Delta/U = -4$), steady-state chaos ($\Delta/U = 2.5$) and transient chaos ($\Delta/U = 7$). The initial state in panel (b) is the coherent state $\ket{\psi(0)}=\ket{\alpha}\otimes\ket{\alpha}$ with $\alpha = 2$. The drive strength is $F/U=3$. Other parameters as in Fig.~\ref{fig:steady_vs_transient_chaos}.
}
\label{fig:OTOCs_2}
\end{figure}

Using the procedure detailed in, e.g., Ref. \cite{breuer_theory_2007}, we define 
\begin{equation}\label{Eq:OTOC_def}
    O_t(\tau) = -\langle [\hat{Q}(t+\tau), \hat{P}(t)]^2 \rangle,
\end{equation}
for $\hat{Q}(t) = [\hat{a}(t) + \hat{a}^{\dagger}(t)]/\sqrt{2}$ and $\hat{P}(t) = i[\hat{a}^{\dagger}(t) - \hat{a}(t)]/\sqrt{2}$, where $\hat{a}(t)$ is the annihilation operator in the Heisenberg picture.
To evaluate the OTOCs, one can either use the master equation or quantum trajectories (with the prescription for evaluating correlation functions given in Ref.~\cite{molmer_monte_1993}, see also \cite{dahan_classical_2022}). 
We follow the first approach, adopting the procedure described in Ref.~\cite{breuer_theory_2007} for the computation of general correlations in open quantum systems. The OTOC can be written as:
\allowdisplaybreaks
\begin{equation}\label{eqs:OTOCs}
\begin{aligned}
    &O_t(\tau) = -(\langle \hat{Q}(t+\tau)\hat{P}(t)\hat{Q}(t+\tau)\hat{P}(t)\rangle \\ &+ \langle \hat{P}(t)\hat{Q}(t+\tau)\hat{P}(t)\hat{Q}(t+\tau)\rangle \\ &- \langle \hat{Q}(t+\tau)\hat{P}(t)\hat{P}(t)\hat{Q}(t+\tau) \rangle \\ &- \langle \hat{P}(t)\hat{Q}(t+\tau)\hat{Q}(t+\tau)\hat{P}(t) \rangle)
\end{aligned}
\end{equation}
We start from a given initial density matrix $\hat{\rho}_{\rm in}$ and perform forward and ``backward'' time evolution solving the master equation with ${\hat{H}}$ and $-{\hat{H}}$ respectively (assuming the same jump operators for both time directions) \cite{dahan_classical_2022}. We introduce the superoperators $\mathcal{F}_O \hat{\rho} = \hat{O} \hat{\rho} \hat{O}$, $\mathcal{F}_O^R \hat{\rho}= \hat{O}\hat{\rho}$, $\mathcal{F}_O^L \hat{\rho}= \hat{\rho} \hat{O}$, and the time evolution superoperator $\mathcal{V}(t, 0)\hat{\rho}$, with $\mathcal{V}(t, 0)\equiv \exp{(\LL t)}$. Then, each term in the OTOC can be explicitly rewritten as
\begin{align}
    &\langle \hat{Q}(t+\tau)\hat{P}(t)\hat{Q}(t+\tau)\hat{P}(t)\rangle =\\
    &\Tr{\mathcal{F}_P^R \mathcal{V}(t, t+\tau) \mathcal{F}_Q \mathcal{V}(t+\tau, t) \mathcal{F}_P^R\mathcal{V}(t, 0)\hat{\rho}_{\rm in}}, \nonumber \\
     \, \nonumber  \\
    & \langle \hat{P}(t)\hat{Q}(t+\tau)\hat{P}(t)\hat{Q}(t+\tau)\rangle =\\
    &\Tr{\mathcal{F}_P^R \mathcal{V}(t, t+\tau) \mathcal{F}_Q \mathcal{V}(t+\tau, t) \mathcal{F}_P^L\mathcal{V}(t, 0)\hat{\rho}_{\rm in}}, \nonumber\\
     \, \nonumber  \\
    & \langle \hat{Q}(t+\tau)\hat{P}(t)\hat{P}(t)\hat{Q}(t+\tau) \rangle =
    \\  &\Tr{\mathcal{F}_P \mathcal{V}(t, t+\tau) \mathcal{F}_Q \mathcal{V}(t+\tau, 0) \hat{\rho}_{\rm in}},  \nonumber\\
     \, \nonumber   \\
    & \langle \hat{P}(t)\hat{Q}(t+\tau)\hat{Q}(t+\tau)\hat{P}(t) \rangle = \\
    & \Tr{\mathcal{F}_Q \mathcal{V}(t+\tau, t) \mathcal{F}_P\mathcal{V}(t, 0)\hat{\rho}_{\rm in}}. \nonumber
\end{align}
The OTOC $O_{\rm ss}(\tau) \equiv \lim_{t\to \infty} O_t(\tau)$ is obtained by setting $\hat{\rho}_{\rm in} = \hat{\rho}_{ss}$.

Here we provide an example of two ambiguous behaviors of OTOCs in open systems.
The first model we consider is the one-photon driven-dissipative Kerr resonator whose Hamiltonian reads
\begin{equation}\label{eq:Kerr_resonator}
    \hat{H} = -\Delta\hat{a}^{\dagger}\hat{a} + \frac{1}{2}U\hat{a}^{\dagger}\hat{a}^{\dagger}\hat{a}\hat{a} + F(\hat{a}^{\dagger} + \hat{a}),
\end{equation}\label{eq:Lindblad_kerr}
with Lindblad equation
\begin{equation}
    \frac{\partial\hat{\rho}}{\partial t} = -i[\hat{H}, \hat{\rho}] + \gamma \left(\hat{a}\hat{\rho}\hat{a}^{\dagger} - \frac{1}{2}\acomm{\hat{a}^{\dagger}\hat{a}}{\hat{\rho}}\right).
\end{equation}

The model is known to exhibit a first-order dissipative phase transition in the thermodynamic limit for a critical drive amplitude \cite{minganti_spectral_2018}, and its steady state can be computed analytically \cite{drummond_quantum_1980, drummond_quantum_1981, bartolo_exact_2016}. The order parameter (in this case the photon number $n = \langle \hat{a}^{\dagger}\hat{a}\rangle$) abruptly jumps from zero to a finite value. Close to criticality, the system displays bistability, switching  between two fixed points \cite{drummond_quantum_1980, bartolo_exact_2016}. 

In Fig.~\ref{fig:OTOCs} (a) we numerically show the behavior of the steady-state OTOC \eqref{Eq:OTOC_def}. Away from criticality, the OTOC predicts the model's integrability, while close to the critical point, in the presence of optical bistability, OTOC's behavior becomes ambiguous. Fig.~\ref{fig:OTOCs} (b) shows the full Liouvillian spectrum diagonalized for a cutoff $N_c=30$, which is manifestly regular. Integrability is confirmed by the SSQT criterion, which shows how only a few eigenvalues are involved in steady-state dynamics.

Finally, we show that OTOCs can be not predictive of transient chaos. 
Here, we consider the driven-dissipative Bose-Hubbard dimer discussed in Sec.~\ref{sec:quantum_simulation}.
Results are shown in Fig.~\ref{fig:OTOCs_2} for the mode $\hat{a}_1$. 
$O_{\rm ss}(\tau)$ rapidly increases only in the regions where the SSQT criterion predicts steady-state chaos [see Fig.~\ref{fig:OTOCs_2} (a)].
These findings support the results obtained in the main text. 
When we consider transient chaos, instead, we see  that OTOCs may be non-predictive [c.f.  Fig.~\ref{fig:OTOCs_2} (b)].
For instance, a point characterized by both transient and steady-state DQC displays similar features with respect to one completely integrable.
We thus conclude that OTOCs in Eq.~\eqref{eqs:OTOCs} may not be capable of determining transient chaos in dissipative systems.
We argue that the reason for this lack of correspondence is due to the exponential nature of the Liouvillian map, which may compensate for the diverging growth of the OTOC.

\section{Chaos in the Hamiltonian driven Bose-Hubbard model} \label{sec:Ham}

\begin{figure}[t!]
\includegraphics[width=0.47\textwidth]{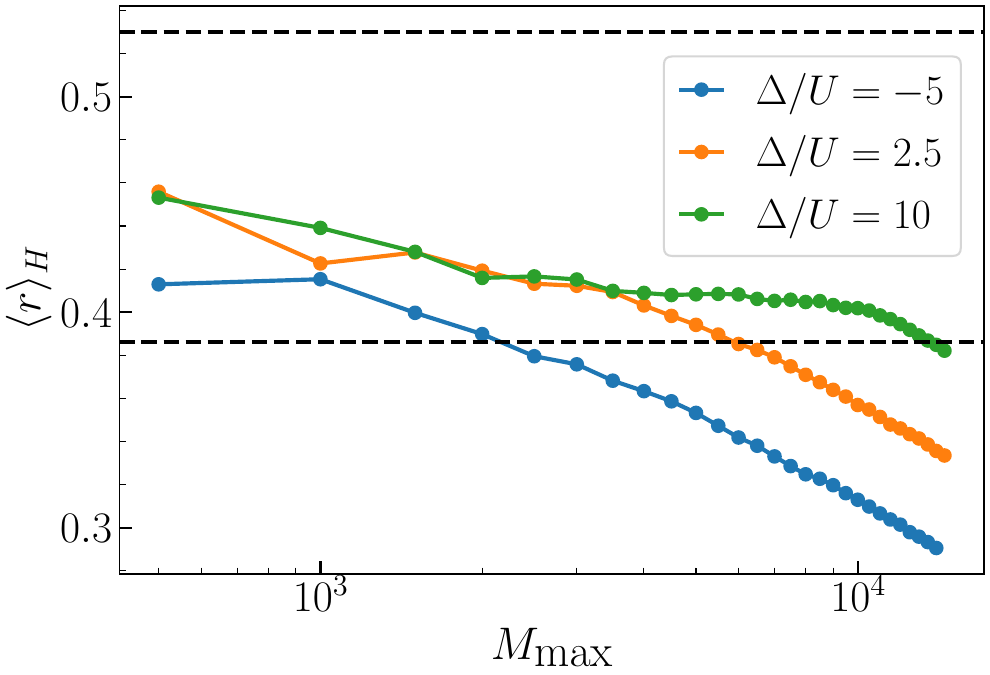}
\caption{Spectral analysis for the model Hamiltonian in Eq.~\eqref{eqs:hamiltonian}. The parameter $\langle r\rangle_H = \sum_{j= 1}^{M_{\rm max}} r_j / M_{\rm max}$, for $r_j$ defined in \eqref{eqs:ratio}, is plotted as a function of $M_{\rm max}$, for three values of detuning $\Delta/U$. Black dashed lines indicate the values of $\langle r \rangle_H$ for uncorrelated random variables and for the Wigner-Dyson distribution. Other parameters set as in Fig.~\ref{fig:steady_vs_transient_chaos}.}
\label{fig:hamiltonian_analysis}
\end{figure}

The goal of this analysis is to provide evidence for the importance of an energy cutoff when studying the distribution of eigenvalue spacings in non-$U(1)$ symmetric systems. 
This plays a role similar to that of $c_{\rm min}$, and shows that an analogous cutoff is not a feature of open systems, but rather of driven ones.

We consider the Hamiltonian version of the boundary-driven Bose-Hubbard dimer, i.e., Eq.~\eqref{eqs:hamiltonian} alone. 
We note, in passing, that the Hamiltonian problem with a driving term but no corresponding dissipation is unphysical, as in typical experimental platforms if a channel is open for input, the same channel will also lead to a dissipation mechanism. 

As a diagnostic of chaos, we consider the single number indicator $\langle r \rangle_H$ \cite{atas_distribution_2013}.
Let $s_j = E_{j+1}-E_j$ be the $j$-th spacing.
Then,
\begin{equation}\label{eqs:ratio}
    r_j = \frac{\min(s_j, s_{j-1})}{\max(s_j, s_{j-1})}
\end{equation}
and $\langle r \rangle_H $ is the average of $r_j$.
One finds $\langle r \rangle_H=0.386$ for the 1D Poisson distribution and $\langle r \rangle_H=0.53$ for the Wigner-Dyson distribution. 
Values of $\langle r \rangle_H$ between these two limits are a signature of a hybrid regime between integrability and chaos \cite{brody_statistical_1973, prosen_energy_1993, prosen_semiclassical_1994, bogomolny_models_1999}. 
Systems with a regularly spaced eigenspectrum are exceptions to this criterion, e.g., the single harmonic oscillator that, even if integrable, recovers $\langle r \rangle_H=1$. 

We introduce the energy cutoff $M_{\rm max}$, and compute the level statistics over the set $\{E_{j}, j \leq M_{\rm max}\}$ \footnote{$M_{\rm max}$ is an energy cutoff, and is distinct from the cutoff $N_c$ adopted for numerical diagonalization in
Fock space. For each value of $M_{\rm max}$ we varied $N_c>M_{\rm max}$ to ensure convergence of all eigenvalues $E_{j<M_{\rm max}}$.}. 
This procedure is justified by the fact that, for an initial state with energy $E_{\rm in} = \langle \psi_{\rm in}|\hat{H}|\psi_{\rm in} \rangle$, the eigendecomposition of $\ket{\psi_{\rm in}}$ yields  $\ket{\psi_{in}} = \sum_j c_j \ket{\psi_j}$, with  $\sum_j |c_j|^2 =1$ and $\hat{H}\ket{\psi_j} = E_j \ket{\psi_j}$. Therefore $E_{\rm in} = \sum_j |c_j|^2 E_j$ and, under reasonable physical assumptions, the initial state has finite energy and relevant components only on eigenstates within a finite energy window.

\begin{figure*}[t]
\includegraphics[width=1\textwidth]{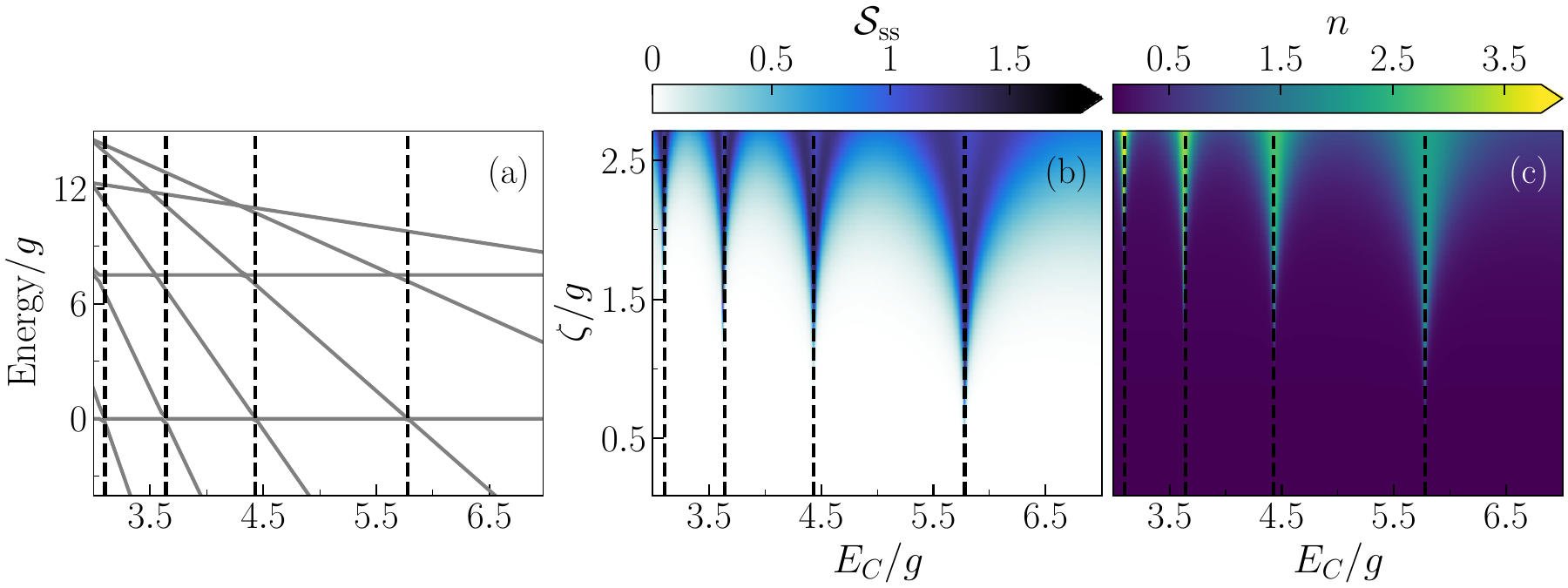}
\caption{Characterization of the driven-dissipative transmon described by Eqs.~\eqref{eqs:driven_transmon} and \eqref{eqs:lindblad_tranmson}. 
(a) Energy levels of the undriven transmon Hamiltonian [cf Eq.~\eqref{eqs:driven_transmon} with $F/g=0$] as a function of the transmon charging energy $E_C/g$.
Black vertical dashed lines indicate the multiphoton resonances between $\ket{0}$ and a higher photon-number state $\ket{n}$.
(b) Von Neumann entropy of the steady state $\mathcal{S}_{\textrm{ss}} = -\operatorname{Tr}(\hat{\rho}_{\textrm{ss}}\log\hat{\rho}_{\textrm{ss}})$ as a function of the effective drive amplitude $\zeta/g$ and the charging energy $E_C/g$. 
(c) Transmon's photon number $n$ as a function of the same parameters. Other parameters are fixed to $\Delta_t/g = -7.5$, $\gamma_t/g = 0.009$, $E_J = 50E_C$.}
\label{fig:transmon_characterization}
\end{figure*}

In Fig. \ref{fig:hamiltonian_analysis} we show the prediction of chaos or integrability according to $\langle r \rangle_H$ for various choices of parameters.
We notice that $\langle r \rangle_H$ strongly depends on the energy cutoff.
As discussed also in Sec.~\ref{Sec:Chaos_in_bosons} and Sec.~\ref{sec:model}, the eigenenergies at large occupation numbers are dominated by the nonlinear term, and roughly scale as $E \simeq U(n_1^2 + n_2^2)$, where $n_1$ ($n_2$) is the number of photons in the first (second) cavity.
As the driving and detuning terms asymptotically become negligible, the  spectrum becomes that of two decoupled anharmonic oscillators. It is then straightforward to show that $r_j\sim1/j$, leading to $\langle r \rangle_H\to0$ if the average is taken over the spectrum.
The presence of a driving field therefore introduces an issue in the characterization of quantum chaos from the eigenvalue statistics. 

\section{Derivation and characterization of the transmon Hamiltonian} \label{sec:Transmon_appendix}

\begin{widetext}
We derive here the transmon Hamiltonian appearing in Eq.~\eqref{eqs:hamiltonian_circuitQED}. 
Inserting into Eq.~\eqref{eqs:transmon_full} the definition of $\hat{b}$ and $\hat{b}^{\dagger}$ in Eq.~\eqref{eqs:bosonic_fields}, one gets
\begin{equation}
\begin{split}
    \hat{H}_t &= -\sqrt{E_JE_C/2}\,(\hat{b}^\dagger - \hat{b})^2 + \sqrt{E_JE_C/2}\,(\hat{b}^\dagger + \hat{b})^2 - E_J\sum_{k>1}^{+\infty}\frac{(-1)^k}{(2k)!}\left(\frac{2E_C}{E_J}\right)^{k/2}(\hat{b}^{\dagger} + \hat{b})^{2k},
\end{split}
\end{equation}

We now expand the binomials $(\hat{b}^{\dagger} \pm \hat{b})^{2}$ and $(\hat{b}^{\dagger} + \hat{b})^{2k}$  considering only the terms where $\hat{b}^{\dagger}$ and $\hat{b}$ appear an equal number of times.
This coincides with dropping all the counter-rotating terms.
For an accurate description, we truncate the series up to order $k=5$. 
We finally obtain
\begin{equation}\label{eqs:transmon_bosonic}
\begin{split}
    \hat{H}_t &= \left[\sqrt{8E_CE_J} - E_C + \omega_{\textrm{eff}}\right]\hat{b}^{\dagger}\hat{b} - \left(\frac{E_C}{2} - \chi_{\textrm{eff}}^{(2)}\right)\hat{b}^{\dagger 2}\hat{b}^{2}  + \chi_{\textrm{eff}}^{(3)}\hat{b}^{\dagger 3}\hat{b}^{3} + \chi_{\textrm{eff}}^{(4)}\hat{b}^{\dagger 4}\hat{b}^{4} + \chi_{\textrm{eff}}^{(5)}\hat{b}^{\dagger 5}\hat{b}^{5},
\end{split}
\end{equation}
where
\begin{equation}\label{eqs:shift}
\begin{split}
    \omega_{\textrm{eff}}/E_J =& \frac{90}{6!}\left(\frac{2E_C}{E_J}\right)^{3/2} - \frac{840}{8!}\left(\frac{2E_C}{E_J}\right)^{2}  \quad + \frac{9450}{10!}\left(\frac{2E_C}{E_J}\right)^{5/2}, \\ 
    \chi_{\textrm{eff}}^{(2)}/E_J = & \frac{90}{6!}\left(\frac{2E_C}{E_J}\right)^{3/2} - \frac{1260}{8!}\left(\frac{2E_C}{E_J}\right)^{2}  \quad+ \frac{18900}{10!}\left(\frac{2E_C}{E_J}\right)^{5/2},\\
    \chi_{\textrm{eff}}^{(3)}/E_J = & \frac{20}{6!}\left(\frac{2E_C}{E_J}\right)^{3/2} - \frac{560}{8!}\left(\frac{2E_C}{E_J}\right)^{2}   + \quad \frac{12600}{10!}\left(\frac{2E_C}{E_J}\right)^{5/2},\\
     \chi_{\textrm{eff}}^{(4)}/E_J =& - \frac{70}{8!}\left(\frac{2E_C}{E_J}\right)^{2} + \frac{3150}{10!}\left(\frac{2E_C}{E_J}\right)^{5/2}, \\
    \chi_{\textrm{eff}}^{(5)}/E_J =& \frac{252}{10!}\left(\frac{2E_C}{E_J}\right)^{5/2}.
\end{split}
\end{equation}

\begin{figure*}[t]
\includegraphics[width=1\textwidth]{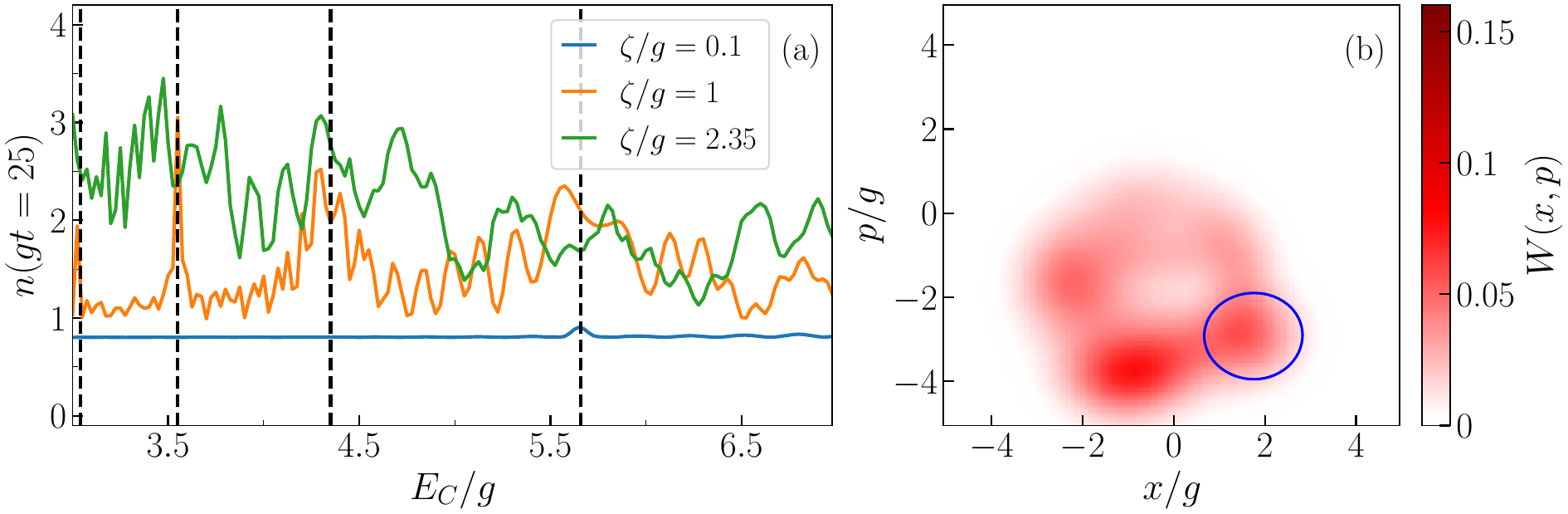}
\caption{Breakdown of the dispersive approximation for the system initialized in the excited transmon state. (a) Using the driven transmon model in Eqs.~\eqref{eqs:driven_transmon} and \eqref{eqs:lindblad_tranmson}, we plot the photon number in the transom at $gt=25$ for different choices of the effective drive amplitude $\zeta/g$. The vertical dashed lines show the position of the resonances between the state $\ket{1}$ and $\ket{n}$, as shown in Fig.~\ref{fig:transmon_characterization} (a).
We observe a large deviation from the initial state even at moderate values of the drive and far from these resonances.
(b) The Wigner function of the readout cavity at $gt =25$ for the full model in Eqs.~\eqref{eqs:lindblad_circuitQED} and \eqref{eqs:hamiltonian_circuitQED}.
$F/g =0.7$ and $E_C/g =4.9$.
The contour plot indicates the value of $W(x,p)$ for an initially excited transmon qubit, while the blue contour line is the $1/3$ height of $W(x,p)$ for the transmon in the ground state.
Other parameters as in Fig.~\ref{fig:transmon_phasediagram}.
}
\label{fig:transmon_nodispersive}
\end{figure*}

In the ideal transmon limit $E_C/E_J \rightarrow 0$ all the terms in Eqs.~\eqref{eqs:shift} disappears, and the transmon Hamiltonian coincides with that of a Kerr resonator. 

\end{widetext}

\subsection{Driven-dissipative transmon}

An intuitive picture of the onset of DQC can be understood through the following simplified model.
Instead of considering the readout cavity, one simply includes its effect in a mean field approximation, both through an effective drive $\zeta \simeq g |\alpha|$ ($\alpha$ representing the coherence of the cavity), and via the induced Purcell decay $\gamma_t = (g/\Delta)^2 \gamma_r$ \cite{blais_circuit_2021}. 
The system becomes then a driven transmon reading
\begin{equation}\label{eqs:driven_transmon}
\begin{split}
    \hat{H} &= -\Delta\hat{b}^{\dagger}\hat{b} + \zeta(\hat{b} + \hat{b}^{\dagger}) - \left(\frac{E_C}{2} - \chi_{\textrm{eff}}^{(2)}\right)\hat{b}^{\dagger 2}\hat{b}^{2}\\ & + \chi_{\textrm{eff}}^{(3)}\hat{b}^{\dagger 3}\hat{b}^{3} + \chi_{\textrm{eff}}^{(4)}\hat{b}^{\dagger 4}\hat{b}^{4} + \chi_{\textrm{eff}}^{(5)}\hat{b}^{\dagger 5}\hat{b}^{5},
\end{split}
\end{equation}
with Lindblad equation
\begin{equation}\label{eqs:lindblad_tranmson}
    \frac{\partial \hat{\rho}}{\partial t} = -i[\hat{H}, \hat{\rho}] + \gamma_t\left(\hat{b}\hat{\rho}\hat{b}^{\dagger} - \frac{1}{2}\acomm{\hat{b}^{\dagger}\hat{b}}{\hat{\rho}}\right).
\end{equation}

Results are presented in Fig.~\ref{fig:transmon_characterization}. 
Panel (a) shows the energy levels of the undriven transmon Hamiltonian [c.f. Eq.~\eqref{eqs:driven_transmon} with $\zeta=0$] for the first $10$ transmon's eigenstates.
We notice that, in the correspondence of the black dashed vertical lines, we have a multiphoton resonance between the state $\ket{0}$ and a higher photon-number state $\ket{n}$.
The presence of such multiphoton resonances has visible effects on the driven-dissipative transmon (and on the driven-dissipative circuit QED setup discussed in the main text).
Panel (b) shows the Von Neumann entropy of the steady state $\hat{\rho}_{\rm ss}$. 
A structure similar to the one in Fig.~\ref{fig:transmon_phasediagram} can be observed. 
Panel (c) shows the transmon's photon number $n$, where we observe that the transmon's population exceeds one photon (so the picture of a two-level system breaks down) exactly when the system's steady state becomes entropic.
Notably, the position of the four observed peaks in both the steady-state entropy and the transmon photon number, exactly coincides with the multiphoton resonances highlighted in panel (a).

Having understood where the two-level system and the dispersive approximation ``break'' for the system initialized in the ground state, we then investigate if they remain valid far from the multiphoton resonances, for the system initialized in the excited state.
In Fig.~\ref{fig:transmon_nodispersive} (a) we study the photon number in the driven transmon initialized in the excited state for different values of $\zeta$ and for a short time evolution.
In this case, we do not see the clear features of multiphoton resonance.
Considering $E_C/g = 4.9$ considered in the main text, far from any resonance between states $\ket{1}$ and $\ket{n}$ we still observe a significant deviation of the system's population from the prediction of the two-level approximation.

We then consider the full quantum model including the dispersive readout as in Eqs.~\eqref{eqs:lindblad_circuitQED} and \eqref{eqs:hamiltonian_circuitQED}.
For $E_C/g = 4.9$, we fix the drive to $F/g=0.7$.
We then plot the Wigner function for the readout cavity, having traced out the qubit's state.
For the dispersive approximation to be valid, we require the readout's state to be almost coherent.
While this is the case for the qubit initialized in the ground state, the Wigner function plotted in Fig.~\ref{fig:transmon_nodispersive} (b) shows that the readout's state is highly noncoherent.

\section{Additional details on transmon readout} \label{sec:Readout_appendix}

We provide here more details on the transmon readout by varying two parameters that we kept constant in the analysis in Sec.~\ref{sec:quantum_information}: the cavity dissipation rate $\gamma_r$ and the cavity-to-transmon detuning $\Delta$.

\subsection{Effect of the cavity loss rate}

\begin{figure}[t!]
\includegraphics[width=0.46\textwidth]{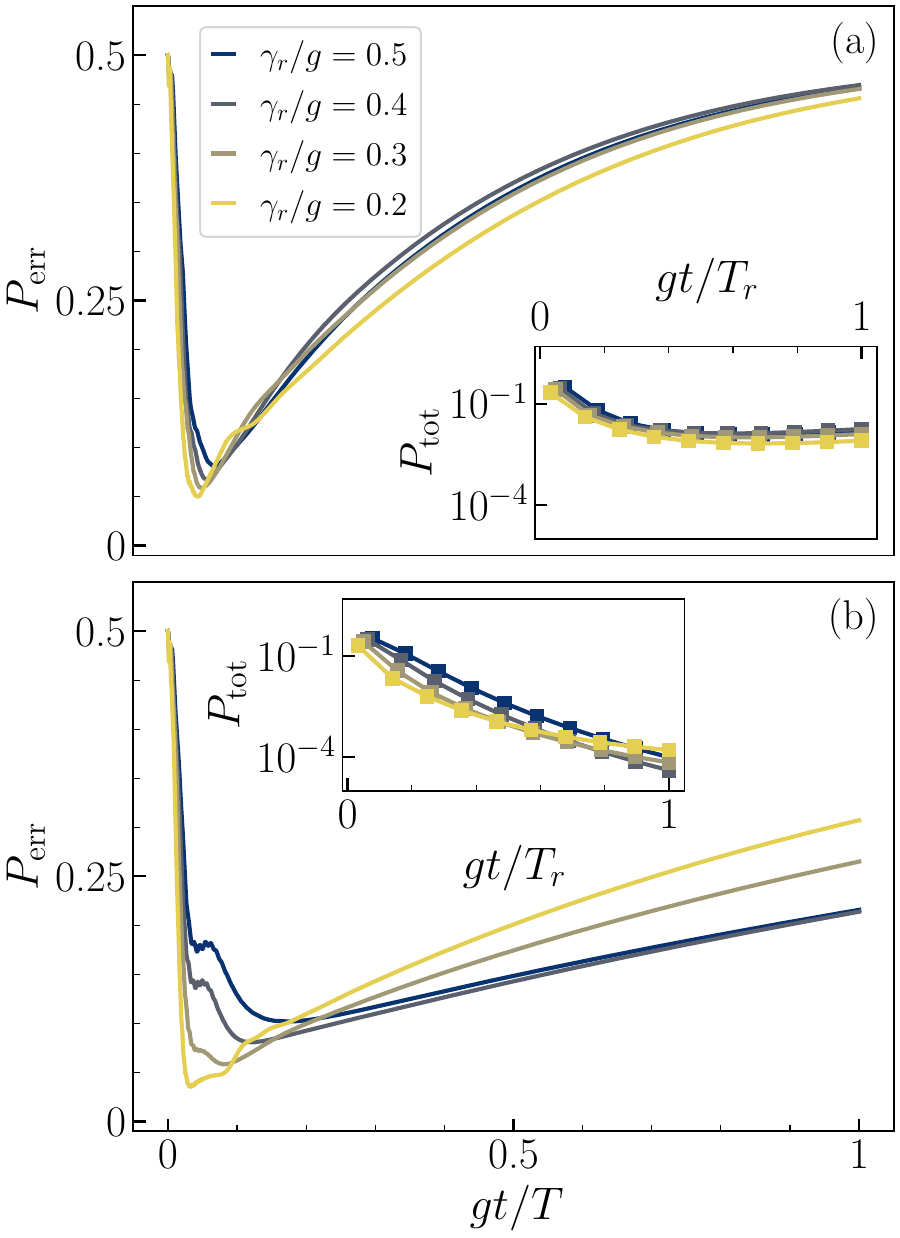}
\caption{Instantaneous error probabilities for $E_C/g=5.7$ [panel (a)] and $E_C/g=4.9$ [panel (b)] for the circuit QED setup described by Eqs.~\eqref{eqs:lindblad_circuitQED} and \eqref{eqs:hamiltonian_circuitQED}. We consider four values of resonator dissipation rate $\gamma_r/g = 0.5, 0.4, 0.3, 0.2$ and we time eveolve the master equation up to $T=1/\gamma_t$. Results are plotted as a function of the rescaled time $gt/T$. The insets of panels (a) and (b) report the total error probability $P_{\rm tot}$ computed with the method described in Sec.~\ref{sec:readout_transmon}, sampling $P_{\rm err}$ each $2/\gamma_r$ up to $T_r = 78g, 96g, 123g, 186g$ with $T_r<1\gamma_t$. The drive amplitudes are fixed such that the resonator's steady state always contains 4-6 photons. Other parameters as in Fig.~\ref{fig:transmon_phasediagram}.
}
\label{fig:smaller_gamma}
\end{figure}

\begin{figure}[t!]
\includegraphics[width=0.46\textwidth]{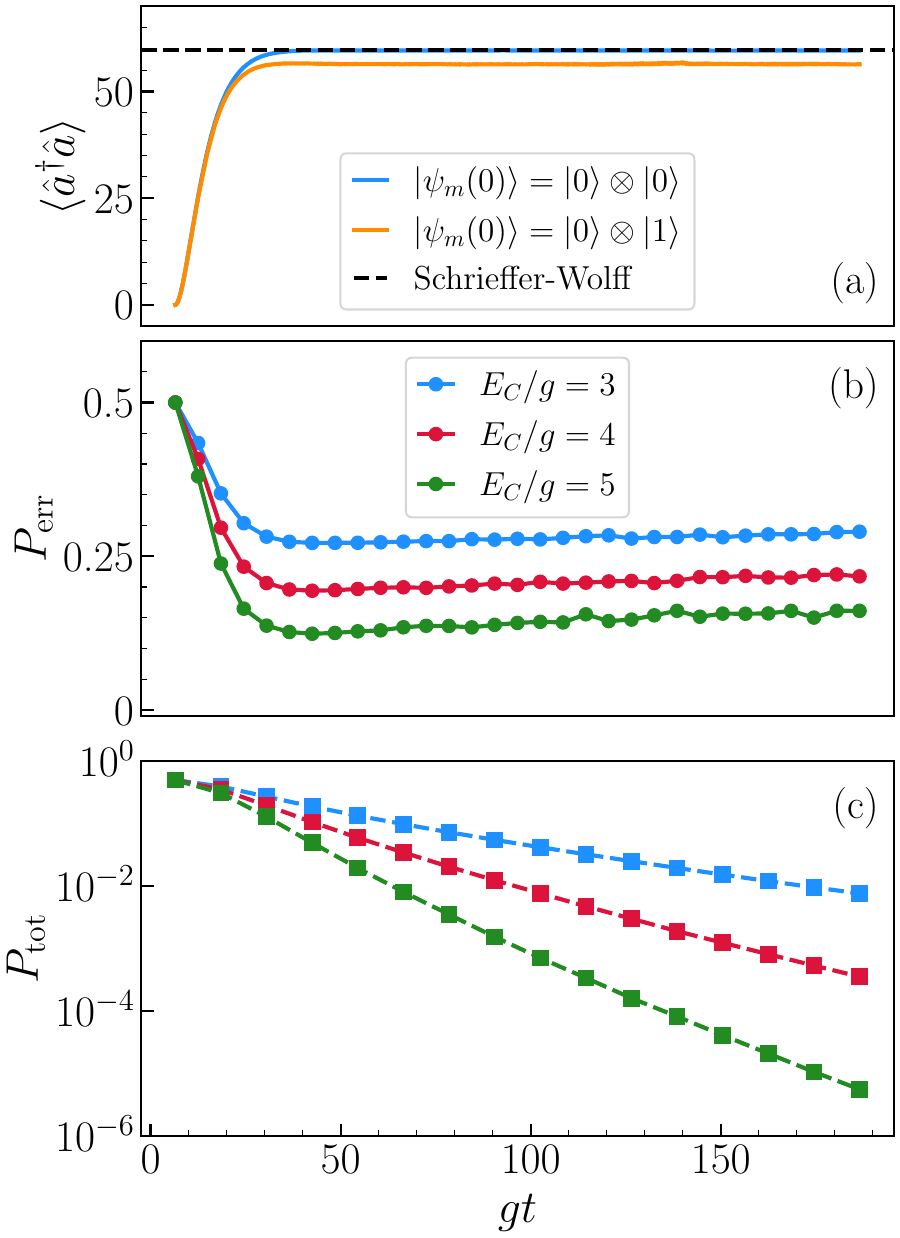}
\caption{Dispersive readout at large detuning. 
(a) Photon number in the resonator along a single quantum trajectory for initial state $\ket{\psi_m(0)} = \ket{0}\otimes\ket{0}$ (orange curve) and $\ket{\psi_m(0)} = \ket{0}\otimes\ket{1}$ (blue curve). 
The black-dashed line indicates the Schrieffer-Wolff prediction given by Eq.~\eqref{eqs:quasi_steady_states}.
We set $E_C/g=4$.
(b) Instantaneous error probability $P_{\rm err}$ at different charging energies, $E_C/g=3, 4, 5$.
(c) Total error probability computed with the method described in Sec.~\ref{sec:readout_transmon}, sampling $P_{\rm err}$ each $2/\gamma_r$ for the same values of $E_C/g$ used for panel (b).
Parameters are fixed to $\Delta/g=-25$, $\gamma_r/g=0.3$, $E_J/E_C=50$, $F/g=1.2$.
Results of (b) and (c) have been obtained upon averaging over 100 quantum trajectories.
}
\label{fig:larger_delta}
\end{figure}

We study the occurrence of DQC for values of $\gamma_r/g$ lower than the value considered above.
In the absence of an intrinsic transmon loss mechanism, the qubit decay rate is determined by the readout cavity losses through the Purcell effect \cite{blais_circuit_2021}, according to $\gamma_t = (g/\Delta)^2\gamma_r$. 
The readout time must be shorter than $1/\gamma_t$, in order to avoid readout-induced qubit's decoherence.
At the same time, the readout dissipation rate $\gamma_r$ measures the rate at which information can be extracted from the system. 

In Fig.~\ref{fig:smaller_gamma} we plot the instantaneous error probability $P_{\rm err}$ for varying $\gamma_r$. 
At the multiphoton resonance, all values of $P_{\rm err}$ collapse on the same curve, suggesting that DQC affect similarly all studied cases. Outside the multiphoton resonance, different albeit similar values of $P_{\rm err}$ are found, thus leading to similar conclusions about the effect of DQC.

In the insets of Figs.~\ref{fig:smaller_gamma} (a) and (b) we report the total error probability $P_{\rm tot}$ obtained with the procedure described in Sec.~\ref{sec:readout_transmon}.
Data are obtained by sampling each time step $2/\gamma_r$ up to $T_r = 78g, 96g, 123g, 186g$ respectively, always fulfilling the condition $T_r<1/\gamma_t$. The results support the conclusions of the analysis of $P_{\rm err}$.

This analysis suggests that dissipative quantum chaos manifests also with smaller resonator's decay rates. 
An open question is how quantum chaos and readout are qualitatively and quantitatively modified when a third linear \cite{blais_circuit_2021} or nonlinear \cite{sunada_photon-noise-tolerant_2024} bosonic mode is added in the circuit QED setup.
This additional bosonic mode, the Purcell filter, is a routinely used strategy in circuit QED to maintain a relatively large cavity dissipation rate and induce a much smaller qubit's decoherence time.

\subsection{Readout in the large-detuning limit}

In Sec.~\ref{sec:quantum_information} we established that DQC arises in the parameter region where the transmon multiphoton resonance structure emerges.
Here, we show that increasing the detuning between the transmon and the readout cavity can be a promising way to avoid DQC.
Multiphoton resonances are still present up to $|\Delta|/2\pi=1.2$GHz (not shown). 
We consider here still larger values of $\Delta$, and investigate the hypothesis that a large positive detuning may offer a favorable tradeoff by eliminating the detrimental effects of chaos on the readout process. Let us consider $\Delta/2\pi=-2.5$GHz and correspondingly assume a significantly larger number of photons in the readout resonator.  
The critical photon number in this regime is $n_{\rm crit} \simeq 156$ against a computed cavity photon number $\langle\hat{a}^\dagger\hat{a}\rangle\simeq 60$.

In Fig.~\ref{fig:larger_delta} (a) we plot the dynamics of single quantum trajectories with the transmon qubit initialized in $\ket{0}$ or $\ket{1}$. 
The dynamics is now regular and the cavity state is well described by a coherent state.
Our results in this regime of large detuning lead to the same conclusions when still increasing the photon number while keeping it below $n_{\rm crit}$, and varying other system parameters.
Besides, we find that the steady-state photon number coincides with the Schrieffer-Wolff prediction in Eq.~\eqref{eqs:quasi_steady_states}.
In Fig.~\ref{fig:larger_delta} (b) we plot the instantaneous error probability $P_{\rm err}$ for three values of $E_C/g$. 
We notice that increasing the charging energy leads to decreasing $P_{\rm err}$.
After a transient, $P_{\rm err}$ converges to a constant value, again suggesting a completely regular dynamics.
The behavior of the instantaneous error probability is reproduced by the total error probability, that we report in Fig.~\ref{fig:larger_delta} (c).

These results suggest that the transmon readout with large values of detuning, can avoid both the occurrence of DQC and the detrimental effect of multiphoton resonances, at the cost of a larger occupation of the readout cavity, thereby constituting a very favorable tradeoff for improving the readout performance. 
Nonetheless, we point out that, in this regime with large occupation of the readout cavity, our model in the rotating-wave approximation may have a limited predictivity, and a more accurate description including the counter-rotating terms may be needed to confirm our prediction.

\newpage

\end{document}